\RequirePackage{rotating}
\documentclass[sigconf]{acmart}

\AtBeginDocument{%
  }

\setcopyright{acmcopyright}
\copyrightyear{2024}
\acmYear{2024}
\acmDOI{XXXXXXX.XXXXXXX}

\acmConference[Chinese CHI 2024]{Chinese CHI 2024}{Nov 22--25,
  2024}{Shenzhen, China}
\acmPrice{15.00}
\acmISBN{978-1-4503-XXXX-X/18/06}

\acmSubmissionID{23}

\usepackage{wrapfig}
\usepackage{xcolor}
\definecolor{verbal}{RGB}{164, 224, 224}
\definecolor{nonverbal}{RGB}{178, 220, 253}
\definecolor{visual}{RGB}{203, 188, 230}
\usepackage{paralist,bbding,pifont}
\usepackage{multirow}
\usepackage{multicol}
\usepackage{rotating}
\usepackage{adjustbox}
\usepackage{graphicx}
\usepackage{soul}
\usepackage{tikz}
\usepackage{ragged2e}
\usepackage{ulem}
\usetikzlibrary{calc}
\usetikzlibrary{decorations.pathmorphing}

\makeatletter

\newcommand{\defhighlighter}[3][]{%
  \tikzset{every highlighter/.style={color=#2, fill opacity=#3, #1}}%
}

\defhighlighter{yellow}{.5}

\newcommand{\highlight@DoHighlight}{
  \fill [ decoration = {random steps, amplitude=1pt, segment length=15pt}
        , outer sep = -15pt, inner sep = 0pt, decorate
        , every highlighter, this highlighter ]
        ($(begin highlight)+(0,8pt)$) rectangle ($(end highlight)+(0,-3pt)$) ;
}

\newcommand{\highlight@BeginHighlight}{
  \coordinate (begin highlight) at (0,0) ;
}

\newcommand{\highlight@EndHighlight}{
  \coordinate (end highlight) at (0,0) ;
}

\newdimen\highlight@previous
\newdimen\highlight@current

\DeclareRobustCommand*\highlight[1][]{%
  \tikzset{this highlighter/.style={#1}}%
  \SOUL@setup
  \def\SOUL@preamble{%
    \begin{tikzpicture}[overlay, remember picture]
      \highlight@BeginHighlight
      \highlight@EndHighlight
    \end{tikzpicture}%
  }%
  \def\SOUL@postamble{%
    \begin{tikzpicture}[overlay, remember picture]
      \highlight@EndHighlight
      \highlight@DoHighlight
    \end{tikzpicture}%
  }%
  \def\SOUL@everyhyphen{%
    \discretionary{%
      \SOUL@setkern\SOUL@hyphkern
      \SOUL@sethyphenchar
      \tikz[overlay, remember picture] \highlight@EndHighlight ;%
    }{%
    }{%
      \SOUL@setkern\SOUL@charkern
    }%
  }%
  \def\SOUL@everyexhyphen##1{%
    \SOUL@setkern\SOUL@hyphkern
    \hbox{##1}%
    \discretionary{%
      \tikz[overlay, remember picture] \highlight@EndHighlight ;%
    }{%
    }{%
      \SOUL@setkern\SOUL@charkern
    }%
  }%
  \def\SOUL@everysyllable{%
    \begin{tikzpicture}[overlay, remember picture]
      \path let \p0 = (begin highlight), \p1 = (0,0) in \pgfextra
        \global\highlight@previous=\y0
        \global\highlight@current =\y1
      \endpgfextra (0,0) ;
      \ifdim\highlight@current < \highlight@previous
        \highlight@DoHighlight
        \highlight@BeginHighlight
      \fi
    \end{tikzpicture}%
    \the\SOUL@syllable
    \tikz[overlay, remember picture] \highlight@EndHighlight ;%
  }%
  \SOUL@
}
\makeatother

\begin{document}
\title{Trinity: Synchronizing Verbal, Nonverbal, and Visual Channels to Support Academic Oral Presentation Delivery}

\author{Yuchen Wu}
\orcid{0009-0005-8333-4405}
\affiliation{%
  \institution{ShanghaiTech University}
  \city{Shanghai}
  \country{China}
}
\email{wuych3@shanghaitech.edu.cn}

\author{Shengxin Li}
\orcid{0009-0006-4117-3954}
\affiliation{%
  \institution{ShanghaiTech University}
  \city{Shanghai}
  \country{China}}
\email{lishx1@shanghaitech.edu.cn}

\author{Shizhen Zhang}
\affiliation{%
  \institution{ShanghaiTech University}
  \city{Shanghai}
  \country{China}
}
\email{zhangshzh@shanghaitech.edu.cn}

\author{Xingbo Wang}
\orcid{0000-0001-5693-1128}
\affiliation{%
 \institution{Weill Cornell Medicine, Cornell University}
 \city{New York}
 \state{New York}
 \country{USA}}
 \email{xiw4011@med.cornell.edu}

\author{Quan Li}
\authornote{The corresponding author.}
\orcid{0000-0003-2249-0728}
\affiliation{%
  \institution{ShanghaiTech University}
  \city{Shanghai}
  \country{China}
}
\email{liquan@shanghaitech.edu.cn}

\renewcommand{\shortauthors}{Wu et al.}

\newcommand{\yuchen}{\textcolor[RGB]{0, 150, 174}}
\newcommand{\shengxin}{\textcolor[RGB]{171, 71, 188}}
\newcommand{\xingbo}{\textcolor{orange}}
\newcommand{\prefix}{\textcolor[RGB]{0, 102, 204}}
\newcommand{\shizhen}{\textcolor[RGB]{85, 187, 138}}
\begin{abstract}
  Academic Oral Presentation (AOP) allows English-As-Foreign-Language (EFL) students to express ideas, engage in academic discourse, and present research findings. However, while previous efforts focus on training efficiency or speech assistance, EFL students often face the challenge of seamlessly integrating verbal, nonverbal, and visual elements into their presentations to avoid coming across as monotonous and unappealing. Based on a need-finding survey, a design study, and an expert interview, we introduce \textit{Trinity}, a hybrid mobile-centric delivery support system that provides guidance for multichannel delivery on-the-fly. On the desktop side, \textit{Trinity} facilitates script refinement and offers customizable delivery support based on large language models (LLMs). Based on the desktop configuration, \textit{Trinity} App enables a remote mobile visual control, multi-level speech pace modulation, and integrated delivery prompts for synchronized delivery. A controlled between-subject user study suggests that \textit{Trinity} effectively supports AOP delivery and is perceived as significantly more helpful than baselines, without excessive cognitive load.
\end{abstract}

\ccsdesc[500]{Human-centered computing~Interactive systems and tools}

\keywords{Multichannel Communication, Academic Oral Presentation, Delivery Support}


\begin{teaserfigure}
    \includegraphics[width=\textwidth]{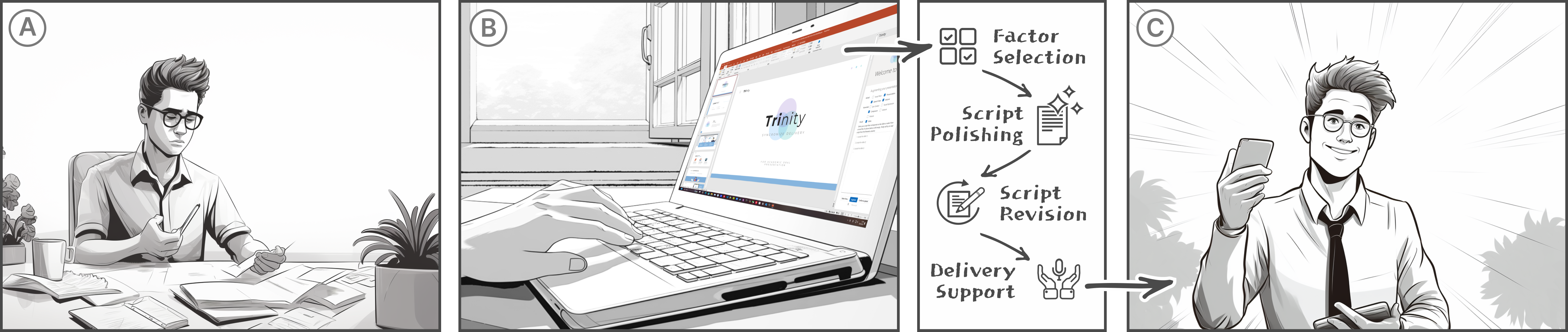}
    \vspace{-6mm}
    \caption{Example scenario of using \textit{Trinity}. A) The user, lacking comprehensive knowledge and sufficient experiences, struggles to find a effective way to deliver his presentation. B) Seeking assistance, the user opens the \textit{Trinity} PowerPoint add-in on their laptop. The user customizes the supportive features based on the specific needs, refines the presentation script, makes necessary revisions, and finally uploads the improved script to receive delivery support. C) The user confidently and smoothly delivers the presentation. \textit{Trinity} App provides features like remote mobile visual control, multilevel speech pace modulation, and integrated delivery prompts, ensuring synchronized multichannel delivery and enhancing the overall effectiveness of the presentation.}
    \label{fig:Teaser}
\end{teaserfigure}

\maketitle

\section{Introduction} 
\par Academic Oral Presentation (AOP) is frequently applied in universities to showcase students' understanding and mastery of a subject before an audience~\cite{kaur2018exploring}. Particularly prevalent in subjects within institutions of higher education, AOP holds significant importance as it allows students to effectively communicate their ideas to the audience~\cite{wan2016english}, assess and socialize them in genre‐specific academic discourse~\cite{duff2007second,morita2000discourse}, and present research findings to peers or academic communities~\cite{zareva2011and}. The omnipresence of AOPs in various courses and disciplines provides students with platforms to showcase their knowledge~\cite{levrai2015developing}, foster self-assurance~\cite{king2002preparing}, and cultivate indispensable skills for academic and future professional pursuits~\cite{alexander2008eap,zappa2007academic,morita2000discourse}.

\begin{table*}[]
\caption{Communication channels and delivery factors in presentation.}
\vspace{-3mm}
\label{tab:factors}
\resizebox{\textwidth}{!}{%
\begin{tabular}{ccl}
\hline
\textbf{Channel} & \textbf{Factor} & \multicolumn{1}{c}{\textbf{Utility}}                                                  \\ \hline
\multirow{3}{*}{\highlight[verbal]{Verbal}}    & Vocal Pitch       & To place emphasis,convey emotions, demonstrate authority, or add vocal variety.~\cite{glasgow1952semantic,hincks2005measures}                           \\ \cline{2-3} 
                 & Speech Rate     & To capture attention, emphasize points, adjust to time, or match the content.~\cite{hincks2005measures,hoogterp2014your}         \\ \cline{2-3} 
                 & Volume          & To attract attention, emphasize points, adjust to environment, or match the content.~\cite{hoogterp2014your}  \\ \hline
\multirow{5}{*}{\highlight[nonverbal]{Nonverbal}} & Eye Contact       & To establish connection, convey message, project authority and confidence, facilitate engagement.~\cite{senju2009eye,beebe1976effects}           \\ \cline{2-3} 
                           & Facial Expression & To enhance message, engage audience, show credibility, or communicate attitudes and intentions.~\cite{wortwein2015multimodal} \\ \cline{2-3} 
                 & Composure       & To overcome nervousness, handle unexpected situations, and project authority.~\cite{amalia2020anxiety}       \\ \cline{2-3} 
                 & Gesture         & To enhance verbal message, engage audience, or show confidence and credibility.~\cite{wortwein2015multimodal,austin2014presentation}        \\ \cline{2-3} 
                 & Posture         & To show confidence, enhance voice and breathing, or engage audience and create rapport.~\cite{beebe1976effects, mehrabian1968relationship} \\ \hline
\highlight[visual]{Visual}                     & Slides            & To organize the content, communicate message clearly and persuasively, or attract attention.~\cite{duarte2008slide,carter2012designing} \\ \hline
\end{tabular}%
}
\vspace{-3mm}
\end{table*}

\par As outlined in \autoref{tab:factors}, previous literature has categorized the presentation delivery into three communication channels: \highlight[verbal]{verbal} (e.g., vocal pitch), \highlight[nonverbal]{nonverbal} (e.g., gesture), and \highlight[visual]{visual} (e.g., slides)~\cite{peeters2010standardized,hertz2015spotlight}.
As indicated by Hunyadi et al.~\cite{hunyadi2020temporal}, effective communication requires congruence among all paraverbal dimensions. Similarly, Hentz~\cite{hentz2006enhancing} argues that presenters should examine how different modes of communication interact to create meaning. However, many presenters, especially EFL students, lack comprehensive knowledge and awareness of utilizing diverse communication channels in AOPs. Aside from watching presentation examples, current supportive tools only provide presentation assistance in certain channels (e.g.,~\cite{wang2020voicecoach,zeng2022gesturelens}). They lack the tools to understand and effectively utilize diverse communication channels for delivering AOPs. As a result, EFL students still encounter various obstacles when dealing with diverse communication channels in AOPs to effectively communicate with the audience. 
\textbf{First, Absence of Certain Channels in Multichannel Delivery.} Effective presentations derive their potency from adequately utilizing various communication channels for delivery~\cite{dolan2017effective,cheung2008teaching}. However, due to inadequate preparation and unbalanced training, the act of EFL students delivering AOPs often regresses into a monotonous recitation or reading of content directly from slides or scripts, lacking engagement or dynamism~\cite{tareen2022investigating,imaniah2018students}. 
\textbf{Second, Inconsistencies and Discrepancies in Channel Integration.}
Despite endeavors to exploit multiple communication channels, disparities in the execution of these channels by EFL students emerge~\cite{de2014differential}, resulting in damages to the audience's perceptions of their attitudes~\cite{argyle1971communication} and even undermining the credibility of their presentation content~\cite{gillis2017consistency}. For instance, an EFL student may highlight a point on a slide without adjusting their tone of voice and maintain an inappropriate hand gesture with crossed hands. 
\textbf{Third, Linguistic Hurdles for EFL Students.} Language proficiency is another factor that adds to the difficulties faced by EFL students during oral presentations~\cite{whai2015causes}. Specifically, having a limited vocabulary and struggling with sentence structure can hinder comprehension and make the presentation confusing, even when the topic is engaging and effectively delivered~\cite{tareen2022investigating}. 

\par Our goal is to offer guidance for synchronizing multiple communication channels during presentations. However, offering such guidance encounters three significant research gaps:
\textbf{G1. Empirical Gaps in Presentation Delivery Factors.} 
The existing literature has proposed and examined various factors that impact presentation delivery based on communication theories~\cite{giles2007communication,dennis1998testing,brehmer1976social}. In the field of Human-Computer Interaction (HCI), several works aim to offer effective training or learning experience to improve presentation delivery~\cite{wang2020voicecoach,trinh2014pitchperfect,oh2020scriptfree}. However, there remains a notable lack of empirical investigation into the factors that EFL students need the most support for and the factors that most affect the perception of AOPs. 
\textbf{G2. Limitations of Predefined Design.} 
Existing works often directly provide delivery guidance for a predetermined channel in a pre-defined design, leaving presenters with limited control over what prompts are presented. This lack of adaptability hinders the accommodation of presenters' diverse needs regarding delivery support. Furthermore, audience's perceptions and concerns towards these tools were not comprehensively taken into account from the start of system design in previous works. 
\textbf{G3. Unexplored Synchronization Approach.} 
Previous studies~\cite{tanveer2015rhema,asadi2017intelliprompter,tanveer2016automanner} have primarily focused on enhancing presentation delivery from the perspective of certain channels (e.g., verbal delivery). However, how to synchronize diverse communication channels to facilitate presentation delivery effectively and appropriately remains unexplored.

\par Therefore, our study aims to investigate the following five research questions (\textbf{RQ1-RQ5}). First, to investigate EFL students' actual needs and preferences and the suitable approach of providing such support during AOPs, we have formulated two fundamental research questions: \textbf{RQ1:} \textit{What factors do EFL students need the most support for and What factors most affect the perception of AOPs (\textbf{G1})?} and \textbf{RQ2:} \textit{What \textbf{methods} would be deemed acceptable by both EFL students and audience to provide on-the-fly delivery support and effectively synchronize diverse communication channels (\textbf{G2, G3})?}
To answer \textbf{RQ1}, we designed and launched a survey with $49$ EFL students and $36$ instructors from different majors in a local university. 
To address \textbf{RQ2}, we conducted a design study after identifying crucial AOP delivery factors and their ranking through a literature review and a survey. In the design study, we initially employed these factors to facilitate an exploratory design brainstorming session. Subsequently, we utilized storyboards to visually represent the design concepts generated from the brainstorming process. To further explore more detailed designs, we organized a design workshop to elaborate on the supportive approach. Additionally, to obtain professional perspectives on the output results and findings, we conducted interviews with five presentation experts. Moreover, during these interviews, we also performed a pilot test on potential design implementations. Considering LLMs have harvested rich knowledge from large-scale diverse corpus (e.g., website, books) and have demonstrated promising capabilities in understanding, generating, and expressing human languages~\cite{xu2023does}, our experts were asked to evaluate the applicability of LLMs as an implementation choice.
Incorporating feedback from instructors, EFL students, and experts, we identified six design goals and proposed \textit{Trinity}, a hybrid mobile-centric system. On the PC end, \textit{Trinity} refines scripts and provides customizable delivery prompts using GPT-4. Based on the PC configuration, \textit{Trinity} App enables remote visual control, speech pace modulation, and integrated delivery prompts for synchronized delivery.

\par With the proposed research prototype, we further investigate the following research questions: \textbf{RQ3:} \textit{How are the \textbf{usability and effectiveness} of the supportive system?} Due to the lack of evaluation regarding usage patterns and influences on students, we further investigate \textbf{RQ4:} \textit{How will EFL students \textbf{interact with and be influenced} by the system?} Since the system provides delivery support on-the-fly, we are interested in investigating \textbf{RQ5:} \textit{How will students \textbf{trust and collaborate} with our system?} 
To answer \textbf{RQ3-RQ5}, we conducted a controlled between-subject user study with $33$ EFL student presenters and $21$ audience members, comparing our proposed \textit{Trinity} with two baselines: 1) \textit{IntelliPrompter}, a PC-based system with features like speech tracking and dynamic script display, and 2) \textit{OfficeRemote}, a mobile office subsystem offering slide navigation, script reference, a timer, and a laser pointer. We used mixed-methods analyses, combining in-task and post-task questionnaires, and interviews. The results consistently showed that \textit{Trinity} effectively supports multichannel presentation delivery and is perceived as significantly more useful than the baselines.
The development and evaluation of \textit{Trinity} contribute to our understanding of AOP factors. It offers valuable insights into system design considerations, usage patterns, and its overall impact. The successes and challenges associated with the design choices made in \textit{Trinity} provide valuable guidance for the development of future presentation support tools within educational contexts and for enhancing public speaking tools beyond the classroom. In summary, this study makes four key contributions:
\begin{compactenum}[(1)]
    \item \textbf{Surveying EFL Students and Instructors:} We surveyed $49$ EFL students to understand their specific AOP delivery needs and $36$ instructors to identify factors significantly influencing the perception of AOP delivery.
    \item \textbf{Design Study:} We held a design study with $9$ EFL students and $4$ instructors, aiming to explore the design space and refine our approach iteratively. The focus was on providing on-the-fly delivery support and synchronizing diverse communication channels.
    \item \textbf{Trinity System:} We developed \textit{Trinity}, a comprehensive delivery support system designed to enhance academic oral presentation delivery based on a hybrid infrastructure.
    \item \textbf{User Study:} We conducted a controlled between-subject user study involving $33$ EFL students and $21$ audience to systematically investigate the usability, effectiveness, interaction, influence, trust, and collaboration of \textit{Trinity}, as well as trade-offs in design choices.
\end{compactenum}


\section{Related Work}

\subsection{Context and Challenges in AOPs for EFL Students}
\par AOP, a well-established educational strategy, employs various platforms, from traditional mediums like chalkboards to modern tools like presentation software~\cite{kaur2018exploring, paltridge2013handbook}. In the implementation of AOPs, slideware tools such as Microsoft PowerPoint, Google Slides, and Apple Keynote have become predominant. These tools allow students to create and display slides, offering real-time aids like markers, laser pointers, and timers~\cite{PowerPointFunc}. However, challenges persist, especially regarding the suitability of slide-based AOPs for EFL students.

\par The primary concern of EFL students is the lack of sufficient preparation~\cite{tareen2022investigating}. Despite their substantial effort, many EFL students are plagued by apprehensions about memory lapses and organizing ideas effectively, as thorough preparation is essential for effective oral presentations. Whai and Mei~\cite{whai2015causes} found that inadequate preparation is the main challenge in oral presentations, which may be due to students prioritizing core subjects such as engineering and commerce. As a result, they tend to assign less importance to oral presentation tasks because they perceive them as irrelevant to their field of study~\cite{whai2015causes}. Consequently, presentation preparation is compromised, negatively impacting delivery. Students often rely on notes or scripts for coherence~\cite{razawi2019anxiety}. However, this dependence leads EFL students to simply read from slides or scripts instead of engaging with the audience and maintaining eye contact~\cite{tareen2022investigating,imaniah2018students,el2011difficulties}. These challenges arise from EFL students' apprehension and vulnerability when speaking in front of a large audience~\cite{tareen2022investigating}. This study focuses on the issue of suboptimal presentation delivery among EFL students, where reading from slides or scripts occurs due to inadequate preparation and lack of effective learning approach. To address this, we propose an auxiliary method to enhance the AOP delivery of EFL students.

\subsection{Requisites for Effective Presentation Delivery}
\par Flawless delivery of a presentation is crucial for its success~\cite{chivers2007student}. Both verbal and nonverbal components play a vital role in the effectiveness of a presentation~\cite{strangert2008makes}. Skillful speakers seamlessly blend spoken content with nonverbal cues, including body language, gestures, and facial expressions. For example, Dolan~\cite{dolan2017effective} emphasized the centrality of nonverbal communication within presentations and provided insights into effective presentation skills. In addition to verbal and nonverbal dimensions, visual aid constitute an additional pivotal avenue for delivery. Peeters et al.~\cite{peeters2010standardized} delved into performance evaluation within an AOP course, culminating in the development of a standarized rubric that integrates visual aids into the evaluation schema for students' presentations. Furthermore, Bourne~\cite{bourne2007ten} developed a set of ten principles for proficient oral presentations, advocating for judicious yet impactful utilization of visual elements. Beyond the individual components, literature unveiled that inconsistencies among verbal, nonverbal, and visual communication channels can significantly affect audience perceptions concerning presenters' attitudes~\cite{argyle1971communication}, emotions~\cite{furnham1981relative}, and credibility~\cite{gillis2017consistency}. These explorations highlight the crucial importance of aligning verbal, nonverbal, and visual elements throughout a presentation.

\par Nevertheless, most prior supportive systems have focused mainly on specific elements of presentation delivery, such as volume and speech tempo. This narrow focus has overlooked the overall cohesion and synchronicity among diverse communication channels. Therefore, in this study, we aim to coordinate verbal, nonverbal, and visual channels to enhance presentation delivery effectiveness.

\subsection{Supportive Tools for Enhanced Presentation Delivery}

\begin{figure}[h]
 \vspace{-3mm}
 \centering 
 \includegraphics[width=\linewidth]{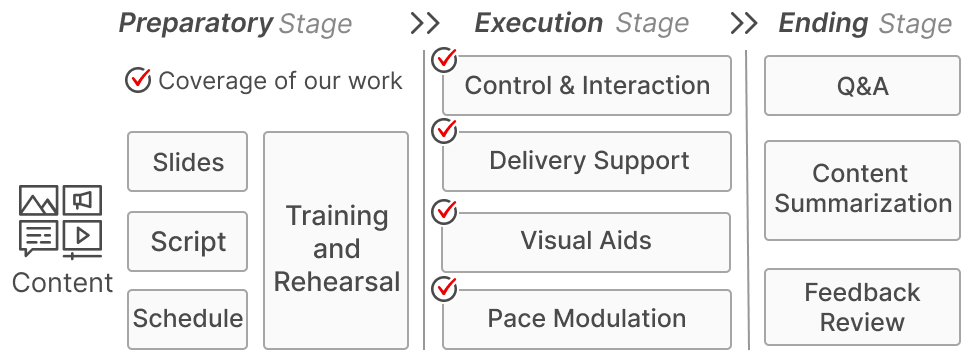}
 \vspace{-6mm}
 \caption{Taxonomy of existing works in presentation domain and coverage of our work.}
 \label{fig:taxonomy}
  \vspace{-3mm}
\end{figure}

\par Numerous systems and tools have been developed to improve presentation performance across various stages, including preparatory, execution, and ending stages. We detail the taxonomy of existing works and our scope in~\autoref{fig:taxonomy}. Careful planning, efficient time management, effective communication skills, and smart technology use, crucial for successful presentations~\cite{chivers2007student}, are primarily influenced by the preparation and execution stages. Hence, we discuss the related work for these stages.

\par During the preparation stage, one line of research has focused on supporting content formulation~\cite{khataei2014personalized,jokela2008mobile,roels2014mindxpres,bubel2016awareme,wang2023slide4n,chi2014demowiz,pschetz2014turningpoint,edge2016slidespace,edge2013hyperslides}, such as slide generation and script creation. For instance, Khataei et al.~\cite{khataei2014personalized} and Jokela et al.~\cite{jokela2008mobile} introduced tools for sharing personalized narratives and creating mobile audio-visual presentations, respectively. Another line of work focuses on presentation skills training and rehearsal~\cite{yi2020presentationtrainer, kurihara2007presentation, trinh2014pitchperfect, wang2020voicecoach, oh2020scriptfree, zeng2022gesturelens}. However, EFL students often struggle with dedicating adequate time for thorough rehearsal, prompting our exploration of on-the-fly support during execution.

\par In the execution stage, existing efforts offer on-the-fly support in control and interaction, speech assistance, visual augmentation, and pace modulation. Liu et al.~\cite{liu2021igscript} proposed \textit{IGScript}, an interaction grammar framework for scientific data presentations, while Cao et al.~\cite{cao2005evaluation} and others explored gesture-based presentation control~\cite{ackovska2013gesture,ram2012remotely}. Some studies focused on enhancing visual aids, such as Kim et al.'s \textit{HoloBox}~\cite{kim2018holobox}, which spatially integrated presenters, and Thanyadit et al.'s mixed reality avatar presentation tools~\cite{thanyadit2023tutor}. Trinh et al.~\cite{trinh2015dynamicduo} introduced \textit{DynamicDuo}, using virtual agents to assist inexperienced speakers. However, these works either require specialized, expensive media or use inflexible visual augmentations lacking adaptivity.

\par Our study is more similar to works in the delivery support domain, which focus on on-the-fly assistant for improved presentation delivery. Tanveer et al.~\cite{tanveer2015rhema} introduced \textit{Rhema} that provides real-time speech pace and volume cues. Microsoft's \textit{Office Remote}~\footnote{https://support.microsoft.com/en-us/topic/office-remote-for-pc-7e3d9342-61c7-4bc2-8bc0-fa47542d1bce} facilitates mobile slide navigation and script referencing for PowerPoint. Asadi et al.~\cite{asadi2017intelliprompter} proposed \textit{IntelliPrompter}, a system that tracks presenter's slide content coverage and dynamically adapts the script display to highlight the next relevant topic, improving presentation coherence.

\par While existing efforts provide foundational contributions, they focus on a limited set of delivery factors within predefined channels, such as loudness and speech rate~\cite{tanveer2015rhema}, without exploring the integration and synchronization of all three communication channels. Our work, shown in \autoref{fig:taxonomy}, covers key aspects of the execution stage and synchronizes three communication channels. We explored using LLMs to optimize linguistic expression, followed a user-centered design approach considering individual preferences and audience perceptions, and conducted a comprehensive user study to evaluate the system's usability, effectiveness, usage patterns, and influence.

\section{Formative Study}
\par Previous research have shown the benefits of on-the-fly support for presentation delivery~\cite{tanveer2015rhema, asadi2017intelliprompter}. However, existing designs face issues like limited audience engagement and fixed delivery factors. The formative study addresses: \textbf{RQ1}, exploring EFL student needs and audience perceptions through surveys, and \textbf{RQ2}, investigating student preferences and audience concerns regarding on-the-fly delivery support and channel synchronization via a design study. We included instructors as audience representatives considering their extensive experience in assessing presentations.

\begin{figure*}
  \centering
  \includegraphics[width=\linewidth]{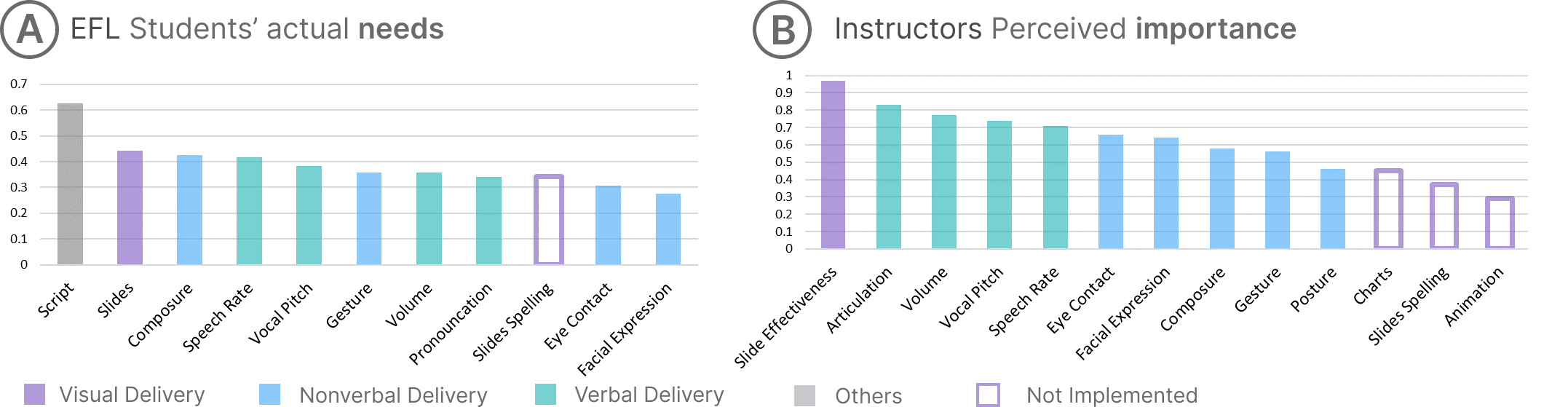}
    \vspace{-6mm}
  \caption{EFL student's needs and perceived importance from instructors based on a normalized weighted importance score (from $0$ to $1$, the closer to $1$, the more important). The information without color filling represents the factor that is not implemented in the system. A) The ranking of students’ actual needs in factors influencing AOP delivery. B) The ranking of instructors’ perceived importance levels of delivery factors.}
  \vspace{-3mm}
  \label{fig:Ratings}
\end{figure*}

\subsection{Exploratory Survey}
\subsubsection{Method}

\par To explore \textbf{RQ1}, we first designed a survey to understand the AOP support needs of EFL students. 
From EFL students' perspective, we used a 3-point scale (\textit{In demand}, \textit{Somewhat need}, \textit{No need}) to assess the level of support required for each AOP delivery factor, considering the previously mentioned communication channels and delivery factors (\autoref{tab:factors}). The survey also collected demographic details, familiarity with AOPs, perceived AOP delivery challenges, and usage of supportive tools. Additionally, participants were invited to offer comments or suggestions on a delivery support system design.

\par Likewise, from the audience's perspective, we designed a survey for instructors to explore significant factors influencing presentation delivery perception, covering demographics, experience in overseeing AOPs, and perceptions of supportive systems during presentations. We assessed instructors' perceived importance of each factor using a 3-point scale (\textit{Very important}, \textit{Somewhat important}, and \textit{Not important}) and solicited their suggestions for a real-time AOP delivery support system. After securing Institutional Review Board (IRB) approval, the surveys were distributed at a local university using snowball sampling via email and social media.

\subsubsection{Results}
\par Here we present ratings gathered from both students and instructors for each factor across the three channels, reflecting their perceived levels of need and importance. Full and detailed survey results can be found in \autoref{appendix:Survey_detail}. Following~\cite{ma2022glancee}, we employed normalized weighted importance scores to rank the supportive and perceptual factors. The resulting rankings are depicted in \autoref{fig:Ratings}, showcasing the final score ranging from $0$ to $1$ in descending order. It's noteworthy that individual preferences vary due to differences in abilities and perceptions (\textbf{\textit{F1}}). Even for the top-ranking factor, some participants might still consider them unnecessary or unimportant. As a result, this study acknowledged the existence of diverse individual perspectives. Notably, we implemented the majority of the highly ranked factors, as indicated by the colored columns in \autoref{fig:Ratings}.

\subsection{Design Study}
\par
\begin{figure*}
  \centering
  \includegraphics[width=\linewidth]{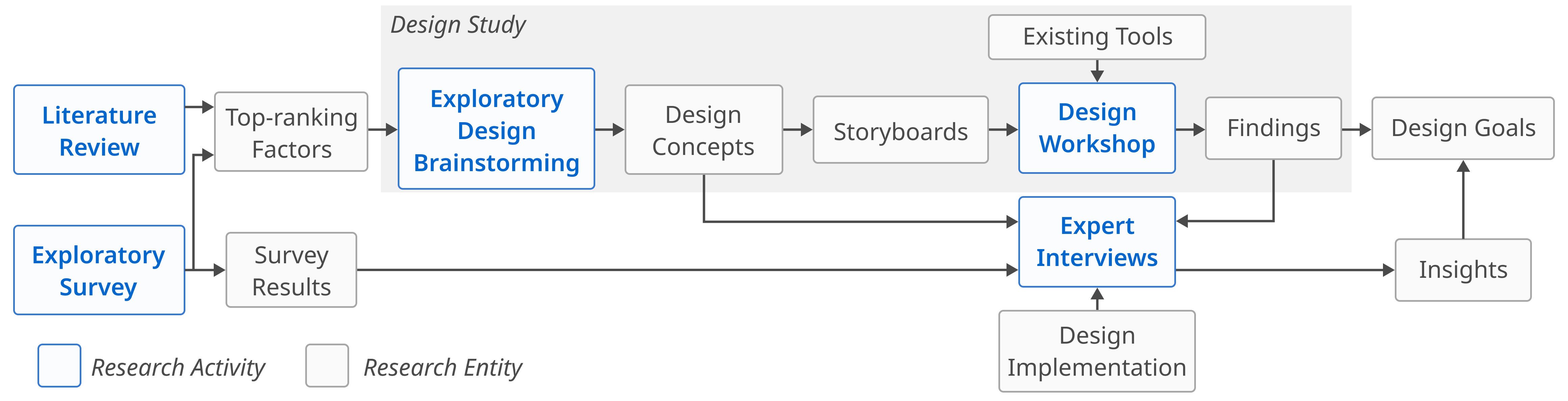}
  \vspace{-6mm}
  \caption{The design development process in the Formative Study, where the blue boxes are different research activities while the gray boxes are the research entities.}
  \label{fig:Design_Schema}
  \vspace{-3mm}
\end{figure*}

 To address \textbf{RQ2}, we conducted a design study to explore methods for synchronizing communication channels and providing on-the-fly support. With this purpose in mind, we elaborate on the design development process in this subsection following our schema in~\autoref{fig:Design_Schema}. According to the schema, we initially identified crucial factors and their ranking through a literature review and an exploratory survey. These factors served as the foundation for coordinating the exploratory design brainstorming. The brainstorming session produced design concepts as its output. Subsequently, we employed storyboards to visualize these design concepts. To explore more fine-grained designs, we organized a design workshop to elaborate on the supportive approach. In this workshop, both storyboards and designs of existing tools were utilized. Notably, both the design brainstorming and the design workshop were conducted through Zoom.

\subsubsection{Participants and Power Dynamics}
\par We recruited a group of $9$ EFL students, all with a background in interaction design (S1-9; $6$ male, $3$ female, average age $22$), by advertising within a local university's design course. Complementing this, we also recruited $4$ instructors (I1-I4; $3$ male, $1$ female, average age $38.6$) to serve as representatives of the audience. Notably, I1 and I2, among the instructors, were actively involved in teaching design courses at the local university's school of creativity and art. In our design study, our primary goal was to establish a balanced power dynamic between EFL students and instructors. We aimed to empower students to express their ideas and creativity while benefiting from the guidance of instructors with valuable experience and expertise. To achieve this equilibrium, we implemented several strategies. First, we encouraged both students and instructors to engage in open communication by asking questions, seeking clarification, and sharing their ideas. The initial step laid the groundwork for a collaborative environment. Building on this, we promoted peer feedback and assigned the role of facilitators to instructors, fostering a cooperative atmosphere throughout the design study. To address potential power imbalances, we considered Lukes's three dimensions of power~\cite{lukes2021power} and implemented intentional interventions:
1) \textit{Observable conflicts of interests:} Recognizing that conflicts might arise, particularly between students' focus on user experience and functionality and instructors' concerns about audience perception, we employed a role reversal approach~\cite{johnson1971role}. This involved asking students and instructors to switch roles (presenter and audience) until a consensus was reached. 2) \textit{Hidden omission of possible alternatives:} This is about choices and agendas are narrowed in favour of more powerful parties’ interests. In our scenario, acknowledging that certain design concepts might overshadow others, we diligently documented all proposed concepts. During the design study, we deliberately brought up any unmentioned concepts for discussion to ensure a comprehensive exploration of ideas. 3) \textit{Deceptive and indiscernible characteristics of power:} This is about a more powerful person's interests may be internalised and sought after as one's own. In our scenario, recognizing the risk that students might overly align with instructors' comments and concepts, we closely monitored the design development process. If the design appeared to predominantly converge in a certain direction, we reminded both students and instructors to maintain a diverse range of perspectives. Given that two instructors possessed extensive design experience, we specifically tasked them with promptly correcting the direction if any deviations were identified.

\subsubsection{Exploratory Design Brainstorming}
\par The goal of our brainstorming session is to gather a substantial number of design concepts addressing the challenges associated with multichannel AOP delivery for EFL students. To achieve this, we adhered to three fundamental principles for group brainstorming~\cite{wilson2013brainstorming}: 1) Aim for sheer quantity. Our focus was on generating a large number of concepts without any limitations. Participants were discouraged from taking personal notes during the brainstorming to ensure their full engagement in idea generation. Additionally, we requested that they refrain from checking emails or engaging in social media, minimizing distractions and fostering a more concentrated brainstorming environment. 2) Defer judgment about the quality of concepts. Participant were explicitly instructed not to criticize others' concepts either implicitly (through facial expressions or other nonverbal cues) or explicitly during the brainstorming session. 3) Encourage new and wild ideas. Participants were encouraged to propose concepts without considering feasibility, allowing for the inclusion of imaginative and unconventional ideas, even those reminiscent of science fiction or movies. During the brainstorming, we initiated the discussion by exploring common presentation delivery challenges for EFL students, drawing insights from survey results. Subsequently, participants were prompted to suggest design concepts for the top-ranking factors identified in the survey, with an explicit directive to withhold criticism and avoid restricting the type or number of concepts. Active documentation of all emerging design concepts occurred throughout the session. Following the brainstorming, both authors and participants collaboratively reviewed all proposed concepts, refining them into a subset deemed potentially applicable to the identified problem.

\subsubsection{Design Concepts and Storyboards}
\par After obtaining the coarse-grained design concepts, we proceeded to enhance and refine these ideas through a design workshop. To facilitate this exploration, we employed the use of storyboards, providing a visual representation to enhance the clarity and depth of the design concepts. Storyboards, chosen for their ability to visualize design concepts quickly~\cite{truong2006storyboarding}, were iteratively refined from broad design concepts to encapsulate different scale ideas. Our design workshop included $10$ storyboards (exemplified in \autoref{fig:Example_Board}B), each illustrating an AOP delivery problem, a potential solution, and hypothetical outcomes, and a summarized version is available in \autoref{fig:Summary_board} in Appendix.

\subsubsection{Design Workshop}
\par
\begin{figure*}
  \centering
  \includegraphics[width=0.95\linewidth]{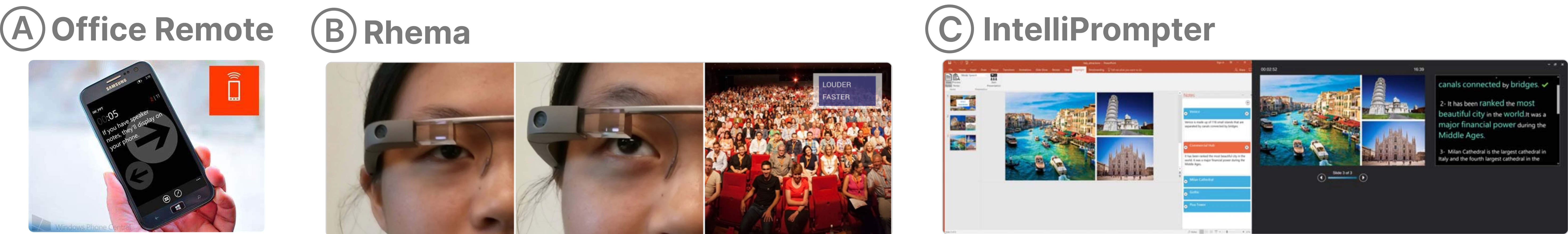}
  \vspace{-3mm}
  \caption{Designs of previous delivery support tools. A) \textit{Office Remote}, an Office subsystem that turns phone into a smart remote that interacts with Microsoft Office on PC. Picture adopted from~\cite{officeRemotePicture}. B) \textit{Rhema}, an intelligent user interface for Google Glass to help people with public speaking. Picture adopted from~\cite{tanveer2015rhema}. C) \textit{IntelliPrompter}, a speech-based note display system that automatically tracks a presenter’s coverage of each slide’s content. Picture adopted from~\cite{asadi2017intelliprompter}.}
\vspace{-3mm}
  \label{fig:Previous_Tools}
\end{figure*}

 During the design workshop, we initiated the session by reviewing the design concepts derived from the earlier brainstorming session. Following~\cite{chen2023meetscript}, to provide context, we showcased previous designs of delivery support tools, as depicted in \autoref{fig:Previous_Tools}. This approach aimed to offer participants a concrete understanding of how a delivery support system could be structured and function. The intention was to facilitate participants' comprehension of subsequent design concepts and, in turn, encourage their active contribution to design ideas during the ensuing sessions.

\par
\begin{figure*}
  \centering
  \includegraphics[width=\linewidth]{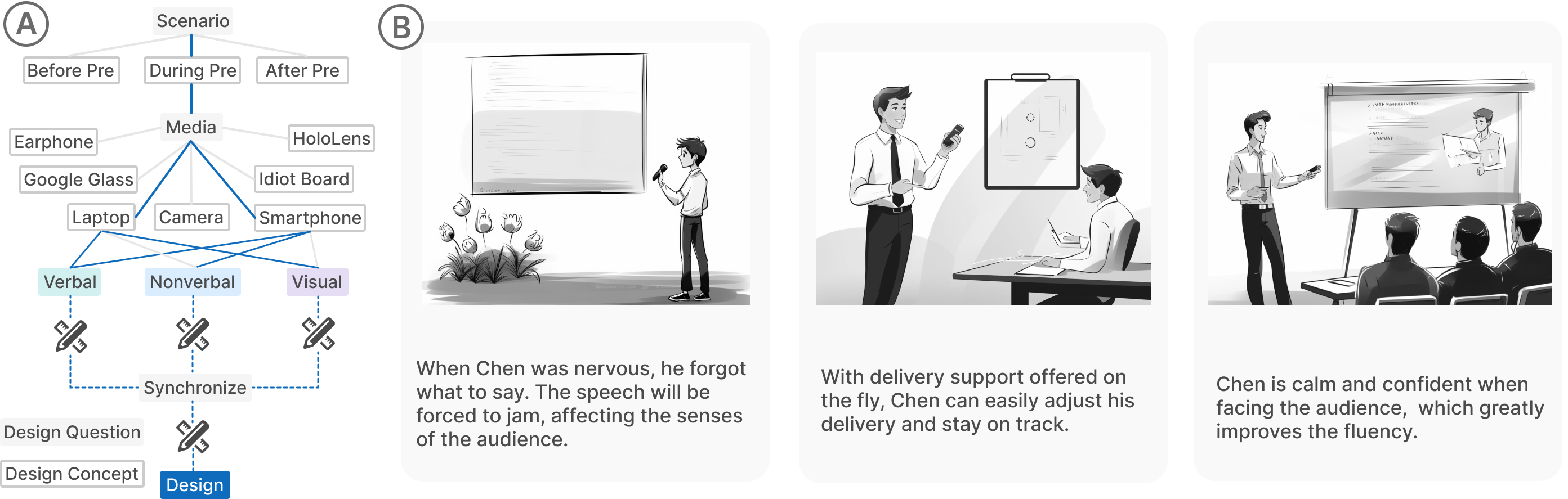}
  \vspace{-6mm}
  \caption{Flow diagram of design formation and Storyboard No.2. A) The design iteration started from storyboarding for scenario and media, then spanned to sketch out designs for verbal, nonverbal, and visual delivery support, ending up with a design for synchronizing diverse communication channels. B) Storyboard No.2: Offer delivery support on the fly. participants thought this would enhance their performance with reduced preparation effort.}
\vspace{-3mm}
  \label{fig:Example_Board}
\end{figure*}

 During the workshop, as illustrated in \autoref{fig:Example_Board}A, one of the authors employed screen sharing to present the storyboards in a randomized sequence. Each storyboard was accompanied by an introduction from the author, providing details about both the scenario and the proposed solution. This presentation format allowed participants ample time to read and comprehend the information presented. Following the introduction of each storyboard, participants were encouraged to assess the respective advantages and disadvantages of the proposed solutions. Their insights were solicited, specifically seeking feedback on how each solution could be effectively applied to the given scenario. Building on these insights, participants were then tasked with the challenge of sketching out a comprehensive design that addressed all relevant delivery factors. Emphasis was placed on encouraging participants to articulate how their designs intricately integrated multiple factors. Additionally, participants were prompted to iteratively refine their designs based on feedback received from their peers throughout the workshop. Throughout this collaborative process, the authors actively documented design insights that emerged from the participants' contributions, providing valuable, unintended insights into the design process.

\par The speed-dating interviews, averaging around $62.4$ minutes, were conducted following IRB guidelines. Participants were compensated with $\$20$ USD. The interviews were transcribed and analyzed using an Affinity Diagram approach, with the authors collaboratively organizing quotes into themes, refining and structuring the data through iterative discussions.

\subsubsection{Findings}
\par Regarding the media aspect, participants favored using smartphones and laptops to improve their presentation delivery. For the approach aspect, they preferred inline prompts with scripts for delivery support. The study also highlighted participants' concerns about potential cognitive load challenges and managing different presentation genres.

\par \textbf{\textit{F2}: Employing on-the-fly delivery support facilitates multichannel communication but requires careful management of cognitive load.} All participants agreed that on-the-fly delivery support would enhance their AOPs performance. S6 noted, ``\textit{Some presentation topics can completely throw me off track... having access to such delivery support would truly have been a game-changer.}'' Participants also saw this support as a way to reduce performance anxiety, as S2 explained, ``\textit{At times, you might possess a deep understanding of the subject matter, but the pressure of the moment can lead to forgetting key details or losing track of progression. Having support to rely on would mitigate that pressure and elevate the overall presentation experience.}'' This aligns with research indicating immediate feedback improves presenter awareness and behavior adjustment~\cite{kelch1985modified,damian2015augmenting,tanveer2015rhema,bubel2016awareme,schneider2015presentation,parmar2020making,peng2021say}. However, I1 and I2 cautioned, ``\textit{while delivery support provides valuable guidance, excessive cognitive burden may in turn affect the intended delivery. Therefore, on-the-fly patterns always require a careful management of cognitive load.}''

\par \textbf{\textit{F3}: Smartphones and laptops emerge as the more feasible and fitting choices for delivery media.} 
S2, S4, and I2 found AR/VR devices like HoloLens unsuitable for delivery support due to weight and audience perceptions. Tripod-mounted cameras and idiot boards faced disapproval due to site constraints, but some saw potential in adapting their features. Smartphones and laptops were favored for AOP delivery for their accessibility and capabilities. S3 called it a "\textit{low learning cost}" method, and I1 found it "\textit{visually unintrusive}." S5 envisioned integrating teleprompter features into smartphones, comparing it to "\textit{holding cue cards like hosts do.}"

\par \textbf{\textit{F4}: In-line delivery prompts may enhance my perception of support.} 
In the workshop, students favored designs with in-line delivery prompts over separate sections, citing prompt recognition as the main reason. S1 noted that "\textit{integrated prompts help me scan the script seamlessly, preventing prompt omission}" (S1). Interest was also expressed in how in-line prompts could aid script refinement. I2 suggested that "\textit{in-line prompts and script refinement suggestions could encourage presenters to improve their content, thereby enhancing presentation quality}".

\begin{figure*}[h]
\vspace{-3mm}
  \centering
  \includegraphics[width=\linewidth]{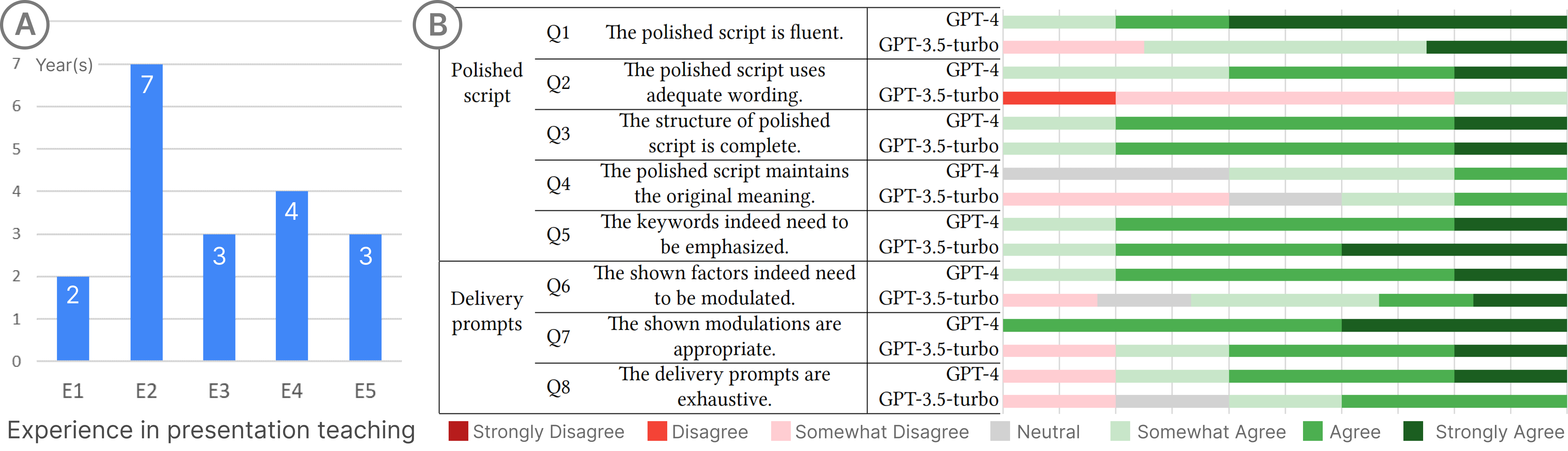}
  \vspace{-6mm}
  \caption{A) Experience of experts in presentation teaching. B) Questions (Q1 - Q8) and experts' feedback on the quality of output script and delivery prompts for GPT-4 and GPT-3.5-turbo.}
\vspace{-3mm}
  \label{fig:Expert_Interview}
\end{figure*}

\par \textbf{\textit{F5}: Polished scripts may have a double-edged sword effect.} 
Instructors (I1, I3, I4) highlighted concerns about students' AOPs, citing excessive use of slang and improper language. They acknowledged the benefits of script refinement for presentations, such as improved fluency and syntax. However, I1 pointed out a potential issue: "\textit{We have certain speech styles of our own; however, a refined script may `standardize' our speech style, making us feel awkward to use them.}" Additionally, students like S9 (male, age: 22) worried about the challenges of understanding and pronouncing unfamiliar words in polished scripts, saying, "\textit{While the idea of having a polished script is appealing, what if it includes challenging or unfamiliar words?}"

\par \textbf{\textit{F6}: Enhanced speech progress monitoring and hints could assist me in maintaining better pacing.} 
Participants (S2, S4, S5, S7, S8, I3, and I4) frequently recognized time management as a common issue in AOPs, especially for inexperienced presenters. The idea of features to track delivery progress and provide real-time timing hints was widely supported for its practicality.
S7 highlighted the potential benefits, stating, ``\textit{Such hints would encourage me to focus on the overall pace of delivery, which is crucial when adhering to strict time limits.}''

\par \textbf{\textit{F7}: Intuitive prompts may enhance my comprehension speed.} 
All the participants unanimously agreed on the importance of quick response for fluent AOP and emphasized reducing cognitive recognition demands in design. S2, S8, and I2 preferred \textit{Rhema}'s design approach, using large font sizes and distinct color blocks for intuitiveness.
S8 shared, ``\textit{Initially, my design for verbal factors resembled a fountain-like graph. However, feedback from others highlighted its lack of user-friendliness. Subsequently, I developed a sketch incorporating icons to signify delivery prompts, which proved to be a much more effective approach.}''

\par \textbf{\textit{F8}: Enhanced control over visuals on smartphone may facilitate my delivery.} 
S3, S4, and I2 were concerned about the lack of dynamism in AOP delivery and the challenge of maintaining engagement. They recognized the value of nonverbal elements like gestures and movement but faced difficulties synchronizing them with the presentation due to the need to stay near the laptop. S4 proposed a solution where "the tool can automatically add pictures, charts, and animations into the slides while adjusting them based on your speech," which was criticized for the potential loss of control. S5 countered with the idea of using smartphones for visual control, acting as "a handheld teleprompter providing both \highlight[verbal]{verbal} and \highlight[nonverbal]{nonverbal} delivery prompts... allowing free movement, gesture, and audience interaction. A simple click on the phone would seamlessly control the \highlight[visual]{visuals} without concerns regarding smartphone weight or size."

\begin{figure*}[h]
  \centering
  \includegraphics[width=\linewidth]{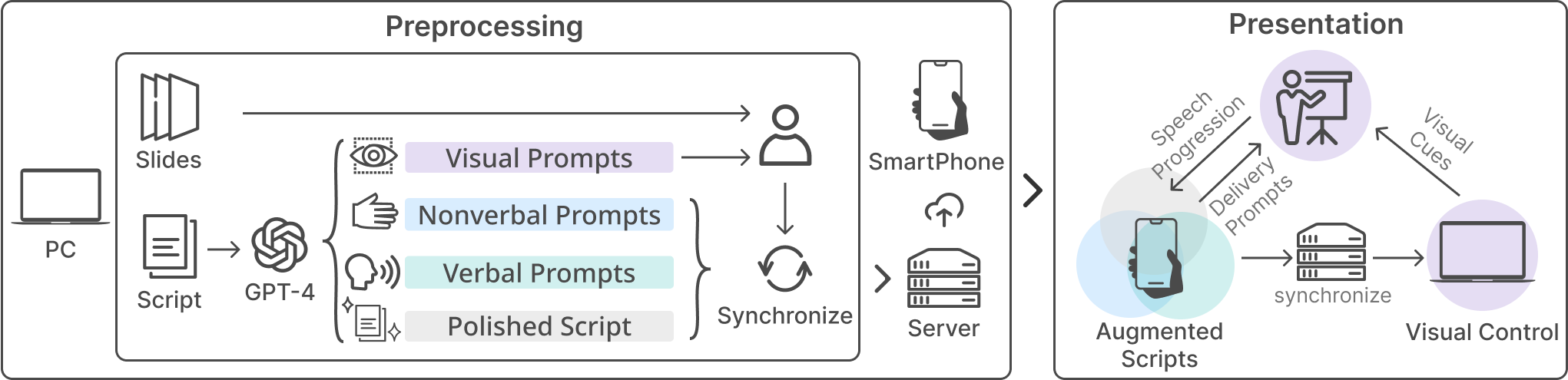}
  \vspace{-6mm}
  \caption{Overview workflow of \textit{Trinity}, including a Preprocessing phase and a Presentation phase.}
\vspace{-3mm}
  \label{fig:Approach_Overview}
\end{figure*}

\subsection{Expert Interview}
\par We conducted semi-structured interviews with five experts in presentation instruction (E1-5), and their teaching experiences are detailed in \autoref{fig:Expert_Interview}A. A comprehensive version of the expert interview is available in \autoref{appendix:Interview_detail}. Our expert interviews had two primary objectives. First, we aimed to gather their professional perspectives on the outcomes of our initial exploratory surveys and design study, thereby enhancing our understanding of the potential applications and implications of the technology. Second, we sought to conduct a pilot test on potential design implementations. since LLMs, like ChatGPT, store diverse and vast knowledge about different areas and are capable of understanding, generating, and expressing human languages flexibly, we were particularly interested in exploring the feasibility of using ChatGPT for refining presentation scripts and providing delivery prompts. We evaluated two LLMs, GPT-3.5-turbo and GPT-4 (both March 14 versions), using $10$ EFL student presentation scripts collected from social networks. We adopted a 7-point Likert scale to assess the quality of the LLMs' output scripts and delivery prompts, and open-ended questions to capture insights from the experts' experiences and perceptions about our survey and design outcomes (\autoref{fig:Expert_Interview}B). We present primary insights obtained from the interviews. 
\par \textbf{\textit{Insight1}: GPT-4 outperformed GPT-3.5-turbo on all metrics, suggesting potential for integration into practice.} GPT-3.5-turbo's complex language and sentence structure were criticized. Both LLMs were favorably rated for fluency and meaning preservation, with GPT-4 slightly leading. The LLMs were praised for enhancing script clarity, coherence, and organization. The in-line delivery prompts by the GPT-4 were also positively rated. Experts generally preferred GPT-4 for factor judgment and modulation during presentations.
\par \textbf{\textit{Insight2}: Incorporating visual cues like arrows or highlights to guide students' focus.} Experts also proposed design suggestions for an LLM-based tool to aid EFL students in real-time presentations, such as integrating multimodal feedback. E1 suggested using auditory cues like beeps or pitches to adjust speech volume or speed, and somatosensory cues like vibrations for emphasizing points. E3 proposed incorporating visual cues like arrows or highlights to guide students' focus on relevant slides or audience members.
\par \textbf{\textit{Insight3}: Pronunciation of unfamiliar or polysyllabic words can pose challenges for EFL students.} E4 mentioned that EFL students often struggle more with verbal communication due to limited opportunities to enhance these skills in academic or professional contexts. E5 added that pronunciation of unfamiliar or polysyllabic words can pose significant challenges for EFL students during presentations.

\subsection{Design Goals}
\par Based on the findings and insights derived, we summarized the following design goals for a novel delivery support system. In this system, the smartphone takes on the role of an augmented prompter, while the laptop functions as the responder. The aim is to synchronize diverse communication channels:
\par \textbf{[D1] Customizable Delivery Factors (\textit{F1})}: Users should have the ability to personalize delivery factors based on their individual preferences and needs.
\par \textbf{[D2] Moderate Script Polishing (\textit{F5})}: The system should enhance the fluency, wording, and sentence structure of users' presentation script while maintaining a balanced level of embellishment.
\par \textbf{[D3] Remote Mobile Visual Control (\textit{F8})}: Users need to be empowered with remote slide control capabilities on their smartphones.
\par \textbf{[D4] Flexible Speech Pace Modulation (\textit{F6})}: The system should allow for multilevel modulation of speech pace to accommodate diverse presentation scenarios.
\par \textbf{[D5] Integrated Delivery Prompts (\textit{F4})}: Essential prompts for both verbal and nonverbal channels, along with corresponding scripts, should be presented to users to minimize cognitive load.
\par \textbf{[D6] Rapid Pronunciation Acquisition (\textit{Insight3})}: The system should aid users in quickly mastering the pronunciation of unfamiliar and polysyllabic words.

\section{Trinity}

\subsection{Approach Overview}
\par Based on the design goals derived from our formative study, we propose \textit{Trinity}, an interactive system that synchronizes verbal, nonverbal, and visual channels to improve users' AOP delivery on the fly. The workflow of using \textit{Trinity} consists of two distinct phases, namely the \textbf{Preprocessing} phase and the \textbf{Presentation} phase. This system comprises three integral components: a smartphone-based prompter, a central server, and a PC-end interface (\textbf{\textit{F3}}) (\autoref{fig:Approach_Overview}).

\par \textbf{1) In the Preprocessing phase}, the PC end uses GPT-4 to process the script, which can be provided separately or written in the remark column. The enhanced script with in-line textual prompts for all three communication channels is generated and sent to the server. Users then add or adjust visual cues on PowerPoint according to the provided prompts. The final package of slides, refined script, and their mappings is uploaded to the smartphone via the server.
\par \textbf{2) In the Presentation phase}, the smartphone-based prompter tracks the speaker's progress and presents the script with \highlight[verbal]{verbal} and \highlight[nonverbal]{nonverbal} prompts for real-time guidance. Users can remotely control slides via the smartphone, with synchronization data transmitted to the PC end through the server, enabling manipulation of slides and \highlight[visual]{visual} cues.

\begin{figure*}
  \centering
  \includegraphics[width=\linewidth]{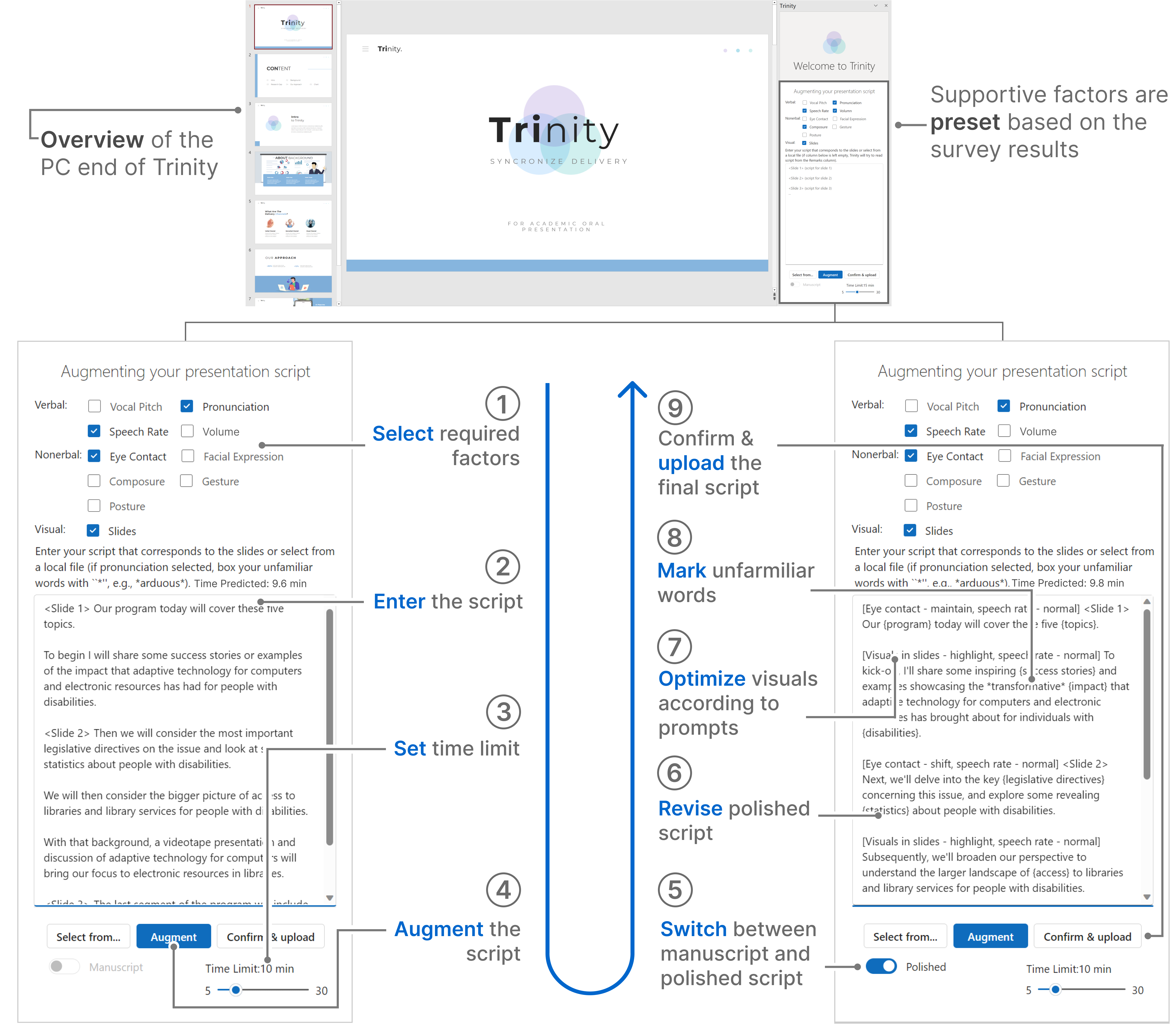}
  \vspace{-6mm}
  \caption{PC end interface of \textit{Trinity}. A complete workflow on PC end encompasses $9$ steps, users 1) select required delivery factors on the checkboxes, 2) enter the script in the input field, 3) set time limit, 4) augment the script, 5) switch between the manuscript and polished script, 6) perform necessary revisions, 7) optimize visuals in the slides according to prompts, 8) mark unfamiliar words, and 9) confirm and upload the final script.}
\vspace{-3mm}
  \label{fig:PC_Design}
\end{figure*}

\subsection{Interface Design}
\subsubsection{Presentation Script Augmentation}
\par The PC interface of \textit{Trinity}, an add-in for PowerPoint, caters to delivery support prerequisites and integrates seamlessly as a sidebar (\autoref{fig:PC_Design}). This add-in, comprising \textit{factor checkboxes}, \textit{script input field}, and \textit{operational buttons}, provides comprehensive support.

\par \textbf{[D1] Customizable Delivery Factors.} Recognizing diverse user preferences for delivery support, \textit{Trinity} enables users to customize their selection of delivery factors with a simple click on the corresponding checkboxes (\autoref{fig:PC_Design}-1). \textit{Trinity} also includes a recommended preset, derived from our extensive survey, for balanced and effective delivery support. It is worth noting that \textit{Trinity} will pop up a reminder when the user selects too many factors (e.g., over $5$), signaling the potential risk of overload (\textbf{\textit{F2}}). Moreover, the selected factors are not obligatory for prompt but rather intended for consideration solely in cases where modulation is deemed necessary.

\par \textbf{[D2] Moderate Script Polishing.} \textit{Trinity} uses GPT-4 (\textbf{\textit{Insight1}}) to refine presentation scripts based on expert interview insights. Users can either input scripts paragraph by paragraph or select a local file by clicking the \raisebox{-1.0ex}{\includegraphics[height=3.5ex]{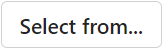}}(\autoref{fig:PC_Design}-2) button. If no input is given, \textit{Trinity} uses the script from the Remark column. Users can set a presentation time limit using the slider \raisebox{-0.5ex}{\includegraphics[height=2.2ex]{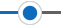}} (\autoref{fig:PC_Design}-3) and refine the script by clicking the \raisebox{-1.0ex}{\includegraphics[height=3.5ex]{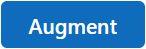}} button (\autoref{fig:PC_Design}-4). The server returns a refined script with in-line prompts, with GPT-4 ensuring no overuse of embellishments or excessive expansion. Being aware of potential issues arising from GPT (e.g., odd words and unfamiliar expressions), we grant users full autonomy to adjust the refined scripts in accordance with their personal preferences and habits. Users can toggle between the original and refined scripts for comparison (\autoref{fig:PC_Design}-5), make revisions (\autoref{fig:PC_Design}-6), optimize slide visuals based on prompts (\autoref{fig:PC_Design}-7), label unfamiliar words (\autoref{fig:PC_Design}-8), and upload the revised script by clicking the \raisebox{-1.0ex}{\includegraphics[height=3.5ex]{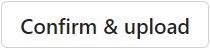}} button (\autoref{fig:PC_Design}-9) for transfer to their smartphones.

\begin{figure*}[h]
  \centering
  \includegraphics[width=\linewidth]{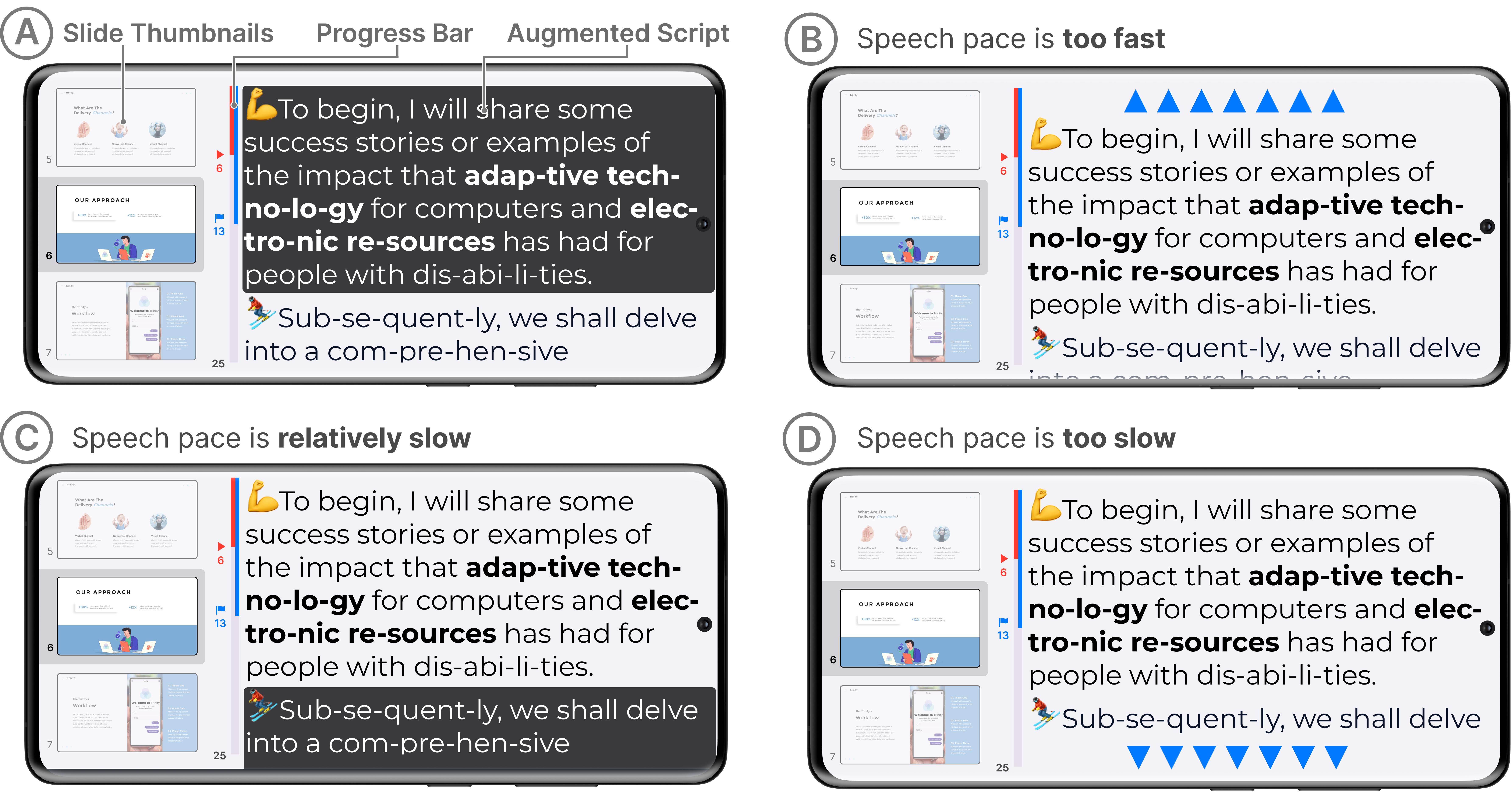}
  \vspace{-6mm}
  \caption{Interface design of \textit{Trinity} App. A) Interface of \textit{Trinity} App consists primarily of three components: \textit{slide thumbnails}, \textit{progress bars,} and the \textit{augmented script}. B) Upward triangle group indicates that speech pace is too fast. C) Underpainting fades out as speech pace is relatively slow. D) Downward triangle group indicates that speech pace is too slow. }
\vspace{-3mm}
  \label{fig:Smartphone}
\end{figure*}

\subsubsection{On-the-fly Delivery Support}
\par The smartphone interface of \textit{Trinity} integrates key delivery support features into an App, including \textit{slide thumbnails}, \textit{progress bars}, and the \textit{augmented script}. Notably, the interface can be adjusted for users' handedness, as illustrated by the left-handed orientation in \autoref{fig:Smartphone}.


\par
\begin{figure}[h]
 \centering 
 \includegraphics[width=\linewidth]{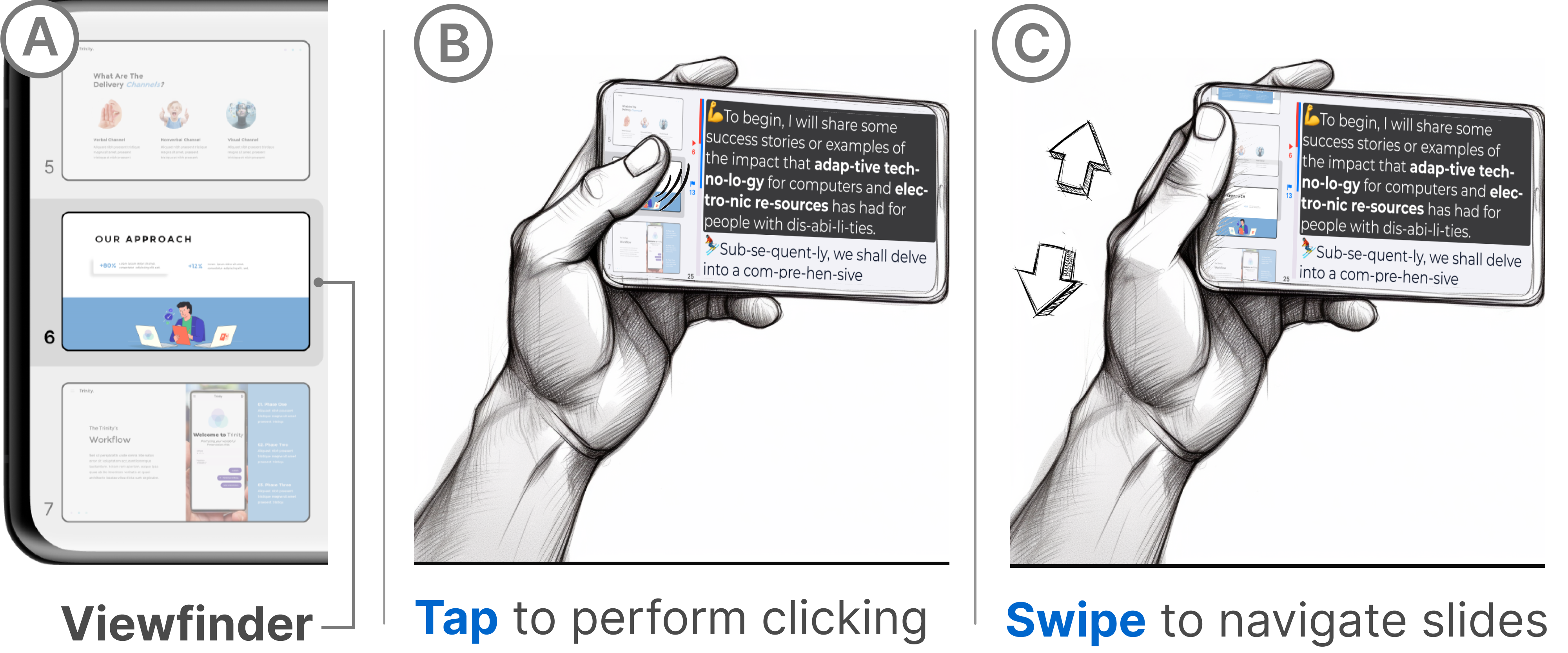}
 \vspace{-6mm}
 \caption{Interactions for \textit{Trinity} App. A) The viewfinder appears as a static gray box to capture the slides intended for display. B) Tapping in the viewfinder to perform clicking. C) Swiping the slide thumbnails to promptly switch among slides.}
 \label{fig:Interaction}
  \vspace{-3mm}
\end{figure}

 \textbf{[D3] Remote Mobile Visual Control.} The smartphone interface of \textit{Trinity} allows filmstrip-style interactions for \highlight[visual]{visual} control. The central static gray box \raisebox{-1.0ex}{\includegraphics[height=3.5ex]{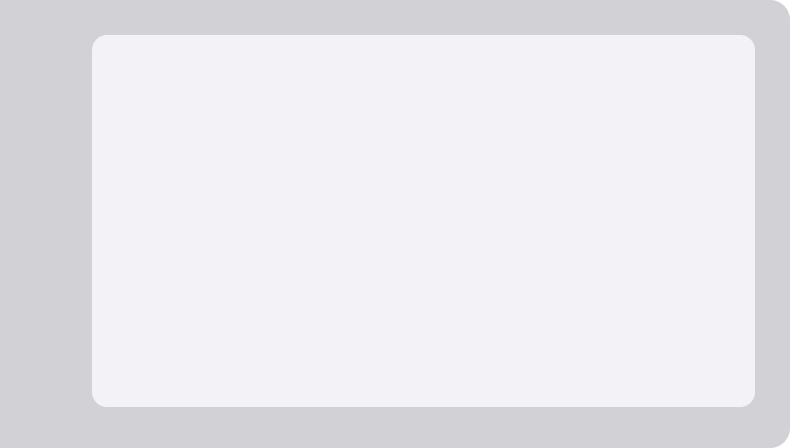}} (\autoref{fig:Interaction}A) serves as a viewfinder to capture slides , and a simple tap (\autoref{fig:Interaction}B) triggers a click operation at the PC end. Users can switch slides by swiping up and down on the thumbnails (\autoref{fig:Interaction}C). To reduce cognitive load, only the most relevant thumbnails are shown: the previous, current, and next slide. The thumbnails of the non-current slides are slightly transparent to maintain user focus.

\par \textbf{[D4] Flexible Speech Pace Modulation.} \textit{Trinity} employs a dual-layered approach to regulate speech pace, addressing both broad and fine-grained aspects. Central progress bars \raisebox{-1.0ex}{\includegraphics[height=3.5ex]{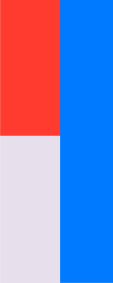}} (\autoref{fig:Smartphone}A) illustrate the overall pace, while the augmented script's underpainting \raisebox{-0.5ex}{\includegraphics[height=2.2ex]{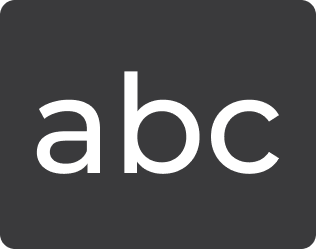}} provides detailed pace control. The left (red) and right (blue) progress bars show current and ideal pace respectively, allowing users to quickly gauge their pace. The underpainting highlights the current sentence for reading at the optimal pace. Fast and slow paces are visualized with upward and downward triangle groupings (\textbf{\textit{Insight2}}) in the underpainting (\autoref{fig:Smartphone}B\includegraphics[height=1.5ex]{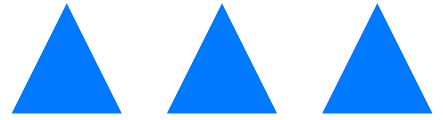}, D\includegraphics[height=1.5ex]{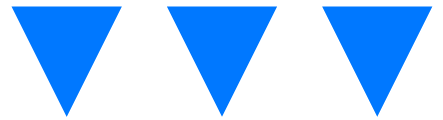}). Moreover, \textit{Trinity} synchronizes the speech with the script displayed at the screen's upper part. If the pace is slightly slow, the underpainting gradually fades (\autoref{fig:Smartphone}C).

\begin{table*}[h]
\caption{Emoji mapping for delivery prompts.}
\label{tab:Emoji}
\vspace{-3mm}
\resizebox{\textwidth}{!}{%
\begin{tabular}{|ccc|ccc|ccc|ccc|}
\hline
 Factor & 
 Modulation & 
 Prompt & 
 Factor &
 Modulation &
 Prompt &
 Factor &
 Modulation &
 Prompt &
 Factor &
 Modulation &
 Prompt \\ \hline
 \multirow{3}{*}{\begin{tabular}[c]{@{}c@{}}Vocal\\ pitch\end{tabular}} &
 High &
 \raisebox{-0.5ex}{\includegraphics[width=1em]{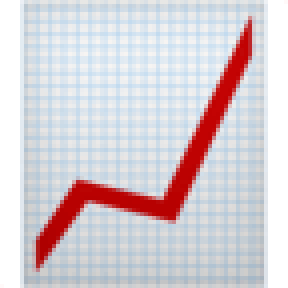}} &
 \multirow{6}{*}{\begin{tabular}[c]{@{}c@{}}Facial\\ expression\end{tabular}} &
 Happy &
 \raisebox{-0.5ex}{\includegraphics[width=1em]{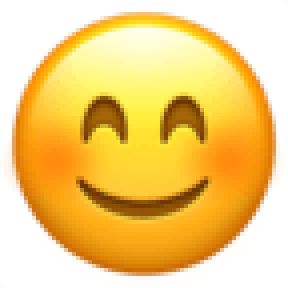}} &
 \multirow{3}{*}{Volume} &
 Loud &
 \raisebox{-0.5ex}{\includegraphics[width=1em]{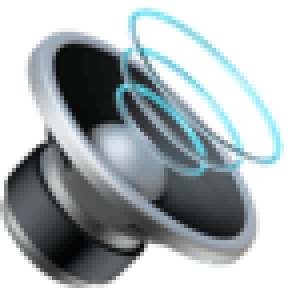}} &
 \multirow{3}{*}{Gesture} &
 Wave &
 \raisebox{-0.5ex}{\includegraphics[width=1em]{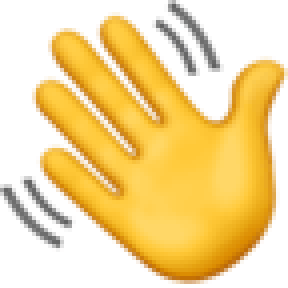}} \\ \cline{2-3} \cline{5-6} \cline{8-9} \cline{11-12} &
 Normal &
 \raisebox{-0.5ex}{\includegraphics[width=1em]{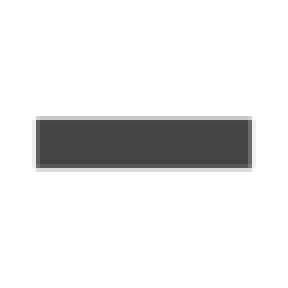}} && 
 Sad &
 \raisebox{-0.5ex}{\includegraphics[width=1em]{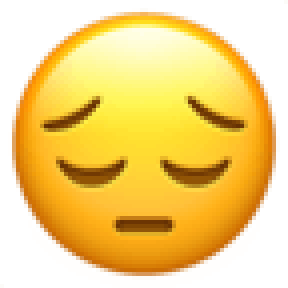}} &&
 Normal &
 \raisebox{-0.5ex}{\includegraphics[width=1em]{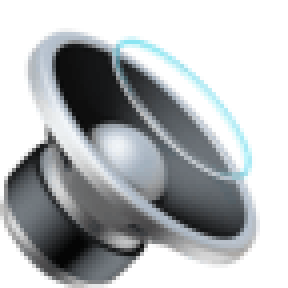}} &&
 Unfold &
 \raisebox{-0.5ex}{\includegraphics[width=1em]{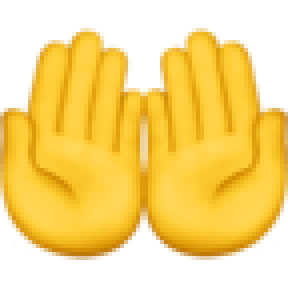}}
 \\ \cline{2-3} \cline{5-6} \cline{8-9} \cline{11-12} &
 Low &
 \raisebox{-0.5ex}{\includegraphics[width=1em]{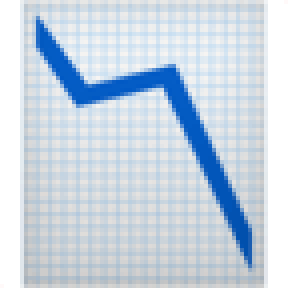}} &&
 Angry &
 \raisebox{-0.5ex}{\includegraphics[width=1em]{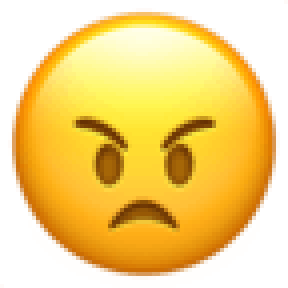}} &&
 Soft &
 \raisebox{-0.5ex}{\includegraphics[width=1em]{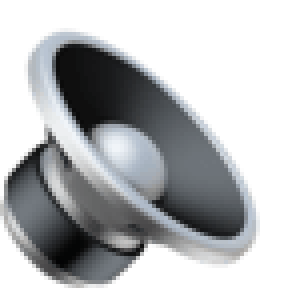}} &&
 Point &
 \raisebox{-0.5ex}{\includegraphics[width=1em]{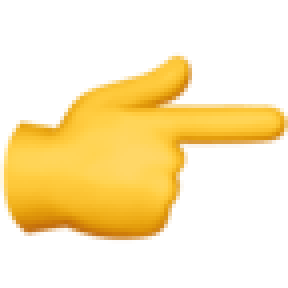}} \\ \cline{1-3} \cline{5-12} 
 \multirow{3}{*}{\begin{tabular}[c]{@{}c@{}}Speech\\ rate\end{tabular}} &
 Fast &
 \raisebox{-0.5ex}{\includegraphics[width=1em]{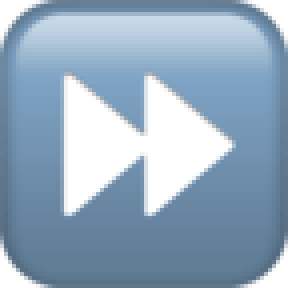}} &&
 Surprised &
 \raisebox{-0.5ex}{\includegraphics[width=1em]{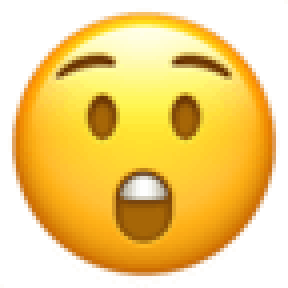}} &
 \multirow{3}{*}{Composure} &
 Calm &
 \raisebox{-0.5ex}{\includegraphics[width=1em]{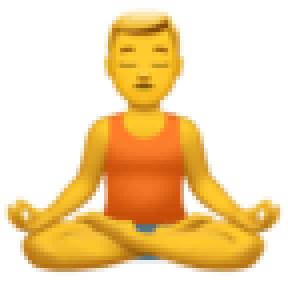}} &
 \multirow{2}{*}{Posture} &
 Stand &
 \raisebox{-0.5ex}{\includegraphics[width=1em]{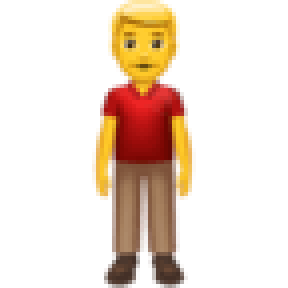}} \\ \cline{2-3} \cline{5-6} \cline{8-9} \cline{11-12} &
 Normal &
 \raisebox{-0.5ex}{\includegraphics[width=1em]{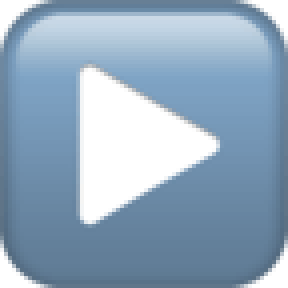}} &&
 Embarrassed &
 \raisebox{-0.5ex}{\includegraphics[width=1em]{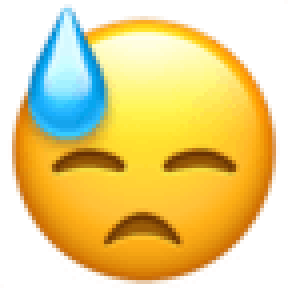}} &&
 Relaxed &
 \raisebox{-0.5ex}{\includegraphics[width=1em]{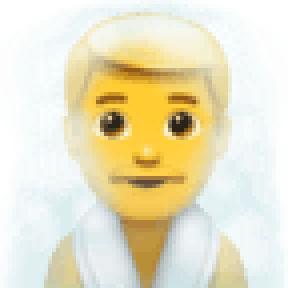}} &&
 Walk &
 \raisebox{-0.5ex}{\includegraphics[width=1em]{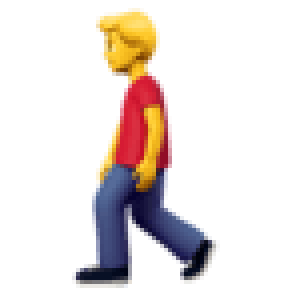}} \\ \cline{2-3} \cline{5-6} \cline{8-12} &
 Slow &
 \raisebox{-0.5ex}{\includegraphics[width=1em]{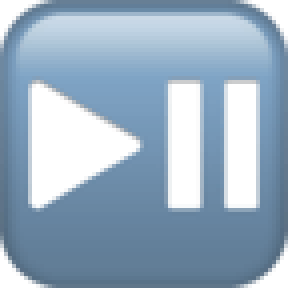}} &&
 Serious &
 \raisebox{-0.5ex}{\includegraphics[width=1em]{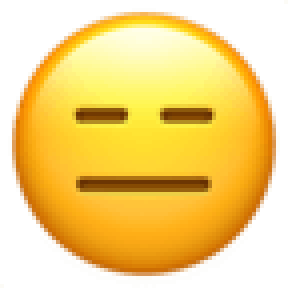}} &&
 Confident &
\raisebox{-0.5ex}{\includegraphics[width=1em]{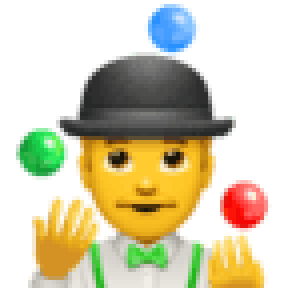}} &
 Eye contact &
 Direct &
 \raisebox{-0.5ex}{\includegraphics[width=1em]{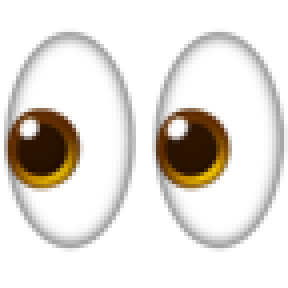}} \\ \hline
\end{tabular}
}
\vspace{-3mm}
\end{table*}

\par \textbf{[D5] Integrated Delivery Prompts.} Drawing from previous work~\cite{wang2020voicecoach, zeng2022gesturelens}, we present \highlight[verbal]{verbal} and \highlight[nonverbal]{nonverbal} delivery prompts using an integrated approach. To avoid overwhelming users with on-the-fly delivery support, we generate prompts at the start of each sentence (e.g., [volume - loud]), using emojis for encoding due to their rich meanings and familiarity~\cite{lo2008nonverbal,thompson2016emotional,skiba2016face}. We enlarge emojis for saliency and resolve any potential ambiguity~\cite{bich2019say,riordan2017communicative} through author discussions, ensuring the selection of suitable emojis for various modulation scenarios (\autoref{tab:Emoji}). Users have a lookup table within the smartphone interface (\autoref{fig:Lookup_table} in Appendix) to aid emoji interpretation and the option to deactivate emoji transcriptions. Moreover, we bold the keywords within the script to further inform users. Notably, the use of GPT-4 for delivering prompts is optional; users can manually input prompts or deactivate this feature by deselecting associated checkboxes. Once the final script with textual delivery prompts is uploaded, the server end transcribes the delivery factors and their modulations generated by GPT-4 into corresponding emojis (e.g., [gesture - point] $\rightarrow$ \raisebox{-0.5ex}{\includegraphics[width=1em]{figures/emoji/point.png}}), and then transfers the script with emojis to the mobile end.


\par \textbf{[D6] Rapid Pronunciation Acquisition.} To improve presentation scripts and mitigate disruptions from unfamiliar or polysyllabic words, we initially considered using phonetic symbols. However, user feedback indicated that while helpful for pronunciation, this could hinder fluent delivery. Instead, we used phonics to segment unfamiliar words marked (\autoref{fig:PC_Design}-8) and polysyllabic words into syllables (e.g., "Subsequently" $\rightarrow$ "Sub-se-quent-ly"). This approach leverages users' existing phonological knowledge, facilitating swift pronunciation acquisition without imposing a significant learning curve.

\subsection{Implementation Details}
\par The PC end of \textit{Trinity} is an office project created with Yeoman generator\footnote{https://learn.microsoft.com/en-us/office/dev/add-ins/develop/yeoman-generator-overview}, designed as a PowerPoint add-in using Office.js for seamless use. It comprises around 1,000 lines of TypeScript code developed within ReactJS. The server side, implemented in about 500 lines of Python code, uses Flask and the OpenAI API\footnote{https://platform.openai.com/docs/api-reference/chat} to refine scripts, transcribe delivery prompts, and connect the PC and smartphone ends.

\par The Smartphone end of \textit{Trinity} is based on Android due to its market dominance and popularity. It captures voice input, converts it to text using Android's SpeechRecognizer\footnote{https://developer.android.com/reference/kotlin/android/speech/SpeechRecognizer}, and processes the text with the \textit{BM25} string matching ranking algorithm for accurate speech progress monitoring and synchronized script scrolling~\cite{Robertson2009BM25}. The Android App comprises around 1,000 lines of Kotlin\footnote{https://kotlinlang.org/} code within the Compose framework\footnote{https://developer.android.com/jetpack/compose}.

\section{User Study}
\par To provide a thorough assessment of the proposed \textit{Trinity} in relation to its \textbf{\textit{usability and effectiveness}} (\textbf{RQ3}), \textbf{\textit{usage patterns and system impact}} (\textbf{RQ4}), as well as \textbf{\textit{trust and collaboration}} (\textbf{RQ5}), we executed a controlled between-subject user study. In this study, EFL students were tasked with delivering AOPs using \textit{Trinity}, while two other baseline systems were employed as control conditions.

\subsection{Conditions}
\par \textit{Trinity} is built upon the most common practice in presentation, PowerPoint, and integrates both mobile and PC devices for support, with the focus of providing on-the-fly guidance on the mobile end. To evaluate the realization of our six design goals, we compared \textit{Trinity} with two baseline systems which are also built upon PowerPoint, \textit{IntelliPrompter} and \textit{OfficeRemote}. 1) \textit{Intelliprompter} ({\autoref{fig:Previous_Tools}C}) enhances PowerPoint's presenter view, automatically tracking the presenter's slide coverage and adjusting the note display on a PC. This baseline was chosen due to its shared features with \textit{Trinity}, such as speech tracking and dynamic script display, facilitating comparison of effects introduced by mobile support. 
2) \textit{OfficeRemote} ({\autoref{fig:Previous_Tools}A}) is a mobile app that connects with PowerPoint on a desktop, designed for slide navigation, script reference, timer control, and laser pointer feature, all accessible through a mobile phone. We selected this baseline because it operates on mobile like \textit{Trinity} App and offers basic on-the-fly delivery supports as well, facilitating the comparison of designed delivery support provided in \textit{Trinity}.
Initially, EFL students were randomly divided into three groups, each using a different system. These groups were then randomly and evenly allocated to three sessions to deliver two $5$-minute AOPs without $Q\&A$ sessions.

\begin{table}[h]
 \centering 
 \caption{Comparison of baseline systems.}
  \vspace{-3mm}
 \resizebox{0.48\textwidth}{!}{%
\begin{tabular}{cccc}
\hline
                        & \textit{IntelliPrompter} & \textit{Office Remote} & \textit{\textbf{Trinity}} \\ \hline
Dynamic Script Dispaly  & $\checkmark$                        & Static                 & $\checkmark$                         \\
Mobile Silde Navigation & ·                        & $\checkmark$                      & $\checkmark$                         \\
Speech Pace Modulation  & ·                        & ·                      & $\checkmark$                         \\
Script Polishing        & ·                        & ·                      & $\checkmark$                         \\
Delivery Prompts        & ·                        & ·                      & $\checkmark$                         \\
Built upon Powerpoint   & $\checkmark$                        & $\checkmark$                      & $\checkmark$                         \\
Media                   & PC                       & Mobile \& PC           & Mobile \& PC              \\ \hline
\end{tabular}%
}
 \vspace{-3mm}
 \label{tab:Conditions}
\end{table}

\subsection{Participants}
\par Following IRB approval, we recruited $65$ EFL student presenters ($29$ females, $36$ males, S1-S65, average age $21.5$, $SD=2.13$) and $25$ audience members ($11$ females, $14$ males, A1-A25, average age $25.4$, $SD=4.83$) from a local university through email and social media. The presenters came from various majors, and all had AOP experience. The audience comprised undergraduates, graduates, and researchers from diverse fields. Five instructors ($2$ female, $3$ males, I1-I5, average age $41.6$, $SD=6.12$) also joined as audience. The presenters were initially divided randomly into three groups of $22$ for the \textit{Trinity} and \textit{IntelliPrompter} conditions, and $21$ for the \textit{OfficeRemote} condition. After three dropouts, \textit{Trinity} and \textit{IntelliPrompter} conditions had $21$ presenters, and \textit{OfficeRemote} conditions had $20$ presenters.

\begin{figure*}[h]
  \centering
  \includegraphics[width=\linewidth]{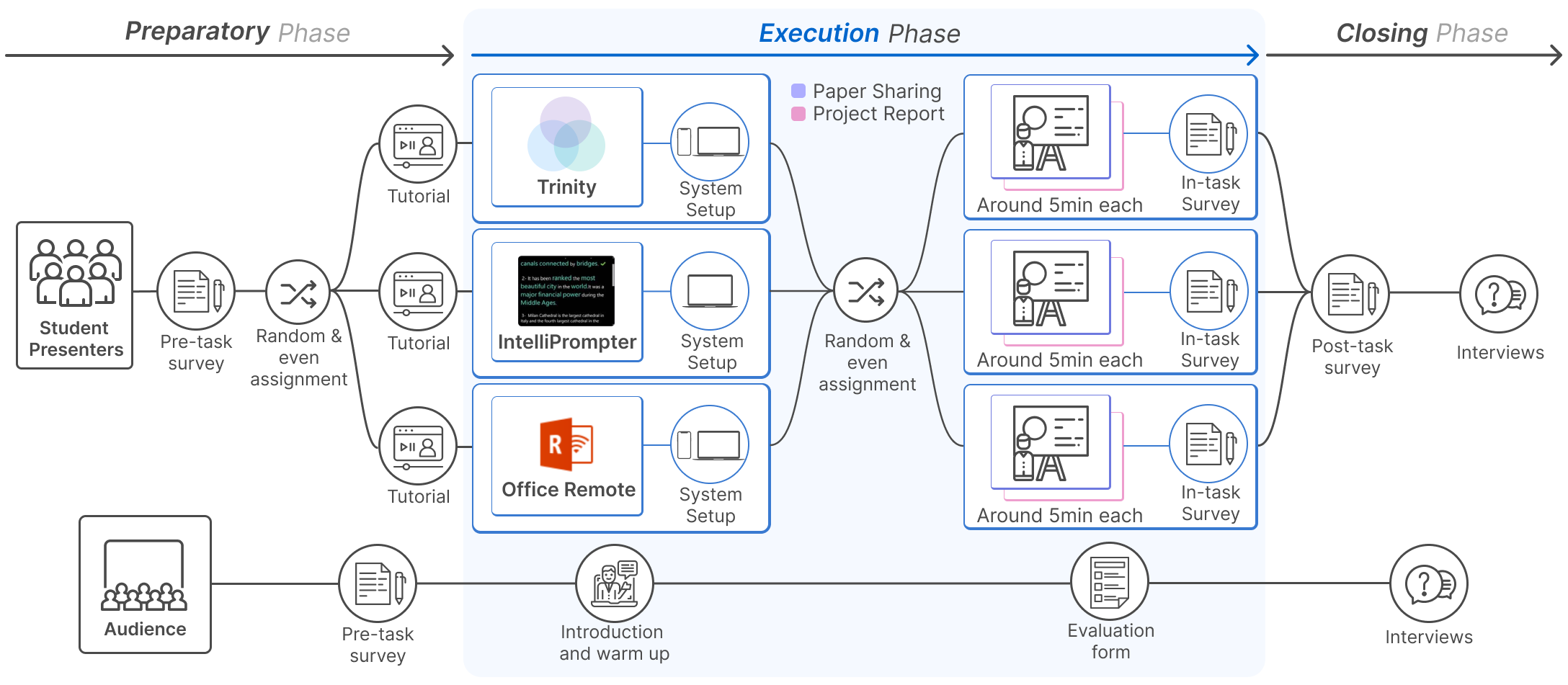}
  \vspace{-6mm}
  \caption{Design and procedure of user study. 1) In the preparatory phase, student presenters and were given a pre-task survey and assigned to one of three conditions randomly and evenly with tutorials. 2) In the execution phase, presenters across three conditions were randomly and evenly assigned to three sessions to give a $5$-minute paper sharing and a $5$-minute project report. An in-task survey was conducted alongside these presentations. Following the warm-up session, audience members evaluated the performance of presenters using evaluation forms. 3) In the closing phase, both presenters and audience members participated separately in post-task surveys and interviews.}
\vspace{-3mm}
  \label{fig:User_Study}
\end{figure*}

\subsection{Procedure}
\par In this study, each EFL student delivered two $5$-minute AOPs in English, separated by a brief break. Students prepared presentation materials, including slides and scripts, with specific criteria: 1) They needed a good understanding of the content without extensive rehearsal, maintaining moderate preparation. 2) The AOPs covered diverse genres: one as a paper sharing session (discussing others' work) and the other as a project report (covering their own work). Materials were submitted at least one day in advance for compliance checks with three criteria: 1) Each presenter must report an approximate preparation time to ensure moderate preparation. 2) The presentation content must be complete (i.e. from introduction to discussion). 3) Materials must be in English and cover both paper sharing and project report. For the \textit{Trinity} group, participants installed the app on their laptops and smartphones, completing preprocessing with screen recording based on provided instructions. Two authors facilitated the experiment and addressed technical issues as needed.

\par The experiment, illustrated in \autoref{fig:User_Study}, had three phases: preparatory, execution, and closing. In the preparatory phase, participants filled out a consent form and a pre-task survey. All presenters received a 5-minute system tutorial. In the execution phase, presenters in the \textit{Trinity}, \textit{IntelliPrompter}, and \textit{OfficeRemote} conditions had $10$ minutes for system setup and familiarization. Presenters were then randomly and evenly assigned to three AOP sessions. A $5$-$10$ minute introduction was given before the AOPs, during which instructors and audience assessed presenter performance using evaluation forms (\autoref{fig:Evaluation_form}). Post-AOP, presenters completed an in-task questionnaire. In the closing phase, a post-task questionnaire was filled by all participants, followed by semi-structured interviews with $12$ audience members and all the presenters in \textit{Trinity} condition separately. To maximize the blindness of the audience, we didn't disclose experimental conditions from the outset of recruitment to the end of the interviews.

\par The whole process of this study took around $11$ hours in total. The AOPs are scheduled with a $15$-minute break inserted every $40$ minutes to mitigate the fatigue effect. Each day, there is an approximate duration of $2.2$ hours dedicated to AOPs. 
We ensured consistency by using a 70-inch monitor for slide projection and a tripod-mounted camera to record presentations, with participants' consent. Presenters were asked to bring their own laptops and smartphones for use, with standby devices available for iOS users and emergencies. Participants were compensated at a rate of $\$15$ USD/hr.

\begin{table*}[h]
\caption{Metrics of student presenter' self-report to investigate \textbf{RQ3-RQ5}, organized by research question, aspect, and approach.}
\label{tab:Metrics}
\vspace{-3mm}
\resizebox{\textwidth}{!}{%
\begin{tabular}{|l|l|l|l|}
\hline
\textbf{Research Question} &
  \textbf{Aspect} &
  \textbf{Approach} &
  \textbf{Metric} \\ \hline
\multirow{2}{*}{\begin{tabular}[c]{@{}l@{}}\textbf{RQ3}. How are the usability \\ and effectiveness of the \\ supportive system?\end{tabular}} &
  Usability &
  \multirow{5}{*}{\begin{tabular}[c]{@{}l@{}}In-task\\ questionnaire\end{tabular}} &
  \begin{tabular}[c]{@{}l@{}}1) ease-of-use, 2) helpfulness in supporting AOP delivery, \\ 3) distraction caused by the system, 4) system satisfaction, \\ 5) likelihood for future use.\end{tabular} \\ \cline{2-2} \cline{4-4} 
 &
  Effectiveness &
   &
  \begin{tabular}[c]{@{}l@{}}1) making the speech more fluid, coherent and modulated, \\ 2) facilitating good eye contacts with audience, 3) facilitating\\  appropriate facial expressions, 4) facilitating natural gestures, \\ 5) enhancing easiness of visual control, 6) enhancing effectiveness\\  of slides, 7) enhancing control of the presentation progress, \\ 8) making the presentation more engaging\end{tabular} \\ \cline{1-2} \cline{4-4} 
\multirow{3}{*}{\begin{tabular}[c]{@{}l@{}}\textbf{RQ4}. How will EFL students\\  interact with and be \\ influenced by the system?\end{tabular}} &
  Interaction &
   &
  \begin{tabular}[c]{@{}l@{}}Collect students’interaction logs on the interface and \\ counted the frequency of students’glancing at the interface.\end{tabular} \\ \cline{2-2} \cline{4-4} 
 &
  \multirow{2}{*}{System Impact} &
   &
  \begin{tabular}[c]{@{}l@{}}Emotions: 1) noticed, 2) excited, 3) exhausted, 4) frustrated, \\ 5) happy, 6) hopeful, 7) overwhelmed, 8) safe, 9) nervous, \\ 10) anxious, 11) confident.\end{tabular} \\ \cline{4-4} 
 &
   &
   &
  \begin{tabular}[c]{@{}l@{}}Load: 1) cognitive load, 2) attentional load, 3) workload, \\ 4) self-perceived performance\end{tabular} \\ \hline
\multirow{2}{*}{\begin{tabular}[c]{@{}l@{}}\textbf{RQ5}. How will students \\ trust and collaborate \\ with our system?\end{tabular}} &
  Trust &
  \begin{tabular}[c]{@{}l@{}}Post-task\\ questionnaire\end{tabular} &
  \begin{tabular}[c]{@{}l@{}}1) perceptions of the usefulness of the provided system features, \\ 2) perceived accuracy of the system and overall trust in the system,\\  3) trust in different features embodied in the system.\end{tabular} \\ \cline{2-4} 
 &
  Collaboration &
  \begin{tabular}[c]{@{}l@{}}Semi-structured\\ interviews\end{tabular} &
  Student presenters' feedback \\ \hline
\end{tabular}%
}
\vspace{-3mm}
\end{table*}

\subsection{Measurement}
\par \textbf{\textit{\textsf{Student Presenters' Self-report.}}} As detailed in \autoref{tab:Metrics}, in our study, we employed a $7$-point Likert in-task questionnaire to gather feedback from student presenters regarding their experiences using different support systems during AOPs~\cite{ma2022glancee,murali2021affectivespotlight}. This questionnaire covered three key aspects: \textbf{RQ3. \textit{Usability and Effectiveness}.} We designed questions on the usability and effectiveness of the systems based on prior research~\cite{murali2021affectivespotlight,murali2018speaker,ma2022glancee,bickmore2021virtual,murali2021affectivespotlight,parmar2020making}; \textbf{RQ4. \textit{Usage Pattern and System Impact}.} We collected interaction logs to track usage patterns and assessed students' emotions and workload using questions inspired by existing studies~\cite{tanveer2015rhema,asadi2017intelliprompter,murali2021affectivespotlight,ma2022glancee} and the NASA-TLX survey~\cite{hart1988development}; \textbf{RQ5. \textit{Trust and Collaboration}.} We administered a post-task questionnaire to gauge presenters' trust in our system~\cite{ma2022glancee}. Additionally, semi-structured interviews were conducted with student presenters, addressing topics related to \textbf{RQ3-RQ5}.

\par \noindent\textbf{\textit{\textsf{Presenter Evaluation by Audience.}}} We employed a standardized rubric~\cite{peeters2010standardized} for the audience to evaluate students' presentation performance. This rubric included assessments of 1) \highlight[verbal]{verbal} delivery, 2) \highlight[nonverbal]{nonverbal} delivery, 3) \highlight[visual]{visual} aids, 4) delivery consistency, and 5) content\&organization. The evaluation form (as detailed in \autoref{fig:Evaluation_form} in Appendix) utilized a $4$-point scale with detailed descriptions for each option. Additionally, we conducted interviews with the audience, inviting them to provide further comments through open-ended questions.

\begin{table*}[h]
\caption{Summary of qualitative insights on design choices of \textit{Trinity}, organized by design goals, to enlighten future presentation support tools.}
\vspace{-3mm}
\label{tab:Pros_Cons}
\resizebox{\textwidth}{!}{
\begin{tabular}{llll}
\hline
\textbf{Design Goal} &
  \textbf{\textit{Trinity} Choice} &
  \textbf{Success} &
  \textbf{Challenges} \\ \hline
\begin{tabular}[c]{@{}l@{}}\textbf{[D1]} Customizable \\ Delivery Factors\end{tabular} &
  \begin{tabular}[c]{@{}l@{}}selectable factors;\\ recommended preset\end{tabular} &
  \begin{tabular}[c]{@{}l@{}}support participants' diverse individual \\ preferences; incorporates \\ recommended preset\end{tabular} &
  \begin{tabular}[c]{@{}l@{}}require to understand self-needs;\\ excessive number of factors can \\ overwhelm presenters\end{tabular} \\ \hline
\begin{tabular}[c]{@{}l@{}}\textbf{[D2]} Moderate \\ Script Polishing\end{tabular} &
  \begin{tabular}[c]{@{}l@{}}script polishing with GPT-4;\\ time limit control; script \\ comparison and revision\end{tabular} &
  \begin{tabular}[c]{@{}l@{}}improve the overall quality and \\ effectiveness; save preparation time; \\ boost the success in conveying \\ messages or ideas.\end{tabular} &
  \begin{tabular}[c]{@{}l@{}}LLM may entail script-related \\ potential misunderstandings;\\ LLM may mismatch with participants' \\ speech style;\end{tabular} \\ \hline
\begin{tabular}[c]{@{}l@{}}\textbf{[D3]} Remote Mobile \\ Visual Control\end{tabular} &
  \begin{tabular}[c]{@{}l@{}}thumb tapping and swiping\\ to navigate slides\end{tabular} &
  \begin{tabular}[c]{@{}l@{}}effortlessly control visuals via \\ smartphone; minimize cognitive load\end{tabular} &
  \begin{tabular}[c]{@{}l@{}}lacks the display of each \\ animation step; limits smartphone \\ holding posture\end{tabular} \\ \hline
\begin{tabular}[c]{@{}l@{}}\textbf{[D4]} Flexible \\ Speech Pace \\ Modulation\end{tabular} &
  \begin{tabular}[c]{@{}l@{}}progress bars;\\ background underpainting;\\ triangle grouping\end{tabular} &
  \begin{tabular}[c]{@{}l@{}}manage and adjust the speaking pace; \\ cater to both macroscopic and \\ microscopic aspects\end{tabular} &
  \begin{tabular}[c]{@{}l@{}}central progress bars are prone to be \\ overlooked; speech recognition jitters\\  may misdirects to incorrect scripts\end{tabular} \\ \hline
\begin{tabular}[c]{@{}l@{}}\textbf{[D5]} Integrated \\ Delivery Prompts\end{tabular} &
  \begin{tabular}[c]{@{}l@{}}emojis; \\ keywords highlighting\end{tabular} &
  \begin{tabular}[c]{@{}l@{}}vividly present emotions and nonverbal \\ cues; ensure a personalized experience\end{tabular} &
  \begin{tabular}[c]{@{}l@{}}nonverbal cues prone to be more \\ irresponsive than verbal cues;\\ response effect is somewhat subpar\end{tabular} \\ \hline
\begin{tabular}[c]{@{}l@{}}\textbf{[D6]} Rapid \\ Pronunciation \\ Acquisition\end{tabular} &
  phonics-based segmentation &
  \begin{tabular}[c]{@{}l@{}}allows participants to acquire correct \\ pronunciation quickly\end{tabular} &
  \begin{tabular}[c]{@{}l@{}}somewhat intensify the workload;\\ presenters exhibit a noticeable \\ spelling process\end{tabular} \\ \hline
\end{tabular}%
}
\vspace{-3mm}
\end{table*}

\subsection{User Study Limitations}
\par \noindent \textbf{\textsf{Variable control.}} Our system is developed upon the new design goals derived in the Formative Study, making it challenging to strictly control the variables for evaluating newly designed features. In the early iteration of our user study, we considered applying a ablation study manner to strictly configure the conditions for newly designed features. However, the total number of conditions ($8$ conditions) makes it unaffordable for researchers to apply in a user study and the fatigue effect becomes inevitable for participants. Therefore, we follow the practices in HCI research (e.g.,~\cite{ma2022glancee,yuan2023critrainer,xia2022persua}) and use two systems with shared features ({~\autoref{tab:Conditions}}) for baseline comparison. In the following, we report the directional results, while more in-depth and fine-grained evaluation of each feature's impact may leave for the future works.
\par \noindent \textbf{\textsf{Audience blindness.}} Presentation is a process through which presenters communicate information to an audience on the spot, making it challenging to strictly ensure the blindness of audience. In our user study, EFL students are expected to employ various communication channels while the audience evaluates their actual performance. In the early iteration of our user study, we considered using certain apparatus (e.g, a board erected in front of the presenter) to mask the experimental conditions; however, this configuration proved impractical and significantly hindered both presentation delivery and evaluation. Therefore, although we did not disclose the experimental conditions throughout recruitment and interviews, the audience may still infer some condition-related information from presenters' usage patterns (i.e., how they interact with devices), potentially influencing their ratings. Future works may focus on addressing the challenge of audience blindness in presentation scenarios to enhance the rigor of subsequent investigations.

\section{Analyses and Results}

\par This section analyzes \textbf{RQ3-RQ5} using a mixed-methods approach. Quantitatively, Kruskal-Wallis with post-hoc Dunn's tests and Bonferroni correction were used for in-task questionnaires, and descriptive statistics for post-task responses. One-way ANOVA with post-hoc Tukey tests studied the impact of delivery support tools on perceived AOP performance. Qualitatively, we reviewed video recordings of AOP sessions with \textit{Trinity}, identified interaction timestamps, and discussed with presenters to understand motivations and reactions. Feedback was audio-recorded, transcribed, and thematically analyzed by two authors until consensus. We present results addressing tensions in tool usage, noting how design choices both facilitated and introduced challenges for \textit{Trinity}. Achievements and obstacles are categorized based on \textit{Trinity}'s design goals, providing insights for future research. Primary findings are presented here, with minor findings in \autoref{appendix:Minor_findings}.

\subsection{RQ3. How are the usability and effectiveness of the supportive system?}
\par We present findings from our mixed-methods analysis, evaluating the usability of \textit{Trinity} for student presenters compared to baseline systems. We also analyse the effectiveness of \textit{Trinity} features as reported by students and assess AOP performance from the audience's perspective.

\begin{figure*}
  \centering
  \includegraphics[width=\linewidth]{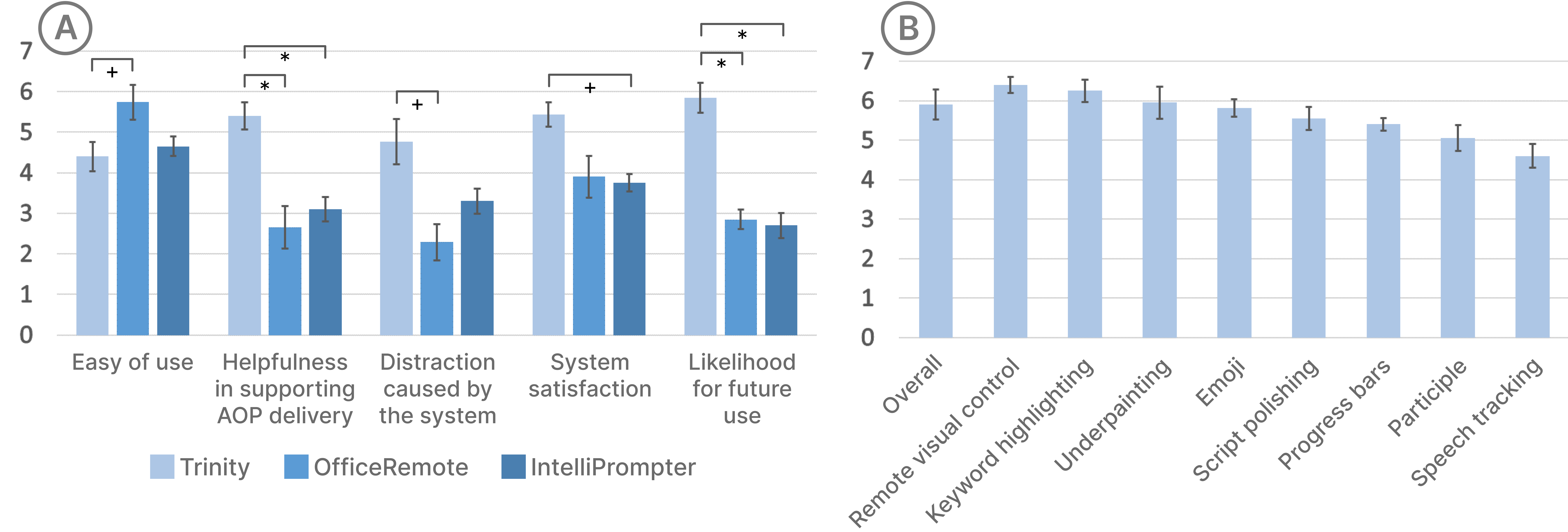}
  \vspace{-6mm}
  \caption{The usability and usefulness of systems. A) The usability of different systems perceived by student presenters. B) The perceived usefulness of different features provide by \textit{Trinity}. The error bars indicate standard errors. ($+:p<.1$; $*:p<.05$; $**:p<.01$; $***:p<.001$).}
\vspace{-3mm}
  \label{fig:Usability_Effectiveness}
\end{figure*}

\par \textbf{i) Usability.} The results of Kruskal-Wallis tests indicate significant differences among the systems in terms of their perceived helpfulness in supporting AOP delivery ($H=9.471$, $p=0.007$) and the likelihood of future use ($H=10.382$, $p=0.005$). However, there is no significant difference in terms of ease-to-use ($H=4.523$, $p>0.05$):
\begin{center}
\begin{minipage}{0.45\textwidth}
\textsf{\textit{``Trinity is not a plug-and-play for me. It took me some time to get familiar with the utilities and figure out how to it works before I put it into practical use.'' (S10, male, age: 21)}}
\end{minipage}
\end{center}
Post-hoc analysis revealed that \textit{Trinity} was significantly more helpful for supporting AOP delivery and received higher ratings for intended future use compared to both \textit{Intelliprompter} ($p=0.032$ and $p=0.010$) and \textit{OfficeRemote} ($p=0.024$ and $p=0.041$). However, there were no significant differences across the three systems in terms of distraction caused by the system and system satisfaction, as indicated by Kruskal-Wallis tests.

\par \textbf{ii) Effectiveness.} \autoref{fig:Usability_Effectiveness} presents participant-perceived usefulness of \textit{Trinity}'s features in descending order. Most features were deemed useful but posed some challenges. Notably, participants highly rated flexible remote visual control, keyword highlighting, and the fine-grained speech pace indicator (underpainting), while also finding integrated delivery prompts helpful.

\begin{figure*}
  \centering
  \includegraphics[width=\textwidth]{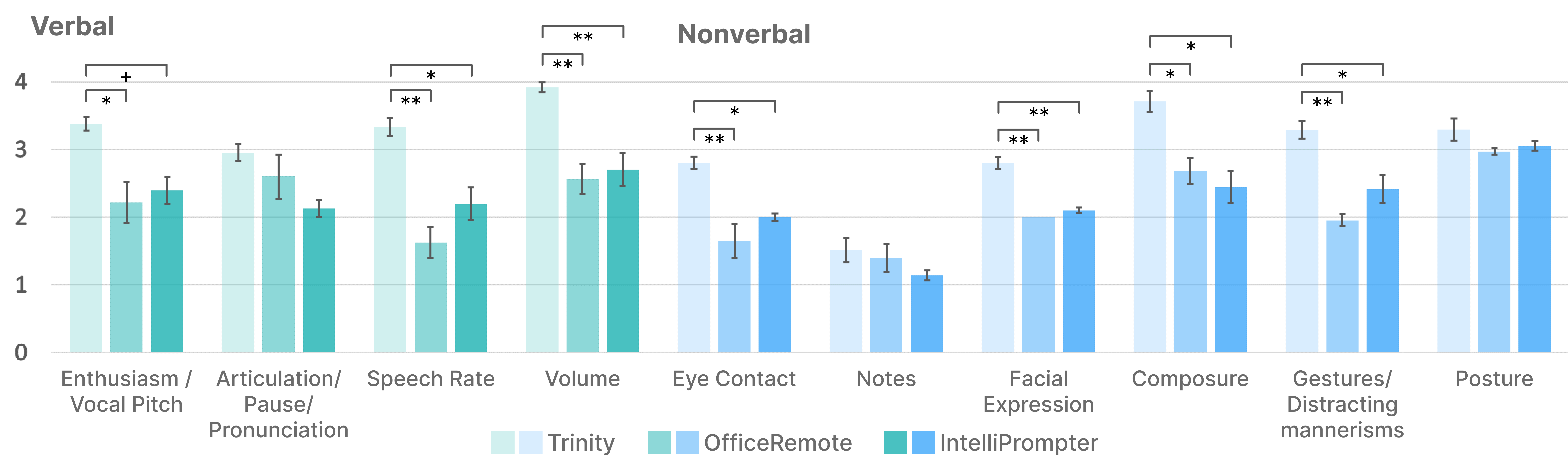}
  \vspace{-6mm}
  \caption{Audience's ratings on verbal and nonverbal delivery of presenters using different systems. The error bars indicate standard errors. ($+:p<.1$; $*:p<.05$; $**:p<.01$; $***:p<.001$).}
\vspace{-3mm}
  \label{fig:Verbal_Nonverbal}
\end{figure*}

\begin{figure}
 \vspace{-3mm}
 \centering 
 \includegraphics[width=\linewidth]{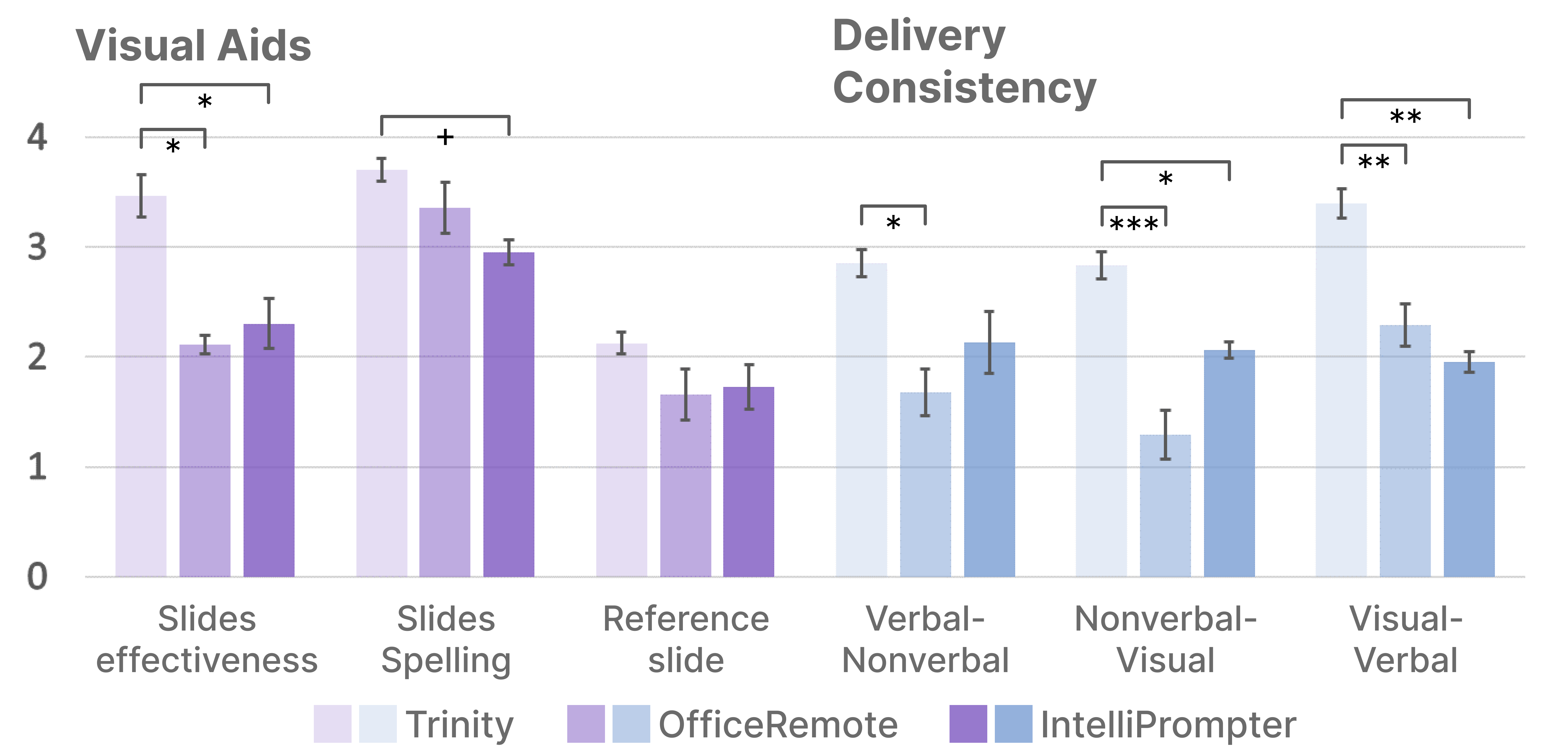}
   \vspace{-6mm}
 \caption{Audience's ratings on visual delivery and delivery consistency of presenters using different systems. The error bars indicate standard errors. ($+:p<.1$; $*:p<.05$; $**:p<.01$; $***:p<.001$).}
 \label{fig:Visual_Consistency}
  \vspace{-3mm}
\end{figure}

\begin{figure*}[h]
  \centering
  \includegraphics[width=0.9\textwidth]{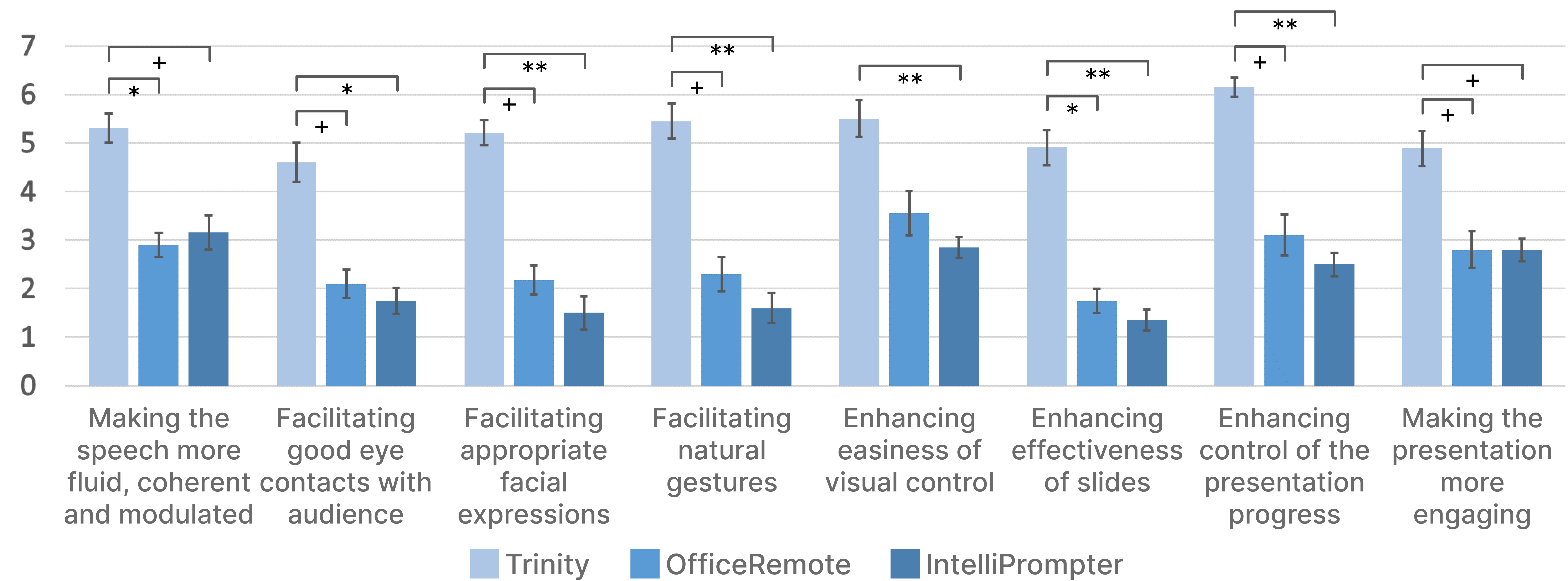}
  \vspace{-3mm}
  \caption{Student presenters' ratings on how different systems support their delivery. The error bars indicate standard errors. ($+:p<.1$; $*:p<.05$; $**:p<.01$; $***:p<.001$).}
\vspace{-3mm}
  \label{fig:Performance}
\end{figure*}

\par Subsequently, we conducted an in-depth analysis of how the \textit{Trinity} system supported EFL student presenters in delivering AOP from various perspectives. Presenter ratings on relevant questions are presented in \autoref{fig:Performance}. Kruskal-Wallis tests with post-hoc tests showed that \textit{Trinity} outperformed both \textit{IntelliPrompter} and \textit{OfficeRemote} in all aspects (\autoref{fig:Performance}).
This findings align with participants' qualitative feedback from interviews. Participants found \textit{Trinity} superior due to its "\textit{comprehensive and customizable delivery support}" (S2, S28, S37, S45), which "\textit{facilitates adaptive delivery}" (S14, S23, S28), "\textit{enriches presentation content}" (S27, S47), and "\textit{enhances control over presentations}" (S3, S31, S39). They also noted the interface as "\textit{easy to understand and interact with}" (S14, S55).

\par Regarding making the speech fluid, coherent and modulated ($H=8.726, p=0.012$, pairwise: $p_{TI}=0.060, p_{TO}=0.025$~\footnote{$p_{TI}$: $p$ value between \textit{Trinity} and \textit{IntelliPrompter}; $p_{TO}$: $p$ value between \textit{Trinity} and \textit{OfficeRemote}.}), facilitating good eye contacts with audience ($H=9.198, p=0.010$, pairwise: $p_{TI}=0.18, p_{TO}=0.060$), facilitating appropriate facial expressions ($H=11.075, p=0.004$, pairwise: $p_{TI}=0.071, p_{TO}=0.005$), facilitating natural gestures ($H=10.104, p=0.006$, pairwise: $p_{TI}=0.009, p_{TO}=0.065$), enhancing easiness of visual control ($H=8.870, p=0.012$, pairwise: $p_{TI}=0.009, p_{TO}=0.316$~\footnote{\textit{OfficeRemote} received higher scores here due to its simplicity.}), enhancing effectiveness of slides ($H=10.824, p=0.004$, pairwise: $p_{TI}=0.010, p_{TO}=0.028$), enhancing control of the presentation progress ($H=10.853, p=0.004$, pairwise: $p_{TI}=0.005, p_{TO}=0.066$), making the presentation more engaging ($H=7.207, p=0.027$, pairwise: $p_{TI}=0.094, p_{TO}=0.005$). These findings align with participants' qualitative feedback obtained through interviews. Participants perceived \textit{Trinity} to be more beneficial than the two baselines due to its provision of "\textit{comprehensive and customizable delivery support}" which effectively ``\textit{facilitates adaptive delivery}'', ``\textit{enriches presentation content}'', and ``\textit{enhances control over presentations}''. Additionally, participants found the interface design to be ``\textit{easy to understand and interact}''.

\par Finally, we assess presenters' AOP performance from the \textbf{audience's perspective}. As depicted in \autoref{fig:Verbal_Nonverbal}, regarding \textbf{\highlight[nonverbal]{nonverbal} delivery}, one-way ANOVA with post-hoc Tukey tests revealed that presenters in the \textit{Trinity} condition received significantly higher ratings for eye contact ($F(2,60)=16.673, p=0.001$, pairwise: $p_{TI}=0.014, p_{TO}=0.001$~\footnote{$p_{TI}$: $p$ value between \textit{Trinity} and \textit{IntelliPrompter}; $p_{TO}$: $p$ value between \textit{Trinity} and \textit{OfficeRemote}.}), facial expression ($F(2,60)=23.416, p<0.001$, pairwise: $p_{TI}=0.001, p_{TO}=0.001$), composure ($F(2,60)=14.728, p=0.003$, pairwise: $p_{TI}=0.015, p_{TO}=0.018$), and gesture ($F(2,60)=13.532, p=0.003$, pairwise: $p_{TI}=0.016, p_{TO}=0.003$). Specifically, seven audience members observed presenters using "\textit{consistent eye contact}" (A3, A5) and a "\textit{sweeping gaze}" (A13) while holding a phone. They showed adaptability with facial expressions, "\textit{frowning when discussing challenges}" (A15, A17) and "\textit{smiling when presenting findings}" (A5). These presenters appeared "\textit{well-prepared and imperturbable}" (A21), demonstrating "\textit{control over their presentations}" (A24). They used active gestures to "\textit{emphasize key points through pointing}" (A15) and "enhance explanations by opening their arms" (A5). Typically, they kept "\textit{one hand in a stationary position}" (A18) to secure the phone and used "\textit{the other hand for gestures and movements}" (A18, A24). Despite differences in eye contact, the audience found no variations in notes usage, as all presenters relied on scripts, "\textit{whether they read them from computers or phones}" (A2, A5, A9, A20). Three audience members also noted little difference in posture, possibly due to minimal posture prompts.

\par In \textbf{\highlight[verbal]{verbal} delivery} (\autoref{fig:Verbal_Nonverbal}), our tests indicate that presenters in \textit{Trinity} condition received significantly higher ratings for vocal pitch ($F(2,60)=10.434, p=0.010$, pairwise: $p_{TI}=0.053, p_{TO}=0.012$), speech rate ($F(2,60)=13.215, p=0.002$, pairwise: $p_{TI}=0.045, p_{TO}=0.002$), and volume ($F(2,60)=15.362, p=0.001$, pairwise: $p_{TI}=0.002, p_{TO}=0.002$). Specifically, nine audience members noted that presenters who "\textit{held the phone sideways}" (A2, A3) showed more vocal variety, using a "\textit{louder voice to emphasize important points}" (A9, A13), speaking "\textit{faster when discussing previous works}" (A15, A17, A18), and "\textit{raising pitch while posing questions}" (A3, A24, A25), unlike those who "\textit{read their notes monotonously}" (A2, A5, A18, A25).

\par In \textbf{\highlight[visual]{visual} delivery} (\autoref{fig:Visual_Consistency}), our tests indicate that \textit{Trinity} user received significantly higher ratings for slides effectiveness ($F(2,60)=7.352, p=0.024$, pairwise: $p_{TI}=0.033, p_{TO}=0.010$), with no significant difference in slide spelling and references. Specifically, seven audience members noted that some presenters overwhelm them with "\textit{excessive text or bullet points}" (A5, A24), leading to questioning the value of attending (A9). However, exceptional presenters use "\textit{highlighting key terms}" (A2, A5, A9) and "\textit{incorporating diagrams}" (A2, A15) to make content "\textit{more accessible}" (A24) and facilitate "\textit{better understanding}" (A18). They create an environment where the audience can "\textit{easily grasp complex concepts}" (A5) without feeling overwhelmed.

\par For \textbf{delivery consistency}, our tests indicate that \textit{Trinity} user received significantly higher ratings for speech-behavior consistency ($F(2,60)=9.532, p=0.018$, pairwise: $p_{TI}=0.143, p_{TO}=0.023$), behavior-visual consistency ($F(2,60)=18.351, p<0.001$, pairwise: $p_{TI}=0.012, p_{TO}<0.001$), and speech-visual consistency ($F(2,60)=17.341, p<0.001$, pairwise: $p_{TI}=0.001, p_{TO}=0.003$). Many presenters "\textit{read from their scripts or slides}" (A3, A20), but some also "\textit{not only present but also coordinate}" (A5), aligning "\textit{what they say, what they do, and what they show}" (A13). Examples include pointing to bold words with higher volume (A18), giving opening statements with a smile and open arms (A9), and pointing and zooming in on mentioned locations (A24).

\par For \textbf{content\&organization} (\autoref{fig:Content_Organization} in appendix), our tests indicate that \textit{Trinity} users received significantly higher ratings for the opening statement ($F(2,60)=12.357, p=0.001$, pairwise: $p_{TI}=0.003, p_{TO}=0.004$), transitions ($F(2,60)=21.628, p<0.001$, pairwise: $p_{TI}<0.001, p_{TO}=0.002$), and organization ($F(2,60)=10.379, p=0.004$, pairwise: $p_{TI}=0.001, p_{TO}=0.021$), with no significant difference in other factors. Six audience members noted that some presenters, unlike others who "\textit{gave a quick hello}" (A9) and "\textit{skimmed through the title}" (A2, A9), used a phone to "\textit{start with the background}" (A2, A5, A9) and "\textit{gradually lead into the main content}" (A5). This approach helped those not "w\textit{ell-versed in a particular field or specialty}" (A21) to "\textit{grasp the main ideas}" (A2, A18) and "\textit{follow along more easily}" (A9). Additionally, these presenters "\textit{provided an outline of content}" (A24) after introduction and "\textit{added smooth transitions}" (A21), avoiding "\textit{jumping next directly}" (A9), making their content "\textit{clear and coherent}" (A9, A18).

\par In summary, audience and student presenter ratings and feedback illustrate \textit{Trinity}'s usability and its effectiveness in supporting verbal, nonverbal, and visual delivery while improving multichannel consistency.

\subsection{RQ4. How will EFL students interact with and be influenced by the system?}
\par Building on the mixed methods mentioned earlier, we begin by exploring how students' interactions with our system affect their AOP delivery, followed by an analysis of how the system impacts students' behaviors, emotions, and cognitive workloads.

\par \textbf{i) Usage pattern.} We share insights into how student presenters use the PC interface to customize delivery factors and improve presentation scripts, their interactions with the \textit{Trinity} App during AOPs, and the motivations driving their actions.

\par \textbf{\prefix{\textsf{F1}}: Presenters generally customize delivery factors to suit their individual needs.} Out of 21 presenters, 17 tailored delivery factors to fit their preferences, reflecting diverse needs. Ten incorporated verbal, nonverbal, and visual factors, emphasizing the importance of content presentation and engagement. Four focused on verbal and nonverbal cues, aiming for a cohesive style where "\textit{body language plays a crucial role}" (S14) and gestures "\textit{aligned with their spoken words}" (S34). Three chose verbal and visual elements, believing visuals "\textit{aid in understanding}" (S27) and make presentations "\textit{more engaging}" (S44, S52). Four presenters used presets, possibly due to time or personal preference.

\par \textbf{\prefix{\textsf{F2}}: Presenters typically adapt their script length to match their allocated time and expected duration.} On the PC end, presenters adjust scripts to fit their time slots, and our PC tool predicts script durations to help. Of 21 presenters, 18 used this feature, with nine replacing text with diagrams, six cutting demos, and three adding discussions. This ensured their presentations "\textit{met the required time limit}" (S3, S27, S34) without "\textit{running out of time or having nothing left to say}" (S14, S62).


\par \textbf{\prefix{\textsf{F3}}: Presenters commonly annotate terminology but substitute embellished words in the polished script.} Presenters often simplify terminology in polished scripts to "\textit{connect better with their audience}" (S23) and "\textit{ensure smooth communication}" (S14). Six emphasized clearer pronunciation for easier concept "\textit{grasp}" (S37, S39). Five noted personal preferences, using transitions for "\textit{more comfort and confidence}" (S21, S37, S47), aiding a "\textit{natural flow}" (S14). Revisions were driven by content confirmation (nine) and personal language alignment (seven).


\begin{figure*}
\vspace{-3mm}
  \centering
  \includegraphics[width=\textwidth]{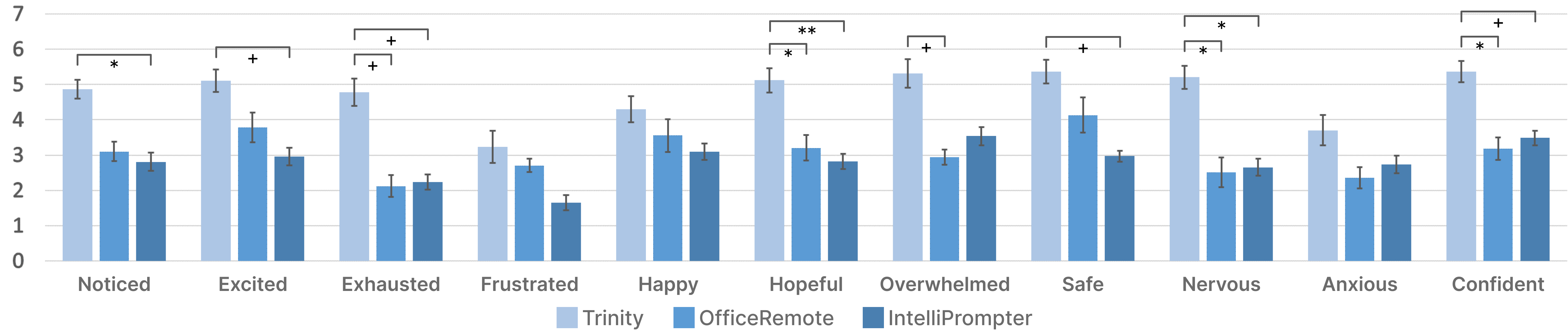}
  \vspace{-6mm}
  \caption{The effects on presenters' emotions and related feelings. The error bars indicate standard errors. ($+:p<.1$; $*:p<.05$; $**:p<.01$; $***:p<.001$).}
\vspace{-3mm}
  \label{fig:Emotion}
\end{figure*}

\par \textbf{\prefix{\textsf{F4}}: Presenters manually scroll through the script when they experience speech recognition jitters.} Despite speech recognition's help in delivering speeches "\textit{without having to constantly refer back to physical scripts}" (S17, S31), glitches can occur. Presenters then intervene by scrolling or "\textit{swiping the screen with the other hand}" (S23, S39, S45). One reported a script that "\textit{suddenly jumped down}" (S39), necessitating a scroll back to resume smoothly.

\par \textbf{ii) Effects on behavior and emotions.} We investigated how \textit{Trinity} influenced both students' presentation behaviors and emotions, resulting in the following insights.

\par \textbf{\prefix{\textsf{F5}}: Emojis prompted presenters to adapt their delivery.} Emojis in presentations enhanced delivery and audience engagement in the \textit{Trinity} condition. They helped presenters "\textit{effectively convey emotions}" (S42) and "\textit{emphasize key points}" (S21, S64). Emojis acted as visual cues for "\textit{how to deliver the content at different times}" (S14), prompting appropriate facial expressions and gestures that improved "\textit{communication with the audience}" (S28). This nonverbal layer made presentations "\textit{dynamic and engaging}" (S37), aiding audience understanding and maintaining attention as they "\textit{followed along with both verbal and nonverbal cues}" (S31).


\begin{figure}
\vspace{-3mm}
 \centering 
 \includegraphics[width=0.9\linewidth]{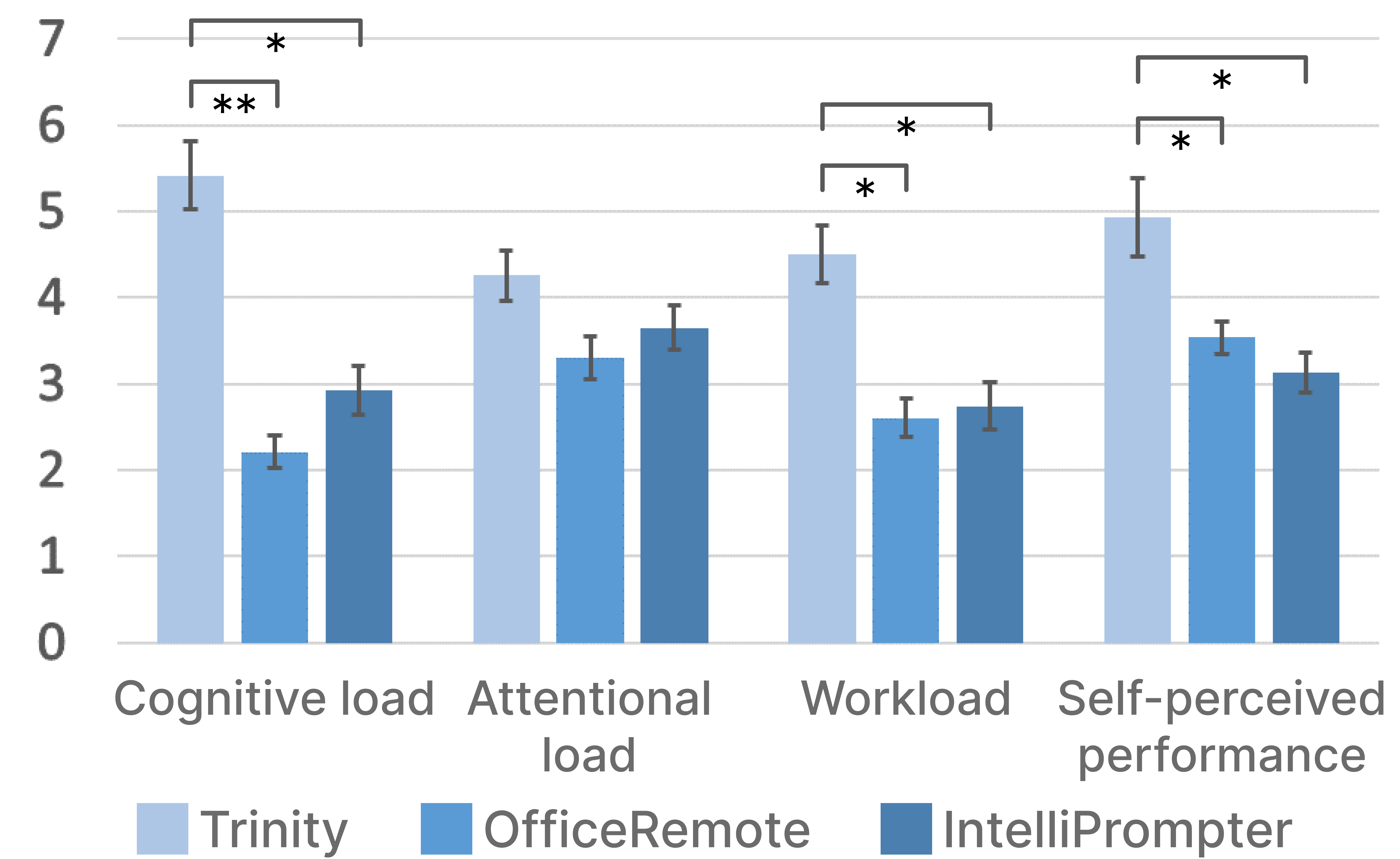}
 \vspace{-3mm}
 \caption{The effects on presenters' cognitive load, attentioal load, workload, and self-perceived performance. The error bars indicate standard errors. ($+:p<.1$; $*:p<.05$; $**:p<.01$; $***:p<.001$).}
 \label{fig:Load}
  \vspace{-3mm}
\end{figure}

\par \textbf{\prefix{\textsf{F6}}: Presenters had a more positive outlook on their presentation experience with \textit{Trinity}.} 
\autoref{fig:Emotion} shows that different presentation support systems significantly affected presenters' feelings of being noticed ($H=10.236, p=0.022$), hopeful ($H=8.431, p=0.034$), nervous ($H=11.237, p=0.013$), and confident ($H=9.842, p=0.019$). Post-hoc tests reveal that \textit{Trinity} made presenters feel more noticed than \textit{IntelliPrompter} (all $p<0.1$) and potentially increased engagement compared to \textit{OfficeRemote}. Presenters felt more hopeful using \textit{Trinity} than \textit{IntelliPrompter} ($p=0.003$) and \textit{OfficeRemote} ($p=0.017$). Also, \textit{Trinity} made presenters feel less nervous and more confident than \textit{IntelliPrompter} (nervous: $p=0.015$, confident: $p=0.540$) and \textit{OfficeRemote} (nervous: $p=0.028$, confident: $p=0.022$). No significant differences were found across the systems regarding feelings of excitement, exhaustion, frustration, happiness, being overwhelmed, or anxiety.

\par \textbf{\prefix{\textsf{F7}}: Presenters in-situ emotions during presentation can be influenced by the system's performance.} 
In retrospective reviews of the \textit{Trinity} condition, most presenters (18 out of 21) saw on-the-fly delivery support as "\textit{a reliable backup for presentation}" (S17) and "\textit{pillar to learn on}" (S2). During smooth system operation, they felt confident and more engaged with their audience, "\textit{feeling more relaxed and perform better}" (S34). However, technical glitches like an "\textit{unresponsive App}" (S7, S42) led to "\textit{feeling flustered and frustrated}" (S7).

\par \textbf{iii) Effects on workload, attentional load, and cognitive load.} 
\autoref{fig:Load} displays average ratings of student presenters' workload, attentional load, and cognitive load. Kruskal-Wallis tests reveal significant differences in workload ($H=9.395, p=0.010$), cognitive load ($H=12.365, p=0.004$), and self-perceived performance ($H=11.545, p=0.010$) across different support systems. Post-hoc tests indicate \textit{Trinity} incurs significantly higher workload ($M=4.54, SD=0.83$, $p=0.025$ with \textit{IntelliPrompter}, $p=0.034$ with \textit{OfficeRemote}), cognitive load ($M=5.68, SD=1.24$, $p=0.036$ with \textit{IntelliPrompter}, $p=0.007$ with \textit{OfficeRemote}), and improves self-perceived performance ($M=5.89, SD=0.94$, $p=0.021$ with \textit{IntelliPrompter}, $p=0.014$ with \textit{OfficeRemote}). No significant difference in attentional load implies similar attention allocation across conditions. Evidently, the baselines impose less workload and cognitive load as they only involve script reading without additional operations.

\begin{figure}[h]
 \vspace{-3mm}
 \centering 
 \includegraphics[width=0.95\linewidth]{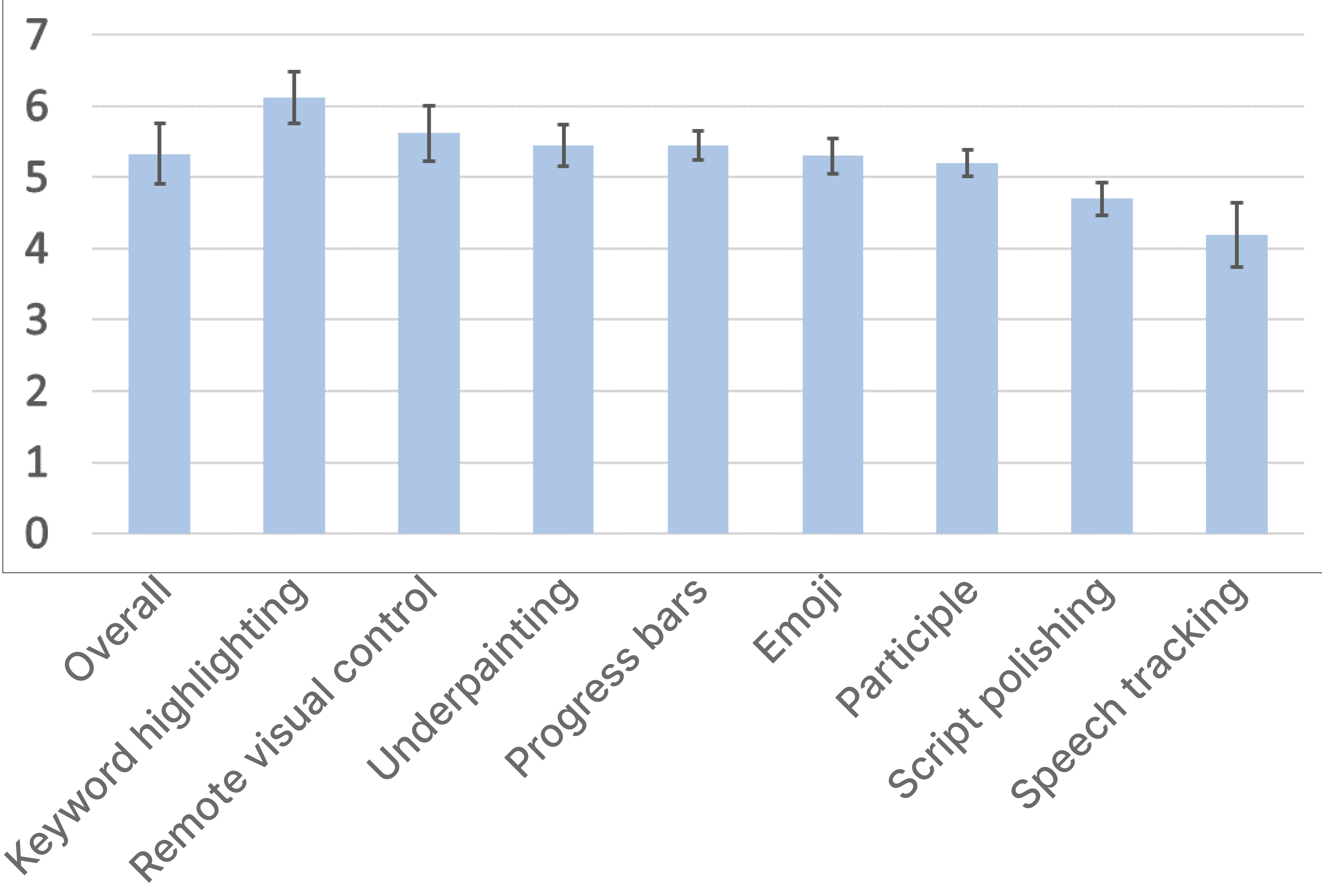}
 \vspace{-4mm}
 \caption{Student presenters' trust levels (in $7$-point Likert) on different system features.}
 \label{fig:Trust}
  \vspace{-3mm}
\end{figure}

\par \textbf{\prefix{\textsf{F8}}: Presenters primarily concentrate on their scripts and briefly glance at emojis during presentations.} 
Presenters prioritized scripts for fluency and confidence, believing it helped them deliver messages effectively without "\textit{missing any important points}" (S21). Many admitted to being "\textit{not fully prepared}" (S33) and felt "\textit{more comfortable relying on a structured plan}" (S55). In the \textit{Trinity} condition, presenters still focused on scripts but "\textit{briefly glanced at the emojis}" (S23, S37, S39, S64), which did not increase attentional load and seemed to enhance comprehension without "\textit{distracting from following along with the script}" (S14).

\par \textbf{\prefix{\textsf{F9}}: Cognitive load increases when malfunctions or recognition jitters occur.} Cognitive load is the mental effort needed for information processing, which can rise during presentation issues. Problems like the ``\textit{Trinity App got stuck}'' (S7, S42) require presenters to troubleshoot, ``\textit{interrupting the train of thought}'' (S7, S42) and adding cognitive burden as they ``\textit{understand what went wrong}'' (S7) and seek solutions, such as ``\textit{check connectivity}'' (S7) or ``\textit{seek technical assistance}'' (S42). Sudden topic shifts also increase cognitive load, as presenters need to ``\textit{backtrack and find where they left off}'' (S7), requiring effort to ``\textit{recall previous content}'' (S42) for a smooth transition.

\par \textbf{\prefix{\textsf{F10}}: Cognitive load diminishes as presenters become more acquainted with the system.} In the \textit{Trinity} condition, presenters initially felt a cognitive burden, but this lessened with practice. They were instructed to preprocess at least half a day before, and four found that practicing with different content helped them "\textit{better understand its functionalities}" (S21, S45, S52) and adapt their presentations. This familiarity "\textit{reduced the cognitive load}" (S45) and "\textit{improved the overall performance}" (S21).

\subsection{RQ5. How will students trust and collaborate with our system?}
\par Finally, as \textit{Trinity} is a hybrid mobile-centric system offering on-the-fly delivery support, we aim to explore how students trust and engage with our system, as well as the factors influencing their trust and collaboration. 

\par \textbf{i) Trust.} \autoref{fig:Trust} shows presenters' average ratings on a $7$-point Likert scale regarding their trust in the system in the post-task questionnaire. We combine these ratings with qualitative insights from the interviews to present the following the findings.

\par \textbf{\prefix{\textsf{F11}}: Students' inherent perceptions of LLMs influence their trust.} Most student presenters (16 out of 21) trust the system more when they know it uses a high-quality LLM. Initially, students may doubt the effectiveness of an unknown LLM, questioning if the output is "\textit{reliable}" (S28) or if there "\textit{could be potential issues}" (S39). However, once informed of an advanced LLM, they gain a "\textit{sense of confidence and assurance}" (S47), trusting the system to be "\textit{accurate and dependable}" (S28). This leads to increased trust and a stronger bond with the system. The use of an advanced LLM also signifies "\textit{continuous improvements}" (S39), which students see as a "\textit{testament to the reliability and credibility}" (S62) of the system.

\par \textbf{\prefix{\textsf{F12}}: System malfunctions could impair students' trust.} Five presenters worried about system malfunctions affecting their trust, noting it could lead to a "\textit{significant decrease in trust}" (S7, S23, S39) and an "\textit{unmanageable sense of crisis}" (S23, S42). Four shared experiences of voice tracking jitters, which made them "\textit{lose confidence in the voice tracking}" (S23, S45) feature. To avoid issues, they manually "\textit{scrolled through the script}" (S23, S39). This shows minor glitches can erode trust, as seen in \autoref{fig:Trust} where "\textit{Speech Tracking}" had lower trust ratings.

\par \textbf{ii) Collaboration.} Beyond trust assessment, our study uncovered valuable insights into the collaborative dynamics between students and our system, gleaned from their open-ended feedback during the interviews.

\par \textbf{\prefix{\textsf{F13}}: The system could serve as a coach for students, offering guidance on their presentation content.} Of 21 student presenters, 16 found the system's guidance on presentation strategies valuable. One student (S3) noted, "\textit{Trinity illuminated a clear path for me by refining the script and generating prompts for delivery, enabling me to enhance my ability to express myself effectively.}" This indicates the system's role in improving language proficiency and communication skills. Additionally, students like (S34, S47) would "\textit{study the polished script}" and (S17) "\textit{identify expressions and sentence structures worth learning.}" The script offered "\textit{concrete examples}" (S28) of enhanced wording and expression within their "\textit{own context}" (S40), suggesting the system's potential to aid in overall language acquisition.

\par \textbf{\prefix{\textsf{F14}}: Some students rely less on the system when presenting their own work.} Analysis from video replays showed that presenters reporting on their own work tended to refer to scripts less and improvise more: "\textit{I'm way more familiar with the stuff I've worked on, so I don't stick to the script and just throw in my own improvisations}" (S21). This suggests that as they gain confidence in their material, they ``\textit{feel empowered}'' (S31) to deviate from scripts, leading to a more natural and engaging presentation style (S27). Such improvisation ``\textit{tests the ability}'' (S14) to think critically and adapt on the spot, fostering creativity and innovation beyond memorized material (S64).

\section{Discussion and Limitation}

\subsection{Implications for Offering On-the-fly Support in Presentations} 

\par \noindent \textbf{\textsf{\prefix{Implication I}: Emphasizing intuitive and attention-grabbing designs for on-the-fly support tools.}} To ease presenters' cognitive load, \textit{Trinity} offers user-friendly features like thumbnail navigation, larger fonts, emojis for clarity, and a darker underpainting to show progress. We bolded keywords for emphasis. These underscores the importance of helping presenters handle prompts without cluttering their presentations, enhancing delivery quality and emphasizing the need for intuitive, eye-catching real-time support tool designs.


\par \noindent \textbf{\textsf{\prefix{Implication II}: Empowering presenters with autonomy over both content and presentation flow.}} During our research's early stages, we considered creating a fully automated system for script and slide optimization. However, feedback from participants led us to reconsider autonomy in content and control. Our user study showed that many presenters want to review and refine their scripts. Thus, \textit{Trinity} was designed for smooth transitions from manuscript to polished script, allowing for content revisions. Additionally, \textit{Trinity} provides manual control via smartphone interactions like tapping and swiping, ensuring control autonomy. We believe future presentation tools should focus on empowering users with autonomy in content and control to build trust and improve the user experience.

\par \noindent \textbf{\textsf{\prefix{Implication III}: Strengthen system robustness to uphold user trust.}} On-the-fly support systems like \textit{Trinity} face challenges in maintaining user trust due to their fragile nature. Our study showed that presenters who experienced system malfunctions during presentations had a significant drop in trust. This underscores the importance of system robustness to prevent trust erosion. We advise designers to focus on thorough testing to fix potential issues, ensuring a reliable and long-lasting user-trust relationship.

\subsection{Usability of LLMs from the Perception of Linguistic Expression}
\par LLMs like ChatGPT have expanded the potential for real-time support during presentations, demonstrating language comprehension and generation skills, including unfamiliar word association, sentence denoising, and syntactic reuse~\cite{xu2023does}. Expert feedback and user studies have validated these capabilities, highlighting improved fluency and syntax in edited scripts. However, for formal and objective language, the polished scripts align with Mu{\~n}oz et al.'s findings~\cite{munoz2023contrasting}, which is beneficial for academic presentations but requires further investigation for other genres like business talks.

\par When it comes to linguistic expression, there's a clear preference for verbal delivery prompts over nonverbal and visual ones, suggesting LLMs excel at verbal communication. The perception of linguistic expression also varies by the specific LLM, with GPT-3.5-turbo showing less proficiency in nonverbal and visual cues. For tasks involving linguistic expression, we recommend using LLMs with multi-modal training. Improving the consistency between LLM and human perceptions is expected to be a growing area of interest in both machine learning and HCI.

\subsection{Presentation Delivery Learning Potential for EFL Students}
\par Our focus is on the methodological aspect of presentation delivery learning for EFL students, diverging from traditional research that focuses on communication channel training before presentations. \textit{Trinity} aims to empower EFL students to use various communication channels effectively in their presentations, potentially fostering the acquisition of multichannel presentation skills through experiential learning \cite{kolb2014experiential}. However, user study participants suggested that while delivery prompts are effective, their trust and understanding could be improved with explanations of the rationale behind them. This insight points towards a more effective approach to experiential delivery learning for EFL students, one that integrates abstract concepts with concrete experiences enhances experiential learning for EFL students. By reviewing video playbacks and adjusting delivery prompts after presentations, they can better understand the logic behind their performance. This approach helps them build a personalized yet generalizable model of presentation logic.

\subsection{Towards a Comprehensive Ecosystem for Presentation Support}
\label{sec:ecosystem}
\par Students experience multiple stages in presentations, and \textit{Trinity} supports them from preparation to reflection. While it enhances delivery, concerns exist about potential overreliance on its real-time support without structured training. To mitigate this, we suggest \textit{Trinity} be integrated into a comprehensive presentation support system. It could complement tools like \textit{Slide4N}~\cite{wang2023slide4n} and \textit{HyperSlides}~\cite{edge2013hyperslides} by optimizing slides and scripts, reducing the need for extensive tutorials. This allows students to incorporate delivery patterns into their scripts and compare with \textit{Trinity}'s prompts for personalized learning. Additionally, feedback mechanisms such as \textit{SlideSpecs}~\cite{warner2023slidespecs} and \textit{AutoManner}~\cite{tanveer2016automanner} could provide audience feedback for performance reflection, completing the experiential learning cycle. By becoming part of a larger ecosystem, \textit{Trinity} encourages consideration of how tools fit into broader systems and promotes open designs for seamless integration with existing practices.

\subsection{Limitations and Future Work}
\par \textbf{\textsf{Towards Better Hardware Support.}} In our design study, storyboards evaluated design concepts, with the mobile and PC combination standing out. Chosen for their common use and minimal visual distraction, this pairing is not the ultimate solution for on-the-fly delivery support. It must meet three criteria: providing clear support without disrupting presenters' communication, remaining unnoticed by the audience, and being portable. Current media may not fully satisfy these needs. Our user study indicates that mobile devices could still limit presenters' nonverbal cues. Thus, we see the mobile and PC combination as progress towards better hardware support. Future research should seek media that can meet all three requirements.

\par \textbf{\textsf{Adapt LLMs to Personal Contexts.}} \textit{Trinity} uses LLMs to enhance script quality and create delivery cues. Most users in our study appreciated the improvements in wording and syntax, but some faced challenges with unfamiliar language and mismatched preferences. To remedy this, \textit{Trinity} allows users to edit the script freely. Yet, this might not be ideal for all EFL students, who may lack the willingness or patience to review thoroughly. A possible solution is LLM few-shot prompting, where a small set of relevant examples is provided to tailor the model's output. This could involve uploading past scripts as references, similar to custom GPT features. However, it's unclear if this approach can fully adapt the LLM to users' vocabulary and style. Another option is fine-tuning the LLM, which, despite being resource-intensive and potentially causing knowledge loss, could be more effective. Beyond script development, LLMs could also help users refine their presentation delivery styles, drawing inspiration from figures like Obama. Future research could aim to find more efficient ways to adapt LLMs to individual needs, enhancing personalization.

\par \textbf{\textsf{Technical Stability and Extendability.}} The \textit{Trinity} App, which uses real-time speech recognition to track presentations, faced issues with jitters and delays, affecting smooth speech tracking. Future studies could look into content-based navigation as a solution~\cite{myers2018patterns,chang2021rubyslippers}. Additionally, the app's slide thumbnail feature struggled with complex animations not fully visible due to PowerPoint interface constraints. Collaboration with Microsoft to integrate better animation support is suggested. Lastly, \textit{Trinity} enhances scripts based on user manuscripts but found that the manuscript's semantic richness and integrity greatly affect the presentation's perceived "content \& organization." Improving this could involve cross-referencing slides with scripts to optimize content and enhance underdeveloped manuscripts.

\section{Conclusion}

\par In this study, we introduced \textit{Trinity}, a hybrid system designed to synchronize verbal, nonverbal, and visual elements in academic oral presentations (AOPs). Developed through a formative study with a survey, design research, and expert interviews, \textit{Trinity} features a PC PowerPoint add-in for script enhancement and delivery with GPT-4, alongside an Android App for remote visual control, speech pacing, and delivery cues. User studies showed that \textit{Trinity} surpassed two baseline systems. The research deepened our understanding of AOP factors, provided design insights, and highlighted the system's impact on presentation practices. The experiences and challenges in \textit{Trinity}'s design offer key lessons for future presentation support tool development. We see \textit{Trinity} as an innovative step towards future presentation practices, potentially leading to more advanced human-AI collaboration in educational settings.

\begin{acks}
We thank anonymous reviewers for their valuable feedback. This work is supported by grants from the National Natural Science Foundation of China (No. 62372298), Shanghai Frontiers Science Center of Human-centered Artificial Intelligence (ShangHAI), and MoE Key Laboratory of Intelligent Perception and Human-Machine Collaboration (KLIP-HuMaCo).
\end{acks}

\bibliographystyle{ACM-Reference-Format}
\bibliography{sample-base}


\begin{thebibliography}{101}


\ifx \showCODEN    \undefined \def \showCODEN     #1{\unskip}     \fi
\ifx \showDOI      \undefined \def \showDOI       #1{#1}\fi
\ifx \showISBNx    \undefined \def \showISBNx     #1{\unskip}     \fi
\ifx \showISBNxiii \undefined \def \showISBNxiii  #1{\unskip}     \fi
\ifx \showISSN     \undefined \def \showISSN      #1{\unskip}     \fi
\ifx \showLCCN     \undefined \def \showLCCN      #1{\unskip}     \fi
\ifx \shownote     \undefined \def \shownote      #1{#1}          \fi
\ifx \showarticletitle \undefined \def \showarticletitle #1{#1}   \fi
\ifx \showURL      \undefined \def \showURL       {\relax}        \fi
\providecommand\bibfield[2]{#2}
\providecommand\bibinfo[2]{#2}
\providecommand\natexlab[1]{#1}
\providecommand\showeprint[2][]{arXiv:#2}

\bibitem[Ackovska(2013)]%
        {ackovska2013gesture}
\bibfield{author}{\bibinfo{person}{Darko Martinovikj~Nevena Ackovska}.}
  \bibinfo{year}{2013}\natexlab{}.
\newblock \showarticletitle{Gesture recognition solution for presentation
  control}. In \bibinfo{booktitle}{\emph{Proc. of 10th Conference for
  Informatics and Information Technology}}.
\newblock


\bibitem[Alexander et~al\mbox{.}(2008)]%
        {alexander2008eap}
\bibfield{author}{\bibinfo{person}{Olwyn Alexander}, \bibinfo{person}{Sue
  Argent}, {and} \bibinfo{person}{Jenifer Spencer}.}
  \bibinfo{year}{2008}\natexlab{}.
\newblock \bibinfo{booktitle}{\emph{EAP essentials: A teacher's guide to
  principles and practice}}.
\newblock \bibinfo{publisher}{Garnet Education}.
\newblock


\bibitem[Amalia and Ma'mun(2020)]%
        {amalia2020anxiety}
\bibfield{author}{\bibinfo{person}{Nur~Lina Amalia} {and}
  \bibinfo{person}{Nadiah Ma'mun}.} \bibinfo{year}{2020}\natexlab{}.
\newblock \showarticletitle{The anxiety of EFL students in presentation}.
\newblock \bibinfo{journal}{\emph{ELITE JOURNAL}} \bibinfo{volume}{2},
  \bibinfo{number}{1} (\bibinfo{year}{2020}), \bibinfo{pages}{65--84}.
\newblock


\bibitem[Argyle et~al\mbox{.}(1971)]%
        {argyle1971communication}
\bibfield{author}{\bibinfo{person}{Michael Argyle}, \bibinfo{person}{Florisse
  Alkema}, {and} \bibinfo{person}{Robin Gilmour}.}
  \bibinfo{year}{1971}\natexlab{}.
\newblock \showarticletitle{The communication of friendly and hostile attitudes
  by verbal and non-verbal signals}.
\newblock \bibinfo{journal}{\emph{European Journal of Social Psychology}}
  \bibinfo{volume}{1}, \bibinfo{number}{3} (\bibinfo{year}{1971}),
  \bibinfo{pages}{385--402}.
\newblock


\bibitem[Asadi et~al\mbox{.}(2017)]%
        {asadi2017intelliprompter}
\bibfield{author}{\bibinfo{person}{Reza Asadi}, \bibinfo{person}{Ha Trinh},
  \bibinfo{person}{Harriet~J Fell}, {and} \bibinfo{person}{Timothy~W
  Bickmore}.} \bibinfo{year}{2017}\natexlab{}.
\newblock \showarticletitle{IntelliPrompter: speech-based dynamic note display
  interface for oral presentations}. In \bibinfo{booktitle}{\emph{Proceedings
  of the 19th ACM International Conference on Multimodal Interaction}}.
  \bibinfo{pages}{172--180}.
\newblock


\bibitem[Austin and Sweller(2014)]%
        {austin2014presentation}
\bibfield{author}{\bibinfo{person}{Elizabeth~E Austin} {and}
  \bibinfo{person}{Naomi Sweller}.} \bibinfo{year}{2014}\natexlab{}.
\newblock \showarticletitle{Presentation and production: The role of gesture in
  spatial communication}.
\newblock \bibinfo{journal}{\emph{Journal of experimental child psychology}}
  \bibinfo{volume}{122} (\bibinfo{year}{2014}), \bibinfo{pages}{92--103}.
\newblock


\bibitem[Beebe(1976)]%
        {beebe1976effects}
\bibfield{author}{\bibinfo{person}{Steven~A Beebe}.}
  \bibinfo{year}{1976}\natexlab{}.
\newblock \showarticletitle{Effects of Eye Contact, Posture and Vocal
  Inflection upon Credibility and Comprehension.}
\newblock  (\bibinfo{year}{1976}).
\newblock


\bibitem[Bich-Carri{\`e}re(2019)]%
        {bich2019say}
\bibfield{author}{\bibinfo{person}{Laurence Bich-Carri{\`e}re}.}
  \bibinfo{year}{2019}\natexlab{}.
\newblock \showarticletitle{Say it with [a smiling face with smiling eyes]:
  judicial use and legal challenges with emoji interpretation in Canada}.
\newblock \bibinfo{journal}{\emph{International Journal for the Semiotics of
  Law-Revue internationale de S{\'e}miotique juridique}} \bibinfo{volume}{32},
  \bibinfo{number}{2} (\bibinfo{year}{2019}), \bibinfo{pages}{283--319}.
\newblock


\bibitem[Bickmore et~al\mbox{.}(2021)]%
        {bickmore2021virtual}
\bibfield{author}{\bibinfo{person}{Timothy Bickmore}, \bibinfo{person}{Everlyne
  Kimani}, \bibinfo{person}{Ameneh Shamekhi}, \bibinfo{person}{Prasanth
  Murali}, \bibinfo{person}{Dhaval Parmar}, {and} \bibinfo{person}{Ha Trinh}.}
  \bibinfo{year}{2021}\natexlab{}.
\newblock \showarticletitle{Virtual agents as supporting media for scientific
  presentations}.
\newblock \bibinfo{journal}{\emph{Journal on Multimodal User Interfaces}}
  \bibinfo{volume}{15} (\bibinfo{year}{2021}), \bibinfo{pages}{131--146}.
\newblock


\bibitem[Bourne(2007)]%
        {bourne2007ten}
\bibfield{author}{\bibinfo{person}{Philip~E Bourne}.}
  \bibinfo{year}{2007}\natexlab{}.
\newblock \bibinfo{title}{Ten simple rules for making good oral presentations}.
\newblock , \bibinfo{numpages}{e77}~pages.
\newblock


\bibitem[Braun and Clarke(2006)]%
        {braun2006using}
\bibfield{author}{\bibinfo{person}{Virginia Braun} {and}
  \bibinfo{person}{Victoria Clarke}.} \bibinfo{year}{2006}\natexlab{}.
\newblock \showarticletitle{Using thematic analysis in psychology}.
\newblock \bibinfo{journal}{\emph{Qualitative research in psychology}}
  \bibinfo{volume}{3}, \bibinfo{number}{2} (\bibinfo{year}{2006}),
  \bibinfo{pages}{77--101}.
\newblock


\bibitem[Brehmer(1976)]%
        {brehmer1976social}
\bibfield{author}{\bibinfo{person}{Berndt Brehmer}.}
  \bibinfo{year}{1976}\natexlab{}.
\newblock \showarticletitle{Social judgment theory and the analysis of
  interpersonal conflict.}
\newblock \bibinfo{journal}{\emph{Psychological bulletin}}
  \bibinfo{volume}{83}, \bibinfo{number}{6} (\bibinfo{year}{1976}),
  \bibinfo{pages}{985}.
\newblock


\bibitem[Bubel et~al\mbox{.}(2016)]%
        {bubel2016awareme}
\bibfield{author}{\bibinfo{person}{Mark Bubel}, \bibinfo{person}{Ruiwen Jiang},
  \bibinfo{person}{Christine~H Lee}, \bibinfo{person}{Wen Shi}, {and}
  \bibinfo{person}{Audrey Tse}.} \bibinfo{year}{2016}\natexlab{}.
\newblock \showarticletitle{AwareMe: addressing fear of public speech through
  awareness}. In \bibinfo{booktitle}{\emph{Proceedings of the 2016 CHI
  conference extended abstracts on human factors in computing systems}}.
  \bibinfo{pages}{68--73}.
\newblock


\bibitem[Cao et~al\mbox{.}(2005)]%
        {cao2005evaluation}
\bibfield{author}{\bibinfo{person}{Xiang Cao}, \bibinfo{person}{Eyal Ofek},
  {and} \bibinfo{person}{David Vronay}.} \bibinfo{year}{2005}\natexlab{}.
\newblock \showarticletitle{Evaluation of alternative presentation control
  techniques}. In \bibinfo{booktitle}{\emph{CHI'05 Extended Abstracts on Human
  Factors in Computing Systems}}. \bibinfo{pages}{1248--1251}.
\newblock


\bibitem[Carter(2012)]%
        {carter2012designing}
\bibfield{author}{\bibinfo{person}{Matt Carter}.}
  \bibinfo{year}{2012}\natexlab{}.
\newblock \bibinfo{booktitle}{\emph{Designing science presentations: A visual
  guide to figures, papers, slides, posters, and more}}.
\newblock \bibinfo{publisher}{Academic Press}.
\newblock


\bibitem[Chang et~al\mbox{.}(2021)]%
        {chang2021rubyslippers}
\bibfield{author}{\bibinfo{person}{Minsuk Chang}, \bibinfo{person}{Mina Huh},
  {and} \bibinfo{person}{Juho Kim}.} \bibinfo{year}{2021}\natexlab{}.
\newblock \showarticletitle{Rubyslippers: Supporting content-based voice
  navigation for how-to videos}. In \bibinfo{booktitle}{\emph{Proceedings of
  the 2021 CHI Conference on Human Factors in Computing Systems}}.
  \bibinfo{pages}{1--14}.
\newblock


\bibitem[Chen et~al\mbox{.}(2023)]%
        {chen2023meetscript}
\bibfield{author}{\bibinfo{person}{Xinyue Chen}, \bibinfo{person}{Shuo Li},
  \bibinfo{person}{Shipeng Liu}, \bibinfo{person}{Robin Fowler}, {and}
  \bibinfo{person}{Xu Wang}.} \bibinfo{year}{2023}\natexlab{}.
\newblock \showarticletitle{MeetScript: Designing Transcript-based Interactions
  to Support Active Participation in Group Video Meetings}.
\newblock \bibinfo{journal}{\emph{Proceedings of the ACM on Human-Computer
  Interaction}} \bibinfo{volume}{7}, \bibinfo{number}{CSCW2}
  (\bibinfo{year}{2023}), \bibinfo{pages}{1--32}.
\newblock


\bibitem[Cheung(2008)]%
        {cheung2008teaching}
\bibfield{author}{\bibinfo{person}{Yin~Ling Cheung}.}
  \bibinfo{year}{2008}\natexlab{}.
\newblock \showarticletitle{Teaching effective presentation skills to ESL/EFL
  students}.
\newblock \bibinfo{journal}{\emph{The Internet TESL Journal}}
  \bibinfo{volume}{14}, \bibinfo{number}{6} (\bibinfo{year}{2008}),
  \bibinfo{pages}{1--2}.
\newblock


\bibitem[Chi et~al\mbox{.}(2014)]%
        {chi2014demowiz}
\bibfield{author}{\bibinfo{person}{Pei-Yu Chi}, \bibinfo{person}{Bongshin Lee},
  {and} \bibinfo{person}{Steven~M Drucker}.} \bibinfo{year}{2014}\natexlab{}.
\newblock \showarticletitle{DemoWiz: re-performing software demonstrations for
  a live presentation}. In \bibinfo{booktitle}{\emph{Proceedings of the SIGCHI
  Conference on Human Factors in Computing Systems}}.
  \bibinfo{pages}{1581--1590}.
\newblock


\bibitem[Chivers and Shoolbred(2007)]%
        {chivers2007student}
\bibfield{author}{\bibinfo{person}{Barbara Chivers} {and}
  \bibinfo{person}{Michael Shoolbred}.} \bibinfo{year}{2007}\natexlab{}.
\newblock \bibinfo{booktitle}{\emph{A student' s guide to presentations: Making
  your presentation count}}.
\newblock \bibinfo{publisher}{Sage}.
\newblock


\bibitem[Damian et~al\mbox{.}(2015)]%
        {damian2015augmenting}
\bibfield{author}{\bibinfo{person}{Ionut Damian}, \bibinfo{person}{Chiew~Seng
  Tan}, \bibinfo{person}{Tobias Baur}, \bibinfo{person}{Johannes Sch{\"o}ning},
  \bibinfo{person}{Kris Luyten}, {and} \bibinfo{person}{Elisabeth Andr{\'e}}.}
  \bibinfo{year}{2015}\natexlab{}.
\newblock \showarticletitle{Augmenting social interactions: Realtime
  behavioural feedback using social signal processing techniques}. In
  \bibinfo{booktitle}{\emph{Proceedings of the 33rd annual ACM conference on
  Human factors in computing systems}}. \bibinfo{pages}{565--574}.
\newblock


\bibitem[De~Grez et~al\mbox{.}(2014)]%
        {de2014differential}
\bibfield{author}{\bibinfo{person}{Luc De~Grez}, \bibinfo{person}{Martin
  Valcke}, {and} \bibinfo{person}{Irene Roozen}.}
  \bibinfo{year}{2014}\natexlab{}.
\newblock \showarticletitle{The differential impact of observational learning
  and practice-based learning on the development of oral presentation skills in
  higher education}.
\newblock \bibinfo{journal}{\emph{Higher Education Research \& Development}}
  \bibinfo{volume}{33}, \bibinfo{number}{2} (\bibinfo{year}{2014}),
  \bibinfo{pages}{256--271}.
\newblock


\bibitem[Dennis and Kinney(1998)]%
        {dennis1998testing}
\bibfield{author}{\bibinfo{person}{Alan~R Dennis} {and}
  \bibinfo{person}{Susan~T Kinney}.} \bibinfo{year}{1998}\natexlab{}.
\newblock \showarticletitle{Testing media richness theory in the new media: The
  effects of cues, feedback, and task equivocality}.
\newblock \bibinfo{journal}{\emph{Information systems research}}
  \bibinfo{volume}{9}, \bibinfo{number}{3} (\bibinfo{year}{1998}),
  \bibinfo{pages}{256--274}.
\newblock


\bibitem[Dolan(2017)]%
        {dolan2017effective}
\bibfield{author}{\bibinfo{person}{Robert Dolan}.}
  \bibinfo{year}{2017}\natexlab{}.
\newblock \showarticletitle{Effective presentation skills}.
\newblock \bibinfo{journal}{\emph{FEMS microbiology letters}}
  \bibinfo{volume}{364}, \bibinfo{number}{24} (\bibinfo{year}{2017}),
  \bibinfo{pages}{fnx235}.
\newblock


\bibitem[Duarte(2008)]%
        {duarte2008slide}
\bibfield{author}{\bibinfo{person}{Nancy Duarte}.}
  \bibinfo{year}{2008}\natexlab{}.
\newblock \bibinfo{booktitle}{\emph{Slide: ology: The art and science of
  creating great presentations}}. Vol.~\bibinfo{volume}{1}.
\newblock \bibinfo{publisher}{O'Reilly Media Sebastapol}.
\newblock


\bibitem[Duff(2007)]%
        {duff2007second}
\bibfield{author}{\bibinfo{person}{Patricia~A Duff}.}
  \bibinfo{year}{2007}\natexlab{}.
\newblock \showarticletitle{Second language socialization as sociocultural
  theory: Insights and issues}.
\newblock \bibinfo{journal}{\emph{Language teaching}} \bibinfo{volume}{40},
  \bibinfo{number}{4} (\bibinfo{year}{2007}), \bibinfo{pages}{309--319}.
\newblock


\bibitem[Edge et~al\mbox{.}(2013)]%
        {edge2013hyperslides}
\bibfield{author}{\bibinfo{person}{Darren Edge}, \bibinfo{person}{Joan Savage},
  {and} \bibinfo{person}{Koji Yatani}.} \bibinfo{year}{2013}\natexlab{}.
\newblock \showarticletitle{HyperSlides: dynamic presentation prototyping}. In
  \bibinfo{booktitle}{\emph{Proceedings of the SIGCHI Conference on Human
  Factors in Computing Systems}}. \bibinfo{pages}{671--680}.
\newblock


\bibitem[Edge et~al\mbox{.}(2016)]%
        {edge2016slidespace}
\bibfield{author}{\bibinfo{person}{Darren Edge}, \bibinfo{person}{Xi Yang},
  \bibinfo{person}{Yasmine Kotturi}, \bibinfo{person}{Shuoping Wang},
  \bibinfo{person}{Dan Feng}, \bibinfo{person}{Bongshin Lee}, {and}
  \bibinfo{person}{Steven Drucker}.} \bibinfo{year}{2016}\natexlab{}.
\newblock \showarticletitle{Slidespace: Heuristic design of a hybrid
  presentation medium}.
\newblock \bibinfo{journal}{\emph{ACM Transactions on Computer-Human
  Interaction (TOCHI)}} \bibinfo{volume}{23}, \bibinfo{number}{3}
  (\bibinfo{year}{2016}), \bibinfo{pages}{1--30}.
\newblock


\bibitem[El~Enein(2011)]%
        {el2011difficulties}
\bibfield{author}{\bibinfo{person}{Ayman Hassan~Abu El~Enein}.}
  \bibinfo{year}{2011}\natexlab{}.
\newblock \showarticletitle{Difficulties encountering English majors in giving
  academic oral presentations during class at Al-Aqsa University}.
\newblock \bibinfo{journal}{\emph{Unpublished Master’s thesis, Islamic
  University of Gaza}} (\bibinfo{year}{2011}).
\newblock


\bibitem[Furnham et~al\mbox{.}(1981)]%
        {furnham1981relative}
\bibfield{author}{\bibinfo{person}{Adrian Furnham}, \bibinfo{person}{Robert
  Trevethan}, {and} \bibinfo{person}{George Gaskell}.}
  \bibinfo{year}{1981}\natexlab{}.
\newblock \showarticletitle{The relative contribution of verbal, vocal, and
  visual channels to person perception: Experiment and critique}.
\newblock  (\bibinfo{year}{1981}).
\newblock


\bibitem[Giles et~al\mbox{.}(2007)]%
        {giles2007communication}
\bibfield{author}{\bibinfo{person}{Howard Giles}, \bibinfo{person}{Tania Ogay},
  {et~al\mbox{.}}} \bibinfo{year}{2007}\natexlab{}.
\newblock \showarticletitle{Communication accommodation theory}.
\newblock  (\bibinfo{year}{2007}).
\newblock


\bibitem[Gillis and Nilsen(2017)]%
        {gillis2017consistency}
\bibfield{author}{\bibinfo{person}{Randall~L Gillis} {and}
  \bibinfo{person}{Elizabeth~S Nilsen}.} \bibinfo{year}{2017}\natexlab{}.
\newblock \showarticletitle{Consistency between verbal and non-verbal affective
  cues: A clue to speaker credibility}.
\newblock \bibinfo{journal}{\emph{Cognition and Emotion}} \bibinfo{volume}{31},
  \bibinfo{number}{4} (\bibinfo{year}{2017}), \bibinfo{pages}{645--656}.
\newblock


\bibitem[Glasgow(1952)]%
        {glasgow1952semantic}
\bibfield{author}{\bibinfo{person}{George~M Glasgow}.}
  \bibinfo{year}{1952}\natexlab{}.
\newblock \showarticletitle{A semantic index of vocal pitch}.
\newblock \bibinfo{journal}{\emph{Communications Monographs}}
  \bibinfo{volume}{19}, \bibinfo{number}{1} (\bibinfo{year}{1952}),
  \bibinfo{pages}{64--68}.
\newblock


\bibitem[Hart and Staveland(1988)]%
        {hart1988development}
\bibfield{author}{\bibinfo{person}{Sandra~G Hart} {and}
  \bibinfo{person}{Lowell~E Staveland}.} \bibinfo{year}{1988}\natexlab{}.
\newblock \showarticletitle{Development of NASA-TLX (Task Load Index): Results
  of empirical and theoretical research}.
\newblock In \bibinfo{booktitle}{\emph{Advances in psychology}}.
  Vol.~\bibinfo{volume}{52}. \bibinfo{publisher}{Elsevier},
  \bibinfo{pages}{139--183}.
\newblock


\bibitem[Hentz(2006)]%
        {hentz2006enhancing}
\bibfield{author}{\bibinfo{person}{Brian~S Hentz}.}
  \bibinfo{year}{2006}\natexlab{}.
\newblock \showarticletitle{Enhancing presentation narratives through written
  and visual integration}.
\newblock \bibinfo{journal}{\emph{Business Communication Quarterly}}
  \bibinfo{volume}{69}, \bibinfo{number}{4} (\bibinfo{year}{2006}),
  \bibinfo{pages}{425--429}.
\newblock


\bibitem[Hertz(2015)]%
        {hertz2015spotlight}
\bibfield{author}{\bibinfo{person}{Brigitte Hertz}.}
  \bibinfo{year}{2015}\natexlab{}.
\newblock \emph{\bibinfo{title}{Spotlight on the presenter: a study into
  presentations of conference papers with PowerPoint}}.
\newblock \bibinfo{thesistype}{Ph.\,D. Dissertation}.
  \bibinfo{school}{Wageningen University and Research}.
\newblock


\bibitem[Hincks(2005)]%
        {hincks2005measures}
\bibfield{author}{\bibinfo{person}{Rebecca Hincks}.}
  \bibinfo{year}{2005}\natexlab{}.
\newblock \showarticletitle{Measures and perceptions of liveliness in student
  oral presentation speech: A proposal for an automatic feedback mechanism}.
\newblock \bibinfo{journal}{\emph{System}} \bibinfo{volume}{33},
  \bibinfo{number}{4} (\bibinfo{year}{2005}), \bibinfo{pages}{575--591}.
\newblock


\bibitem[Hoogterp(2014)]%
        {hoogterp2014your}
\bibfield{author}{\bibinfo{person}{Bill Hoogterp}.}
  \bibinfo{year}{2014}\natexlab{}.
\newblock \bibinfo{booktitle}{\emph{Your Perfect Presentation: Speak in Front
  of Any Audience Anytime Anywhere and Never Be Nervous Again}}.
\newblock \bibinfo{publisher}{McGraw Hill Professional}.
\newblock


\bibitem[Hunyadi and Szekr{\'e}nyes(2020)]%
        {hunyadi2020temporal}
\bibfield{author}{\bibinfo{person}{Laszlo Hunyadi} {and}
  \bibinfo{person}{Istv{\'a}n Szekr{\'e}nyes}.}
  \bibinfo{year}{2020}\natexlab{}.
\newblock \bibinfo{booktitle}{\emph{The Temporal Structure of Multimodal
  Communication}}.
\newblock \bibinfo{publisher}{Springer}.
\newblock


\bibitem[Imaniah(2018)]%
        {imaniah2018students}
\bibfield{author}{\bibinfo{person}{Ikhfi Imaniah}.}
  \bibinfo{year}{2018}\natexlab{}.
\newblock \showarticletitle{The students’ difficulties in presenting the
  academic speaking presentation}.
\newblock \bibinfo{journal}{\emph{Globish (An English-Indonesian Journal for
  English, Education and Culture)}} (\bibinfo{year}{2018}).
\newblock


\bibitem[Johnson(1971)]%
        {johnson1971role}
\bibfield{author}{\bibinfo{person}{David~W Johnson}.}
  \bibinfo{year}{1971}\natexlab{}.
\newblock \showarticletitle{Role reversal: A summary and review of the
  research.}
\newblock \bibinfo{journal}{\emph{International Journal of Group Tensions}}
  (\bibinfo{year}{1971}).
\newblock


\bibitem[Jokela et~al\mbox{.}(2008)]%
        {jokela2008mobile}
\bibfield{author}{\bibinfo{person}{Tero Jokela}, \bibinfo{person}{Jaakko~T
  Lehikoinen}, {and} \bibinfo{person}{Hannu Korhonen}.}
  \bibinfo{year}{2008}\natexlab{}.
\newblock \showarticletitle{Mobile multimedia presentation editor: enabling
  creation of audio-visual stories on mobile devices}. In
  \bibinfo{booktitle}{\emph{Proceedings of the SIGCHI Conference on Human
  Factors in Computing Systems}}. \bibinfo{pages}{63--72}.
\newblock


\bibitem[Kaur and Ali(2018)]%
        {kaur2018exploring}
\bibfield{author}{\bibinfo{person}{Kuldip Kaur} {and}
  \bibinfo{person}{Afida~Mohamad Ali}.} \bibinfo{year}{2018}\natexlab{}.
\newblock \showarticletitle{Exploring the genre of academic oral presentations:
  A critical review}.
\newblock \bibinfo{journal}{\emph{International Journal of Applied Linguistics
  and English Literature}} \bibinfo{volume}{7}, \bibinfo{number}{1}
  (\bibinfo{year}{2018}), \bibinfo{pages}{152--162}.
\newblock


\bibitem[Kelch(1985)]%
        {kelch1985modified}
\bibfield{author}{\bibinfo{person}{Ken Kelch}.}
  \bibinfo{year}{1985}\natexlab{}.
\newblock \showarticletitle{Modified input as an aid to comprehension}.
\newblock \bibinfo{journal}{\emph{Studies in second language acquisition}}
  \bibinfo{volume}{7}, \bibinfo{number}{1} (\bibinfo{year}{1985}),
  \bibinfo{pages}{81--90}.
\newblock


\bibitem[Khataei and Arya(2014)]%
        {khataei2014personalized}
\bibfield{author}{\bibinfo{person}{Amirsam Khataei} {and} \bibinfo{person}{Ali
  Arya}.} \bibinfo{year}{2014}\natexlab{}.
\newblock \showarticletitle{Personalized presentation builder}.
\newblock In \bibinfo{booktitle}{\emph{CHI'14 Extended Abstracts on Human
  Factors in Computing Systems}}. \bibinfo{pages}{2293--2298}.
\newblock


\bibitem[Kim and Wohn(2018)]%
        {kim2018holobox}
\bibfield{author}{\bibinfo{person}{Minju Kim} {and} \bibinfo{person}{Kwangyun
  Wohn}.} \bibinfo{year}{2018}\natexlab{}.
\newblock \showarticletitle{HoloBox: Augmented visualization and presentation
  with spatially integrated presenter}.
\newblock \bibinfo{journal}{\emph{Interacting with Computers}}
  \bibinfo{volume}{30}, \bibinfo{number}{3} (\bibinfo{year}{2018}),
  \bibinfo{pages}{224--242}.
\newblock


\bibitem[King(2002)]%
        {king2002preparing}
\bibfield{author}{\bibinfo{person}{Jane King}.}
  \bibinfo{year}{2002}\natexlab{}.
\newblock \showarticletitle{Preparing EFL Learners for Oral Presentations}.
\newblock \bibinfo{journal}{\emph{Dong Hwa Journal of Humanistic Studies}}
  \bibinfo{volume}{4} (\bibinfo{year}{2002}), \bibinfo{pages}{401--418}.
\newblock


\bibitem[Kolb(2014)]%
        {kolb2014experiential}
\bibfield{author}{\bibinfo{person}{David~A Kolb}.}
  \bibinfo{year}{2014}\natexlab{}.
\newblock \bibinfo{booktitle}{\emph{Experiential learning: Experience as the
  source of learning and development}}.
\newblock \bibinfo{publisher}{FT press}.
\newblock


\bibitem[Kurihara et~al\mbox{.}(2007)]%
        {kurihara2007presentation}
\bibfield{author}{\bibinfo{person}{Kazutaka Kurihara},
  \bibinfo{person}{Masataka Goto}, \bibinfo{person}{Jun Ogata},
  \bibinfo{person}{Yosuke Matsusaka}, {and} \bibinfo{person}{Takeo Igarashi}.}
  \bibinfo{year}{2007}\natexlab{}.
\newblock \showarticletitle{Presentation sensei: a presentation training system
  using speech and image processing}. In \bibinfo{booktitle}{\emph{Proceedings
  of the 9th international conference on Multimodal interfaces}}.
  \bibinfo{pages}{358--365}.
\newblock


\bibitem[Levine(2014)]%
        {levine2014truth}
\bibfield{author}{\bibinfo{person}{Timothy~R Levine}.}
  \bibinfo{year}{2014}\natexlab{}.
\newblock \showarticletitle{Truth-default theory (TDT) a theory of human
  deception and deception detection}.
\newblock \bibinfo{journal}{\emph{Journal of Language and Social Psychology}}
  \bibinfo{volume}{33}, \bibinfo{number}{4} (\bibinfo{year}{2014}),
  \bibinfo{pages}{378--392}.
\newblock


\bibitem[Levrai and Bolster(2015)]%
        {levrai2015developing}
\bibfield{author}{\bibinfo{person}{Peter Levrai} {and} \bibinfo{person}{Averil
  Bolster}.} \bibinfo{year}{2015}\natexlab{}.
\newblock \showarticletitle{Developing a closer understanding of academic oral
  presentations}.
\newblock \bibinfo{journal}{\emph{Folio}} \bibinfo{volume}{16},
  \bibinfo{number}{2} (\bibinfo{year}{2015}), \bibinfo{pages}{65--72}.
\newblock


\bibitem[Liu et~al\mbox{.}(2021)]%
        {liu2021igscript}
\bibfield{author}{\bibinfo{person}{Richen Liu}, \bibinfo{person}{Min Gao},
  \bibinfo{person}{Shunlong Ye}, {and} \bibinfo{person}{Jiang Zhang}.}
  \bibinfo{year}{2021}\natexlab{}.
\newblock \showarticletitle{IGScript: An interaction grammar for scientific
  data presentation}. In \bibinfo{booktitle}{\emph{Proceedings of the 2021 CHI
  Conference on Human Factors in Computing Systems}}. \bibinfo{pages}{1--13}.
\newblock


\bibitem[Lo(2008)]%
        {lo2008nonverbal}
\bibfield{author}{\bibinfo{person}{Shao-Kang Lo}.}
  \bibinfo{year}{2008}\natexlab{}.
\newblock \showarticletitle{The nonverbal communication functions of emoticons
  in computer-mediated communication}.
\newblock \bibinfo{journal}{\emph{Cyberpsychology \& behavior}}
  \bibinfo{volume}{11}, \bibinfo{number}{5} (\bibinfo{year}{2008}),
  \bibinfo{pages}{595--597}.
\newblock


\bibitem[Lukes(2021)]%
        {lukes2021power}
\bibfield{author}{\bibinfo{person}{Steven Lukes}.}
  \bibinfo{year}{2021}\natexlab{}.
\newblock \bibinfo{booktitle}{\emph{Power: A radical view}}.
\newblock \bibinfo{publisher}{Bloomsbury Publishing}.
\newblock


\bibitem[Ma et~al\mbox{.}(2022)]%
        {ma2022glancee}
\bibfield{author}{\bibinfo{person}{Shuai Ma}, \bibinfo{person}{Taichang Zhou},
  \bibinfo{person}{Fei Nie}, {and} \bibinfo{person}{Xiaojuan Ma}.}
  \bibinfo{year}{2022}\natexlab{}.
\newblock \showarticletitle{Glancee: An adaptable system for instructors to
  grasp student learning status in synchronous online classes}. In
  \bibinfo{booktitle}{\emph{Proceedings of the 2022 CHI Conference on Human
  Factors in Computing Systems}}. \bibinfo{pages}{1--25}.
\newblock


\bibitem[Mehrabian(1968)]%
        {mehrabian1968relationship}
\bibfield{author}{\bibinfo{person}{Albert Mehrabian}.}
  \bibinfo{year}{1968}\natexlab{}.
\newblock \showarticletitle{Relationship of attitude to seated posture,
  orientation, and distance.}
\newblock \bibinfo{journal}{\emph{Journal of personality and social
  psychology}} \bibinfo{volume}{10}, \bibinfo{number}{1}
  (\bibinfo{year}{1968}), \bibinfo{pages}{26}.
\newblock


\bibitem[Miscrosoft({[n.\,d.]})]%
        {PowerPointFunc}
\bibfield{author}{\bibinfo{person}{Miscrosoft}.}
  \bibinfo{year}{[n.\,d.]}\natexlab{}.
\newblock \bibinfo{booktitle}{\emph{Start the presentation and see your notes
  in Presenter view}}.
\newblock
\urldef\tempurl%
\url{https://support.microsoft.com/en-us/office/start-the-presentation-and-see-your-notes-in-presenter-view-4de90e28-487e-435c-9401-eb49a3801257}
\showURL{%
\tempurl}
\newblock
\shownote{(2023, Aug 12)}.


\bibitem[Morita(2000)]%
        {morita2000discourse}
\bibfield{author}{\bibinfo{person}{Naoko Morita}.}
  \bibinfo{year}{2000}\natexlab{}.
\newblock \showarticletitle{Discourse socialization through oral classroom
  activities in a TESL graduate program}.
\newblock \bibinfo{journal}{\emph{Tesol Quarterly}} \bibinfo{volume}{34},
  \bibinfo{number}{2} (\bibinfo{year}{2000}), \bibinfo{pages}{279--310}.
\newblock


\bibitem[Mu{\~n}oz-Ortiz et~al\mbox{.}(2023)]%
        {munoz2023contrasting}
\bibfield{author}{\bibinfo{person}{Alberto Mu{\~n}oz-Ortiz},
  \bibinfo{person}{Carlos G{\'o}mez-Rodr{\'\i}guez}, {and}
  \bibinfo{person}{David Vilares}.} \bibinfo{year}{2023}\natexlab{}.
\newblock \showarticletitle{Contrasting Linguistic Patterns in Human and
  LLM-Generated Text}.
\newblock \bibinfo{journal}{\emph{arXiv preprint arXiv:2308.09067}}
  (\bibinfo{year}{2023}).
\newblock


\bibitem[Murali et~al\mbox{.}(2021)]%
        {murali2021affectivespotlight}
\bibfield{author}{\bibinfo{person}{Prasanth Murali}, \bibinfo{person}{Javier
  Hernandez}, \bibinfo{person}{Daniel McDuff}, \bibinfo{person}{Kael Rowan},
  \bibinfo{person}{Jina Suh}, {and} \bibinfo{person}{Mary Czerwinski}.}
  \bibinfo{year}{2021}\natexlab{}.
\newblock \showarticletitle{Affectivespotlight: Facilitating the communication
  of affective responses from audience members during online presentations}. In
  \bibinfo{booktitle}{\emph{Proceedings of the 2021 CHI Conference on Human
  Factors in Computing Systems}}. \bibinfo{pages}{1--13}.
\newblock


\bibitem[Murali et~al\mbox{.}(2018)]%
        {murali2018speaker}
\bibfield{author}{\bibinfo{person}{Prasanth Murali}, \bibinfo{person}{Lazlo
  Ring}, \bibinfo{person}{Ha Trinh}, \bibinfo{person}{Reza Asadi}, {and}
  \bibinfo{person}{Timothy Bickmore}.} \bibinfo{year}{2018}\natexlab{}.
\newblock \showarticletitle{Speaker hand-offs in collaborative human-agent oral
  presentations}. In \bibinfo{booktitle}{\emph{Proceedings of the 18th
  International Conference on Intelligent Virtual Agents}}.
  \bibinfo{pages}{153--158}.
\newblock


\bibitem[Myers et~al\mbox{.}(2018)]%
        {myers2018patterns}
\bibfield{author}{\bibinfo{person}{Chelsea Myers}, \bibinfo{person}{Anushay
  Furqan}, \bibinfo{person}{Jessica Nebolsky}, \bibinfo{person}{Karina Caro},
  {and} \bibinfo{person}{Jichen Zhu}.} \bibinfo{year}{2018}\natexlab{}.
\newblock \showarticletitle{Patterns for how users overcome obstacles in voice
  user interfaces}. In \bibinfo{booktitle}{\emph{Proceedings of the 2018 CHI
  conference on human factors in computing systems}}. \bibinfo{pages}{1--7}.
\newblock


\bibitem[Oh et~al\mbox{.}(2020)]%
        {oh2020scriptfree}
\bibfield{author}{\bibinfo{person}{Jeungmin Oh}, \bibinfo{person}{Darren Edge},
  {and} \bibinfo{person}{Uichin Lee}.} \bibinfo{year}{2020}\natexlab{}.
\newblock \showarticletitle{ScriptFree: Designing Speech Preparation Systems
  with Adaptive Visual Reliance Control on Script}. In
  \bibinfo{booktitle}{\emph{Extended Abstracts of the 2020 CHI Conference on
  Human Factors in Computing Systems}}. \bibinfo{pages}{1--7}.
\newblock


\bibitem[Paltridge and Starfield(2013)]%
        {paltridge2013handbook}
\bibfield{author}{\bibinfo{person}{Brian Paltridge} {and} \bibinfo{person}{Sue
  Starfield}.} \bibinfo{year}{2013}\natexlab{}.
\newblock \bibinfo{booktitle}{\emph{The handbook of English for specific
  purposes}}. Vol.~\bibinfo{volume}{592}.
\newblock \bibinfo{publisher}{Wiley Online Library}.
\newblock


\bibitem[Parmar and Bickmore(2020)]%
        {parmar2020making}
\bibfield{author}{\bibinfo{person}{Dhaval Parmar} {and}
  \bibinfo{person}{Timothy Bickmore}.} \bibinfo{year}{2020}\natexlab{}.
\newblock \showarticletitle{Making it personal: Addressing individual audience
  members in oral presentations using augmented reality}.
\newblock \bibinfo{journal}{\emph{Proceedings of the ACM on Interactive,
  Mobile, Wearable and Ubiquitous Technologies}} \bibinfo{volume}{4},
  \bibinfo{number}{2} (\bibinfo{year}{2020}), \bibinfo{pages}{1--22}.
\newblock


\bibitem[Peeters et~al\mbox{.}(2010)]%
        {peeters2010standardized}
\bibfield{author}{\bibinfo{person}{Michael~J Peeters}, \bibinfo{person}{Eric~G
  Sahloff}, {and} \bibinfo{person}{Gregory~E Stone}.}
  \bibinfo{year}{2010}\natexlab{}.
\newblock \showarticletitle{A standardized rubric to evaluate student
  presentations}.
\newblock \bibinfo{journal}{\emph{American journal of pharmaceutical
  education}} \bibinfo{volume}{74}, \bibinfo{number}{9} (\bibinfo{year}{2010}).
\newblock


\bibitem[Peng et~al\mbox{.}(2021)]%
        {peng2021say}
\bibfield{author}{\bibinfo{person}{Yi-Hao Peng}, \bibinfo{person}{JiWoong
  Jang}, \bibinfo{person}{Jeffrey~P Bigham}, {and} \bibinfo{person}{Amy
  Pavel}.} \bibinfo{year}{2021}\natexlab{}.
\newblock \showarticletitle{Say It All: Feedback for Improving Non-Visual
  Presentation Accessibility}. In \bibinfo{booktitle}{\emph{Proceedings of the
  2021 CHI Conference on Human Factors in Computing Systems}}.
  \bibinfo{pages}{1--12}.
\newblock


\bibitem[Pschetz et~al\mbox{.}(2014)]%
        {pschetz2014turningpoint}
\bibfield{author}{\bibinfo{person}{Larissa Pschetz}, \bibinfo{person}{Koji
  Yatani}, {and} \bibinfo{person}{Darren Edge}.}
  \bibinfo{year}{2014}\natexlab{}.
\newblock \showarticletitle{TurningPoint: narrative-driven presentation
  planning}. In \bibinfo{booktitle}{\emph{Proceedings of the SIGCHI Conference
  on Human Factors in Computing Systems}}. \bibinfo{pages}{1591--1594}.
\newblock


\bibitem[Ram~Rajesh et~al\mbox{.}(2012)]%
        {ram2012remotely}
\bibfield{author}{\bibinfo{person}{J Ram~Rajesh}, \bibinfo{person}{R
  Sudharshan}, \bibinfo{person}{D Nagarjunan}, {and} \bibinfo{person}{R
  Aarthi}.} \bibinfo{year}{2012}\natexlab{}.
\newblock \showarticletitle{Remotely controlled PowerPoint presentation
  navigation using hand gestures}. In \bibinfo{booktitle}{\emph{Proc. of
  International Conference on Advances in Computer, Electronics and Electrical
  Engineering}}.
\newblock


\bibitem[Razawi et~al\mbox{.}(2019)]%
        {razawi2019anxiety}
\bibfield{author}{\bibinfo{person}{Nurul~Amilin Razawi},
  \bibinfo{person}{Luqmanul~Hakim Zulkornain}, {and}
  \bibinfo{person}{Razifa~Mohd Razlan}.} \bibinfo{year}{2019}\natexlab{}.
\newblock \showarticletitle{Anxiety in oral presentations among ESL students}.
\newblock \bibinfo{journal}{\emph{Journal of Academia}} \bibinfo{volume}{7},
  \bibinfo{number}{1} (\bibinfo{year}{2019}), \bibinfo{pages}{31--36}.
\newblock


\bibitem[Riordan(2017)]%
        {riordan2017communicative}
\bibfield{author}{\bibinfo{person}{Monica~A Riordan}.}
  \bibinfo{year}{2017}\natexlab{}.
\newblock \showarticletitle{The communicative role of non-face emojis: Affect
  and disambiguation}.
\newblock \bibinfo{journal}{\emph{Computers in Human Behavior}}
  \bibinfo{volume}{76} (\bibinfo{year}{2017}), \bibinfo{pages}{75--86}.
\newblock


\bibitem[Robertson and Zaragoza(2009)]%
        {Robertson2009BM25}
\bibfield{author}{\bibinfo{person}{Stephen Robertson} {and}
  \bibinfo{person}{Hugo Zaragoza}.} \bibinfo{year}{2009}\natexlab{}.
\newblock \showarticletitle{The Probabilistic Relevance Framework: BM25 and
  Beyond}.
\newblock \bibinfo{journal}{\emph{Found. Trends Inf. Retr.}}
  \bibinfo{volume}{3}, \bibinfo{number}{4} (\bibinfo{date}{apr}
  \bibinfo{year}{2009}), \bibinfo{pages}{333–389}.
\newblock
\showISSN{1554-0669}
\urldef\tempurl%
\url{https://doi.org/10.1561/1500000019}
\showDOI{\tempurl}


\bibitem[Roels and Signer(2014)]%
        {roels2014mindxpres}
\bibfield{author}{\bibinfo{person}{Reinout Roels} {and} \bibinfo{person}{Beat
  Signer}.} \bibinfo{year}{2014}\natexlab{}.
\newblock \showarticletitle{MindXpres: An extensible content-driven cross-media
  presentation platform}. In \bibinfo{booktitle}{\emph{Web Information Systems
  Engineering--WISE 2014: 15th International Conference, Thessaloniki, Greece,
  October 12-14, 2014, Proceedings, Part II 15}}. Springer,
  \bibinfo{pages}{215--230}.
\newblock


\bibitem[Sabri({[n.\,d.]})]%
        {officeRemotePicture}
\bibfield{author}{\bibinfo{person}{Sam Sabri}.}
  \bibinfo{year}{[n.\,d.]}\natexlab{}.
\newblock \bibinfo{booktitle}{\emph{Microsoft releases Office Remote to allow
  you to control your PowerPoint presentation and more from your Windows
  Phone}}.
\newblock
\urldef\tempurl%
\url{https://www.windowscentral.com/office-remote-windows-phone}
\showURL{%
\tempurl}
\newblock
\shownote{(2023, Aug 30)}.


\bibitem[Schneider et~al\mbox{.}(2015)]%
        {schneider2015presentation}
\bibfield{author}{\bibinfo{person}{Jan Schneider}, \bibinfo{person}{Dirk
  B{\"o}rner}, \bibinfo{person}{Peter Van~Rosmalen}, {and}
  \bibinfo{person}{Marcus Specht}.} \bibinfo{year}{2015}\natexlab{}.
\newblock \showarticletitle{Presentation trainer, your public speaking
  multimodal coach}. In \bibinfo{booktitle}{\emph{Proceedings of the 2015 ACM
  on international conference on multimodal interaction}}.
  \bibinfo{pages}{539--546}.
\newblock


\bibitem[Senju and Johnson(2009)]%
        {senju2009eye}
\bibfield{author}{\bibinfo{person}{Atsushi Senju} {and} \bibinfo{person}{Mark~H
  Johnson}.} \bibinfo{year}{2009}\natexlab{}.
\newblock \showarticletitle{The eye contact effect: mechanisms and
  development}.
\newblock \bibinfo{journal}{\emph{Trends in cognitive sciences}}
  \bibinfo{volume}{13}, \bibinfo{number}{3} (\bibinfo{year}{2009}),
  \bibinfo{pages}{127--134}.
\newblock


\bibitem[Skiba(2016)]%
        {skiba2016face}
\bibfield{author}{\bibinfo{person}{Diane~J Skiba}.}
  \bibinfo{year}{2016}\natexlab{}.
\newblock \showarticletitle{Face with tears of joy is word of the year: are
  emoji a sign of things to come in health care?}
\newblock \bibinfo{journal}{\emph{Nursing education perspectives}}
  \bibinfo{volume}{37}, \bibinfo{number}{1} (\bibinfo{year}{2016}),
  \bibinfo{pages}{56--57}.
\newblock


\bibitem[Strangert and Gustafson(2008)]%
        {strangert2008makes}
\bibfield{author}{\bibinfo{person}{Eva Strangert} {and} \bibinfo{person}{Joakim
  Gustafson}.} \bibinfo{year}{2008}\natexlab{}.
\newblock \showarticletitle{What makes a good speaker? subject ratings,
  acoustic measurements and perceptual evaluations}. In
  \bibinfo{booktitle}{\emph{Ninth Annual Conference of the International Speech
  Communication Association}}.
\newblock


\bibitem[Tanveer et~al\mbox{.}(2015)]%
        {tanveer2015rhema}
\bibfield{author}{\bibinfo{person}{M~Iftekhar Tanveer}, \bibinfo{person}{Emy
  Lin}, {and} \bibinfo{person}{Mohammed Hoque}.}
  \bibinfo{year}{2015}\natexlab{}.
\newblock \showarticletitle{Rhema: A real-time in-situ intelligent interface to
  help people with public speaking}. In \bibinfo{booktitle}{\emph{Proceedings
  of the 20th international conference on intelligent user interfaces}}.
  \bibinfo{pages}{286--295}.
\newblock


\bibitem[Tanveer et~al\mbox{.}(2016)]%
        {tanveer2016automanner}
\bibfield{author}{\bibinfo{person}{M~Iftekhar Tanveer}, \bibinfo{person}{Ru
  Zhao}, \bibinfo{person}{Kezhen Chen}, \bibinfo{person}{Zoe Tiet}, {and}
  \bibinfo{person}{Mohammed~Ehsan Hoque}.} \bibinfo{year}{2016}\natexlab{}.
\newblock \showarticletitle{Automanner: An automated interface for making
  public speakers aware of their mannerisms}. In
  \bibinfo{booktitle}{\emph{Proceedings of the 21st international conference on
  intelligent user interfaces}}. \bibinfo{pages}{385--396}.
\newblock


\bibitem[Tareen(2022)]%
        {tareen2022investigating}
\bibfield{author}{\bibinfo{person}{Hashmatullah Tareen}.}
  \bibinfo{year}{2022}\natexlab{}.
\newblock \showarticletitle{Investigating EFL learners’ perceptions towards
  the difficulties in oral presentation at Kandahar university}.
\newblock \bibinfo{journal}{\emph{ESI Preprints}}  \bibinfo{volume}{11}
  (\bibinfo{year}{2022}), \bibinfo{pages}{535--535}.
\newblock


\bibitem[Thanyadit et~al\mbox{.}(2023)]%
        {thanyadit2023tutor}
\bibfield{author}{\bibinfo{person}{Santawat Thanyadit},
  \bibinfo{person}{Matthias Heintz}, {and} \bibinfo{person}{Effie~LC Law}.}
  \bibinfo{year}{2023}\natexlab{}.
\newblock \showarticletitle{Tutor In-sight: Guiding and Visualizing Students'
  Attention with Mixed Reality Avatar Presentation Tools}. In
  \bibinfo{booktitle}{\emph{Proceedings of the 2023 CHI Conference on Human
  Factors in Computing Systems}}. \bibinfo{pages}{1--20}.
\newblock


\bibitem[Thompson et~al\mbox{.}(2016)]%
        {thompson2016emotional}
\bibfield{author}{\bibinfo{person}{Dominic Thompson}, \bibinfo{person}{Ian~G
  Mackenzie}, \bibinfo{person}{Hartmut Leuthold}, {and} \bibinfo{person}{Ruth
  Filik}.} \bibinfo{year}{2016}\natexlab{}.
\newblock \showarticletitle{Emotional responses to irony and emoticons in
  written language: Evidence from EDA and facial EMG}.
\newblock \bibinfo{journal}{\emph{Psychophysiology}} \bibinfo{volume}{53},
  \bibinfo{number}{7} (\bibinfo{year}{2016}), \bibinfo{pages}{1054--1062}.
\newblock


\bibitem[Trinh et~al\mbox{.}(2015)]%
        {trinh2015dynamicduo}
\bibfield{author}{\bibinfo{person}{Ha Trinh}, \bibinfo{person}{Lazlo Ring},
  {and} \bibinfo{person}{Timothy Bickmore}.} \bibinfo{year}{2015}\natexlab{}.
\newblock \showarticletitle{Dynamicduo: co-presenting with virtual agents}. In
  \bibinfo{booktitle}{\emph{Proceedings of the 33rd Annual ACM Conference on
  Human Factors in Computing Systems}}. \bibinfo{pages}{1739--1748}.
\newblock


\bibitem[Trinh et~al\mbox{.}(2014)]%
        {trinh2014pitchperfect}
\bibfield{author}{\bibinfo{person}{Ha Trinh}, \bibinfo{person}{Koji Yatani},
  {and} \bibinfo{person}{Darren Edge}.} \bibinfo{year}{2014}\natexlab{}.
\newblock \showarticletitle{PitchPerfect: integrated rehearsal environment for
  structured presentation preparation}. In
  \bibinfo{booktitle}{\emph{Proceedings of the SIGCHI Conference on Human
  Factors in Computing Systems}}. \bibinfo{pages}{1571--1580}.
\newblock


\bibitem[Truong et~al\mbox{.}(2006)]%
        {truong2006storyboarding}
\bibfield{author}{\bibinfo{person}{Khai~N Truong}, \bibinfo{person}{Gillian~R
  Hayes}, {and} \bibinfo{person}{Gregory~D Abowd}.}
  \bibinfo{year}{2006}\natexlab{}.
\newblock \showarticletitle{Storyboarding: an empirical determination of best
  practices and effective guidelines}. In \bibinfo{booktitle}{\emph{Proceedings
  of the 6th conference on Designing Interactive systems}}.
  \bibinfo{pages}{12--21}.
\newblock


\bibitem[Wan~Zakaria and Razak(2016)]%
        {wan2016english}
\bibfield{author}{\bibinfo{person}{Wan Nuur~Fazliza Wan~Zakaria} {and}
  \bibinfo{person}{Siti~Shazlin Razak}.} \bibinfo{year}{2016}\natexlab{}.
\newblock \showarticletitle{English as a Second Language (ESL) Learner's
  Perceptions of the Difficulties in Oral Commentary Assessment}.
\newblock \bibinfo{journal}{\emph{Journal of Contemporary Social Science
  Research}} \bibinfo{volume}{1}, \bibinfo{number}{1} (\bibinfo{year}{2016}),
  \bibinfo{pages}{1--15}.
\newblock


\bibitem[Wang et~al\mbox{.}(2023)]%
        {wang2023slide4n}
\bibfield{author}{\bibinfo{person}{Fengjie Wang}, \bibinfo{person}{Xuye Liu},
  \bibinfo{person}{Oujing Liu}, \bibinfo{person}{Ali Neshati},
  \bibinfo{person}{Tengfei Ma}, \bibinfo{person}{Min Zhu}, {and}
  \bibinfo{person}{Jian Zhao}.} \bibinfo{year}{2023}\natexlab{}.
\newblock \showarticletitle{Slide4N: Creating Presentation Slides from
  Computational Notebooks with Human-AI Collaboration}. In
  \bibinfo{booktitle}{\emph{Proceedings of the 2023 CHI Conference on Human
  Factors in Computing Systems}}. \bibinfo{pages}{1--18}.
\newblock


\bibitem[Wang et~al\mbox{.}(2020)]%
        {wang2020voicecoach}
\bibfield{author}{\bibinfo{person}{Xingbo Wang}, \bibinfo{person}{Haipeng
  Zeng}, \bibinfo{person}{Yong Wang}, \bibinfo{person}{Aoyu Wu},
  \bibinfo{person}{Zhida Sun}, \bibinfo{person}{Xiaojuan Ma}, {and}
  \bibinfo{person}{Huamin Qu}.} \bibinfo{year}{2020}\natexlab{}.
\newblock \showarticletitle{Voicecoach: Interactive evidence-based training for
  voice modulation skills in public speaking}. In
  \bibinfo{booktitle}{\emph{Proceedings of the 2020 CHI Conference on Human
  Factors in Computing Systems}}. \bibinfo{pages}{1--12}.
\newblock


\bibitem[Warner et~al\mbox{.}(2023)]%
        {warner2023slidespecs}
\bibfield{author}{\bibinfo{person}{Jeremy Warner}, \bibinfo{person}{Amy Pavel},
  \bibinfo{person}{Tonya Nguyen}, \bibinfo{person}{Maneesh Agrawala}, {and}
  \bibinfo{person}{Bjoern Hartmann}.} \bibinfo{year}{2023}\natexlab{}.
\newblock \showarticletitle{SlideSpecs: Automatic and Interactive Presentation
  Feedback Collation}. In \bibinfo{booktitle}{\emph{Proceedings of the 28th
  International Conference on Intelligent User Interfaces}}.
  \bibinfo{pages}{695--709}.
\newblock


\bibitem[Whai and Mei(2015)]%
        {whai2015causes}
\bibfield{author}{\bibinfo{person}{Marcus Kho~Gee Whai} {and}
  \bibinfo{person}{Leong~Lai Mei}.} \bibinfo{year}{2015}\natexlab{}.
\newblock \showarticletitle{Causes of academic oral presentation difficulties
  faced by students at a polytechnic in Sarawak}.
\newblock \bibinfo{journal}{\emph{The English Teacher}} \bibinfo{volume}{44},
  \bibinfo{number}{3} (\bibinfo{year}{2015}).
\newblock


\bibitem[Wilson(2013)]%
        {wilson2013brainstorming}
\bibfield{author}{\bibinfo{person}{Chauncey Wilson}.}
  \bibinfo{year}{2013}\natexlab{}.
\newblock \bibinfo{booktitle}{\emph{Brainstorming and beyond: a user-centered
  design method}}.
\newblock \bibinfo{publisher}{Newnes}.
\newblock


\bibitem[W{\"o}rtwein et~al\mbox{.}(2015)]%
        {wortwein2015multimodal}
\bibfield{author}{\bibinfo{person}{Torsten W{\"o}rtwein},
  \bibinfo{person}{Mathieu Chollet}, \bibinfo{person}{Boris Schauerte},
  \bibinfo{person}{Louis-Philippe Morency}, \bibinfo{person}{Rainer
  Stiefelhagen}, {and} \bibinfo{person}{Stefan Scherer}.}
  \bibinfo{year}{2015}\natexlab{}.
\newblock \showarticletitle{Multimodal public speaking performance assessment}.
  In \bibinfo{booktitle}{\emph{Proceedings of the 2015 ACM on International
  Conference on Multimodal Interaction}}. \bibinfo{pages}{43--50}.
\newblock


\bibitem[Wu et~al\mbox{.}(2022)]%
        {wu2022ai}
\bibfield{author}{\bibinfo{person}{Tongshuang Wu}, \bibinfo{person}{Michael
  Terry}, {and} \bibinfo{person}{Carrie~Jun Cai}.}
  \bibinfo{year}{2022}\natexlab{}.
\newblock \showarticletitle{Ai chains: Transparent and controllable human-ai
  interaction by chaining large language model prompts}. In
  \bibinfo{booktitle}{\emph{Proceedings of the 2022 CHI conference on human
  factors in computing systems}}. \bibinfo{pages}{1--22}.
\newblock


\bibitem[Xia et~al\mbox{.}(2022)]%
        {xia2022persua}
\bibfield{author}{\bibinfo{person}{Meng Xia}, \bibinfo{person}{Qian Zhu},
  \bibinfo{person}{Xingbo Wang}, \bibinfo{person}{Fei Nie},
  \bibinfo{person}{Huamin Qu}, {and} \bibinfo{person}{Xiaojuan Ma}.}
  \bibinfo{year}{2022}\natexlab{}.
\newblock \showarticletitle{Persua: A visual interactive system to enhance the
  persuasiveness of arguments in online discussion}.
\newblock \bibinfo{journal}{\emph{Proceedings of the ACM on Human-Computer
  Interaction}} \bibinfo{volume}{6}, \bibinfo{number}{CSCW2}
  (\bibinfo{year}{2022}), \bibinfo{pages}{1--30}.
\newblock


\bibitem[Xu et~al\mbox{.}(2023)]%
        {xu2023does}
\bibfield{author}{\bibinfo{person}{Qihui Xu}, \bibinfo{person}{Yingying Peng},
  \bibinfo{person}{Minghua Wu}, \bibinfo{person}{Feng Xiao},
  \bibinfo{person}{Martin Chodorow}, {and} \bibinfo{person}{Ping Li}.}
  \bibinfo{year}{2023}\natexlab{}.
\newblock \showarticletitle{Does Conceptual Representation Require Embodiment?
  Insights From Large Language Models}.
\newblock \bibinfo{journal}{\emph{arXiv preprint arXiv:2305.19103}}
  (\bibinfo{year}{2023}).
\newblock


\bibitem[Yi et~al\mbox{.}(2020)]%
        {yi2020presentationtrainer}
\bibfield{author}{\bibinfo{person}{Shengzhou Yi}, \bibinfo{person}{Hiroshi
  Yumoto}, \bibinfo{person}{Xueting Wang}, {and} \bibinfo{person}{Toshihiko
  Yamasaki}.} \bibinfo{year}{2020}\natexlab{}.
\newblock \showarticletitle{Presentationtrainer: Oral presentation support
  system for impression-related feedback}. In
  \bibinfo{booktitle}{\emph{Proceedings of the AAAI Conference on Artificial
  Intelligence}}, Vol.~\bibinfo{volume}{34}. \bibinfo{pages}{13644--13645}.
\newblock


\bibitem[Yuan et~al\mbox{.}(2023)]%
        {yuan2023critrainer}
\bibfield{author}{\bibinfo{person}{Kangyu Yuan}, \bibinfo{person}{Hehai Lin},
  \bibinfo{person}{Shilei Cao}, \bibinfo{person}{Zhenhui Peng},
  \bibinfo{person}{Qingyu Guo}, {and} \bibinfo{person}{Xiaojuan Ma}.}
  \bibinfo{year}{2023}\natexlab{}.
\newblock \showarticletitle{CriTrainer: An Adaptive Training Tool for Critical
  Paper Reading}. In \bibinfo{booktitle}{\emph{Proceedings of the 36th Annual
  ACM Symposium on User Interface Software and Technology}}.
  \bibinfo{pages}{1--17}.
\newblock


\bibitem[Zappa-Hollman(2007)]%
        {zappa2007academic}
\bibfield{author}{\bibinfo{person}{Sandra~Carolina Zappa-Hollman}.}
  \bibinfo{year}{2007}\natexlab{}.
\newblock \showarticletitle{Academic presentations across post-secondary
  contexts: The discourse socialization of non-native English speakers}.
\newblock \bibinfo{journal}{\emph{Canadian Modern Language Review}}
  \bibinfo{volume}{63}, \bibinfo{number}{4} (\bibinfo{year}{2007}),
  \bibinfo{pages}{455--485}.
\newblock


\bibitem[Zareva(2011)]%
        {zareva2011and}
\bibfield{author}{\bibinfo{person}{Alla Zareva}.}
  \bibinfo{year}{2011}\natexlab{}.
\newblock \showarticletitle{‘And so that was it’: Linking adverbials in
  student academic presentations}.
\newblock \bibinfo{journal}{\emph{RELC Journal}} \bibinfo{volume}{42},
  \bibinfo{number}{1} (\bibinfo{year}{2011}), \bibinfo{pages}{5--15}.
\newblock


\bibitem[Zeng et~al\mbox{.}(2022)]%
        {zeng2022gesturelens}
\bibfield{author}{\bibinfo{person}{Haipeng Zeng}, \bibinfo{person}{Xingbo
  Wang}, \bibinfo{person}{Yong Wang}, \bibinfo{person}{Aoyu Wu},
  \bibinfo{person}{Ting-Chuen Pong}, {and} \bibinfo{person}{Huamin Qu}.}
  \bibinfo{year}{2022}\natexlab{}.
\newblock \showarticletitle{Gesturelens: Visual analysis of gestures in
  presentation videos}.
\newblock \bibinfo{journal}{\emph{IEEE Transactions on Visualization and
  Computer Graphics}} (\bibinfo{year}{2022}).
\newblock


\end{thebibliography}

\appendix

\begin{flushleft}
    \Large
    \textbf{\textsf{Appendix}}
\end{flushleft}

\section{Details of Exploratory Survey Results}
\label{appendix:Survey_detail}
\par We analyzed the quantitative data with descriptive statistics, and two authors coded the responses to open-ended questions using thematic analysis methods~\cite{braun2006using}. Subsequently, the two authors engaged in collaborative discussions to synthesize their findings.

\par In the \textbf{survey conducted among EFL students}, a total of $52$ questionnaires were distributed, yielding $49$ valid responses. The respondents were diverse, including $28$ males, $20$ females, and $1$ prefer not to disclose. The participants, with an average age of $23.4$ ($SD=3.0$), represented various academic disciplines such as computer science, electronic engineering, biology, finance, statistics, forestry, and medicine. They spanned different grade levels and had experience with AOPs, with $15\%$ having presented once, $20\%$ twice, $40\%$ three times, $15\%$ four times, and $10\%$ more than four times. Regarding preparation time for AOPs, $20.4\%$ reported spending less than three days, $57.1\%$ between three days and one week, $16.3\%$ between one and two weeks, and $6.1\%$ more than two weeks. In terms of challenges, $65.7\%$ of students found AOPs to be difficult, and $42.9\%$ expressed that their preparation time was insufficient. In terms of mentality, the majority of EFL students experienced varying feelings towards AOPs: $16.3\%$ felt fear, $36.7\%$ felt anxious, $18.4\%$ were uninterested, $18.4\%$ were confident, and $10.2\%$ were excited. The issues most commonly identified during AOPs were lack of fluency ($36.7\%$), flat intonation ($28.6\%$), and absence of body language ($12.2\%$). Notably, fluency was regarded as more crucial than pronunciation and grammar. Lexical gaps were identified as a challenge, leading to difficulties in word retrieval and conveying intended meaning effectively. Consequently, this posed a formidable obstacle for them to express themselves spontaneously and seamlessly, similar to native speakers.

\par In terms of supportive tools, the majority of participants ($67.3\%$) had experience utilizing such tools to facilitate their AOPs. Among tool users, a significant majority ($54.5\%$) employed these tools frequently. Impressively, the user experience with these supportive tools was generally positive, with $45.6\%$ reporting a good experience, $51.5\%$ describing it as mediocre, and only $3\%$ indicating a negative experience. Participants who had suboptimal experiences highlighted challenges in understanding provided hints, locating required script snippets, excessive focus on the tools, and potential overdependence. Furthermore, the participants attributed several benefits to these tools, including facilitating script composition ($57.1\%$), enhancing slide presentations ($22.4\%$), boosting their confidence and mentality ($20.4\%$), and aiding in controlling speech rate ($18.4\%$). Interestingly, a significant majority ($81.6\%$) perceived the necessity of supportive tools for AOPs. When asked about their hesitation in using such tools, cost of learning and instructional prerequisites emerged as the predominant concerns. Regarding anticipated features, participants expressed interest in all provided factors, with particularly prominent preferences for script display ($73.5\%$), slide previews ($61.2\%$), and support for controlling speech rate ($42.9\%$). This underscores the value participants place on tools that assist in the delivery of effective presentations.

\par Turning to the \textbf{survey of instructors}, we received a total of $36$ responses from instructors, with $24$ males and $12$ females, showcasing an average age of $45.3$ ($SD=12.0$). This group of instructors represented a variety of academic fields, including information science, physics, entrepreneurship and management, mathematics, and humanities. Their teaching experiences included organizing AOPs for EFL students throughout their careers, with $15.5\%$ having done so once, $22.5\%$ twice, $39.4\%$ three times, $14.1\%$ four times, and $8.5\%$ more than four times. Interestingly, $83.3\%$ of instructors expressed dissatisfaction with the AOP delivery of EFL students. Among all instructors, $44.4\%$ felt weary, $38.9\%$ were worried, $11.1\%$ showed interest, and $5.6\%$ were satisfied. The instructors identified the most common challenges that students face, namely the lack of preparation ($55.6\%$) and poor delivery skills ($30.6\%$). Additionally, a majority of instructors ($83.3\%$) believed that incorporating delivery support tools into their courses wouldn't impact grading assessment. Among these instructors, $38.9\%$ were open to allowing students to use such tools, and $30.6\%$ supported encouraging students to utilize them. In response to queries about their reasons, most of them mentioned ``\textit{feeling awkward towards students' poor delivery}'' and ``\textit{recognizing the benefits of improved delivery for both speakers and the audience}''. On the other hand, a closer analysis of open-ended responses revealed that the remaining $30.6\%$ of instructors maintaining a negative perspective were primarily concerned that these tools might hinder audience reception and foster excessive dependence among students.

\begin{figure*}[h]
\vspace{-3mm}
  \centering
  \includegraphics[width=\textwidth]{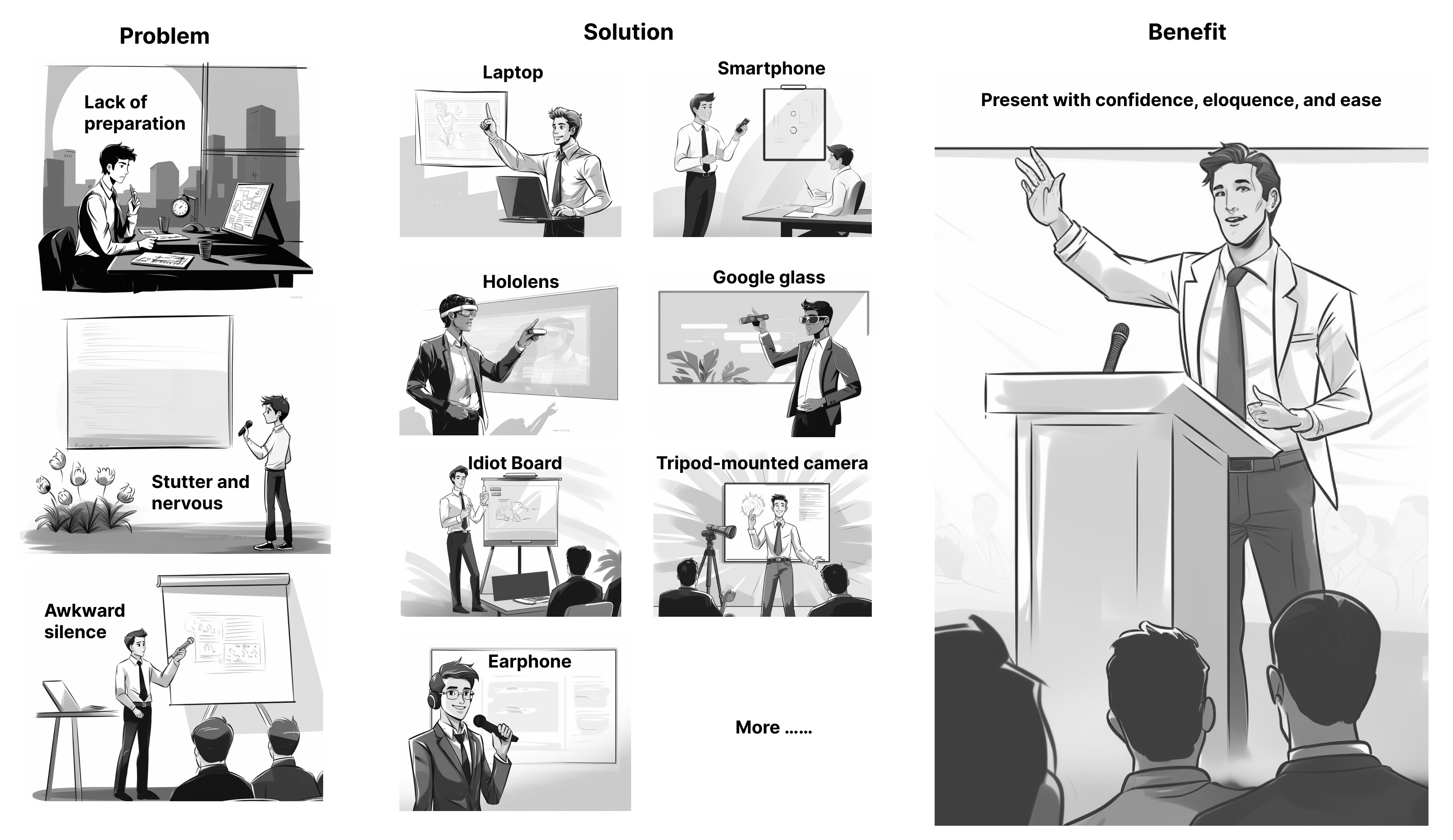}
  \vspace{-6mm}
  \caption{Summarized version of the Storyboards used in our design workshop.}
\vspace{-3mm}
  \label{fig:Summary_board}
\end{figure*}

\section{Details of Expert Interview}
\label{appendix:Interview_detail}
\par LLMs, like ChatGPT, have opened up significant opportunities for human-AI collaboration and demonstrated impressive capabilities~\cite{wu2022ai}. To assess the viability of employing ChatGPT for refining presentation scripts and offering delivery prompts, we conducted semi-structured interviews with five experienced experts (E1-5) in presentation instruction. These interviews aimed to tap into their expertise and gather insights on the feasibility of leveraging ChatGPT in this context. Additionally, we sought their professional perspectives on the outcomes of our initial exploratory surveys and design workshop, enriching our understanding of the potential applications and implications of the technology.

\par \textbf{Participants.} The individuals we interviewed comprise five experts ($2$ males and $3$ females) affiliated with the language teaching group at a local university. Their collective background includes substantial experience in presentation instruction and proficiency in delivering presentations themselves. All these experts possess substantial theoretical knowledge and a wealth of practical experience in the domain of presentations. Their collective expertise (\autoref{fig:Expert_Interview}A) spans a range of $2$ to $7$ years, with an average duration of $4$ years ($SD = 2.78$). Notably, all of these experts have taught presentation courses to EFL students across multiple semesters ($Mean = 6.30$, $SD = 3.01$).

\par \textbf{Preparation.} We selected two prominent LLMs, GPT-3.5-turbo and GPT-4 (both March 14 versions), as our candidates for evaluation. Our process commenced by collecting $10$ practical presentation scripts from EFL students through social networks. These scripts covered a range of topics including HCI, machine learning, sociology, etc. To devise effective prompts for ChatGPT that would yield desired outputs, we initiated the process with a preliminary draft. We iteratively refined this draft by interacting with both GPT-3.5-turbo and GPT-4, ensuring that the generated content and output format aligned with our expectations. Following the completion of this iterative process, we employed the final GPT prompts on both GPT-3.5-turbo and GPT-4 to further polish all the scripts. Additionally, in-line delivery prompts were incorporated into the scripts to enhance their presentation quality and effectiveness.

\par \textbf{Interview Protocol.} Each interview session had a duration of approximately $40$ minutes. Our interview process involved formulating questions based on $7$-point Likert scale to evaluate the quality of output scripts and delivery prompts from LLM candidates (as depicted in \autoref{fig:Expert_Interview}B). These questions were tailored for individual expert interviews. Simultaneously, open-ended questions were crafted to capture insights into their presentation instruction experiences, suggestions on addressing users' needs, and their perceptions of the outcomes from our exploratory survey and design workshop. For example, questions like ``\textit{What strategies do you typically employ to enhance the effectiveness of presentations?}''; ``\textit{what are your suggestions if a LLM-based tool were designed to aid EFL students in improving their presentation delivery on the fly?}''; and ``\textit{Did any particular findings from the survey results stand out to you?}'' Throughout the interviews, we also asked the experts follow-up questions according to their responses and shared specific instances of challenges encountered by EFL students in their presentation deliveries.

\par \textbf{Results and Analysis.} The semi-structured interviews with the expert panel yielded valuable insights regarding the potential of using LLMs, specifically GPT-3.5-turbo and GPT-4, for improving presentation scripts and providing delivery prompts. First, in terms of the quality of output scripts generated by the LLMs, the experts generally provided positive ratings (\autoref{fig:Expert_Interview}B). Specifically, the experts reached a consensus that both LLMs excelled in terms of structural completeness and keyword identification. However, it was noted by E1 and E3 that GPT-3.5-turbo tended to employ overly complex wording and sentence structures, rendering the generated script less suitable for practical uses. In terms of fluency and maintaining meaning, ratings indicated a generally favorable reception for both LLM candidates, with GPT-4 slightly outperforming GPT-3.5-turbo according to the experts' assessments. Qualitative feedback highlighted the strengths of these LLMs in enhancing the clarity, coherence, and overall organization of presentation scripts. E2 illustrated that LLMs have ``\textit{demonstrated an impressive capability in enhancing presentation script.}'' Second, the effectiveness of the in-line delivery prompts provided by the LLMs also received positive ratings from the experts. They generally affirmed the efficacy of GPT-4 in factor judgment and modulation during presentations. In general, the experts agreed that GPT-4 outperformed GPT-3.5-turbo across all metrics. However, concerns were raised regarding the density of these delivery prompts. E4 commented that ``\textit{modulation techniques in presentation training usually focus on details at the phrase level or word level. While these (delivery) prompts are reasonable, I can imagine that such dense prompts might be overwhelming for students if you intend to provide on-the-fly support.}'' Conversely, open-ended responses from the experts indicated that these prompts could assist students in maintaining a smooth flow of speech, adhering to the topic, and effectively engaging the audience. For instance, E3 mentioned that ``\textit{these prompts could serve as a delivery template for students, allowing them to learn from their own presentation scripts.}''

\par Furthermore, during the interviews, our experts shared invaluable insights into their teaching experiences, shedding light on various strategies employed to enhance presentations effectiveness. These strategies included techniques for maintaining eye contact, using body language, and managing nervousness, among others. For instance, E2 mentioned that ``\textit{simply encouraging individuals to keep their lower body motionless can significantly enhance their stage presence.}'' Additionally, the experts provided insightful suggestions for the design of an LLM-based tool to support EFL students in real-time presentation delivery. One innovative approach they proposed involved integrating multimodal feedback to effectively guide students' attention and behavior. To build upon this idea, E1 noted ``\textit{using subtle auditory cues, like a beep or pitch, can signal students to adjust their speech volume or speed. Moreover, incorporating somatosensory cues, such as slight vibrations synchronized with key parts of the speech, can provide haptic feedback when emphasizing specific points.}'' Building upon this concept, E3 emphasized the potential to incorporate visual cues such as arrows or highlights into the tool, acting as a navigational guide to ensure students focus on relevant slides or audience members.

\par Regarding their perspectives on the results obtained from our exploratory surveys and design workshop, the experts highlighted several noteworthy findings. These included self-reported presentation challenges among EFL students, preferences for presentation aid features, and reactions to prototype design concepts. For example, the experts recognized that the discrepancy between verbal and non-verbal communication challenges may be attributed to differences in EFL students' exposure and practice. As E4 pointed out, ``\textit{EFL students tend to acquire and practice non-verbal communication skills through daily experiences like watching videos, playing games, or engaging in interpersonal interactions. However, opportunities to enhance their verbal communication skills, particularly in academic or professional contexts, are more limited, which is where they often face the greatest difficulties.}'' E5 also noted, ``\textit{Especially for them (EFL students), the pronunciation of unfamiliar or polysyllabic words can pose significant challenges during presentations.}''

\par The feedback provided by the experts on these findings will play a pivotal role in refining the development of our tool, ensuring it aligns more closely with user needs. To sum up, our expert interviews have contributed valuable insights into the application of LLMs for presentation enhancement and delivery support. The positive ratings and feedback regarding the quality of output scripts and delivery prompts from both LLMs, especially GPT-4, suggest promising potential for integrating LLMs into practical settings to improve presentation delivery.

\begin{figure*}[h]
\vspace{-3mm}
  \centering
  \includegraphics[width=0.6\textwidth]{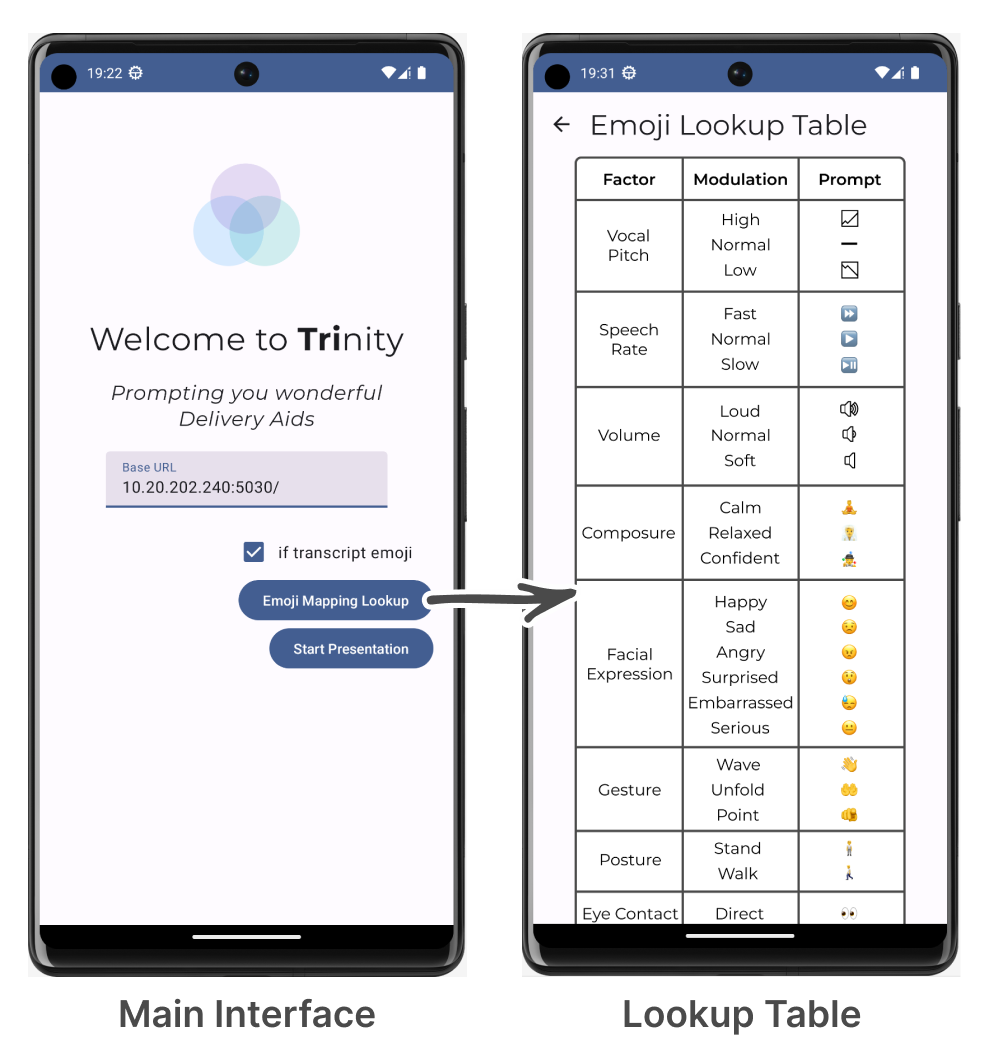}
  \caption{Main interface and built-in emoji lookup table of \textit{Trinity} App. Users have access to the emoji lookup table by tapping the ``Emoji Mapping Lookup'' button in the main interface.}
\vspace{-3mm}
  \label{fig:Lookup_table}
\end{figure*}

\begin{table*}[h]
\caption{Questionnaires for student presenters in $7$-point Likert scale (1: Not at all/very bad, 7: Very Much/very good)}
\begin{tabular}{ll}
\hline
Category                       & Question                                                                           \\ \hline
\multirow{5}{*}{Usability}     & (1) How easy was the system to use?                                                \\
                               & (2) How helpful was the system in presenting?                                      \\
                               & (3) How distracting was the system in presenting?                                  \\
                               & (4) How satisfing was the system?                                                  \\
                               & (5) What is the likelihood of future use?                                          \\ \hline
\multirow{8}{*}{Effectiveness} & (6) How much could the system make the speech more fluid, coherent, and modulated? \\
                               & (7) How much could the system facilitate good eye contacts with audience?          \\
                               & (8) How much could the system facilitate appropriate facial expressions?           \\
                               & (9) How much could the system facilitate natural gestures?                         \\
                               & (10) How much could the system enhance the easiness of visual control?             \\
                               & (11) How much could the system enhance the effectiveness of slides?                \\
                               & (12) How much could the system enhance control of the presentation progress?       \\
                               & (13) How much could the system make the presentation more engaging?                \\ \hline
\multirow{11}{*}{Emotion}      & (14) Did you feel noticed?                                                         \\
                               & (15) Did you feel excited?                                                         \\
                               & (16) Did you feel exhausted?                                                       \\
                               & (17) Did you feel frustrated?                                                      \\
                               & (18) Did you feel happy?                                                           \\
                               & (19) Did you feel hopeful?                                                         \\
                               & (20) Did you feel overwhelmed?                                                     \\
                               & (21) Did you feel sense of safety?                                                 \\
                               & (22) Did you feel nervous?                                                         \\
                               & (23) Did you feel anxious?                                                         \\
                               & (24) Did you feel confident?                                                       \\ \hline
\multirow{4}{*}{Load}          & (25) How is your cognitive load?                                                   \\
                               & (26) How is your attentional load?                                                 \\
                               & (27) How is your workload?                                                         \\
                               & (28) How is your self-perceived performance?                                       \\ \hline
\multirow{9}{*}{Usefulness}    & (29) How do you perceive the usefulness of the system?                             \\
                               & (30) How do you perceive the usefulness of keyword highlighting?                   \\
                               & (31) How do you perceive the usefulness of remote visual control?                  \\
                               & (32) How do you perceive the usefulness of underpainting?                          \\
                               & (33) How do you perceive the usefulness of progress bars?                          \\
                               & (34) How do you perceive the usefulness of emoji?                                  \\
                               & (35) How do you perceive the usefulness of participle?                             \\
                               & (36) How do you perceive the usefulness of script polishing?                       \\
                               & (37) How do you perceive the usefulness of speech tracking?                        \\ \hline
Accuracy                       & (38) How do you perceive the accuracy of the whole system?                         \\ \hline
\multirow{9}{*}{Trust}         & (39) How do you perceive the trust level of the system?                            \\
                               & (40) How do you perceive the trust level of keyword highlighting?                  \\
                               & (41) How do you perceive the trust level of remote visual control?                 \\
                               & (42) How do you perceive the trust level of underpainting?                         \\
                               & (43) How do you perceive the trust level of progress bars?                         \\
                               & (44) How do you perceive the trust level of emoji?                                 \\
                               & (45) How do you perceive the trust level of participle?                            \\
                               & (46) How do you perceive the trust level of script polishing?                      \\
                               & (47) How do you perceive the trust level of speech tracking?                       \\ \hline
\end{tabular}
\end{table*}

\begin{figure*}[h]
  \centering
  \includegraphics[angle=90, height=0.85\textheight]{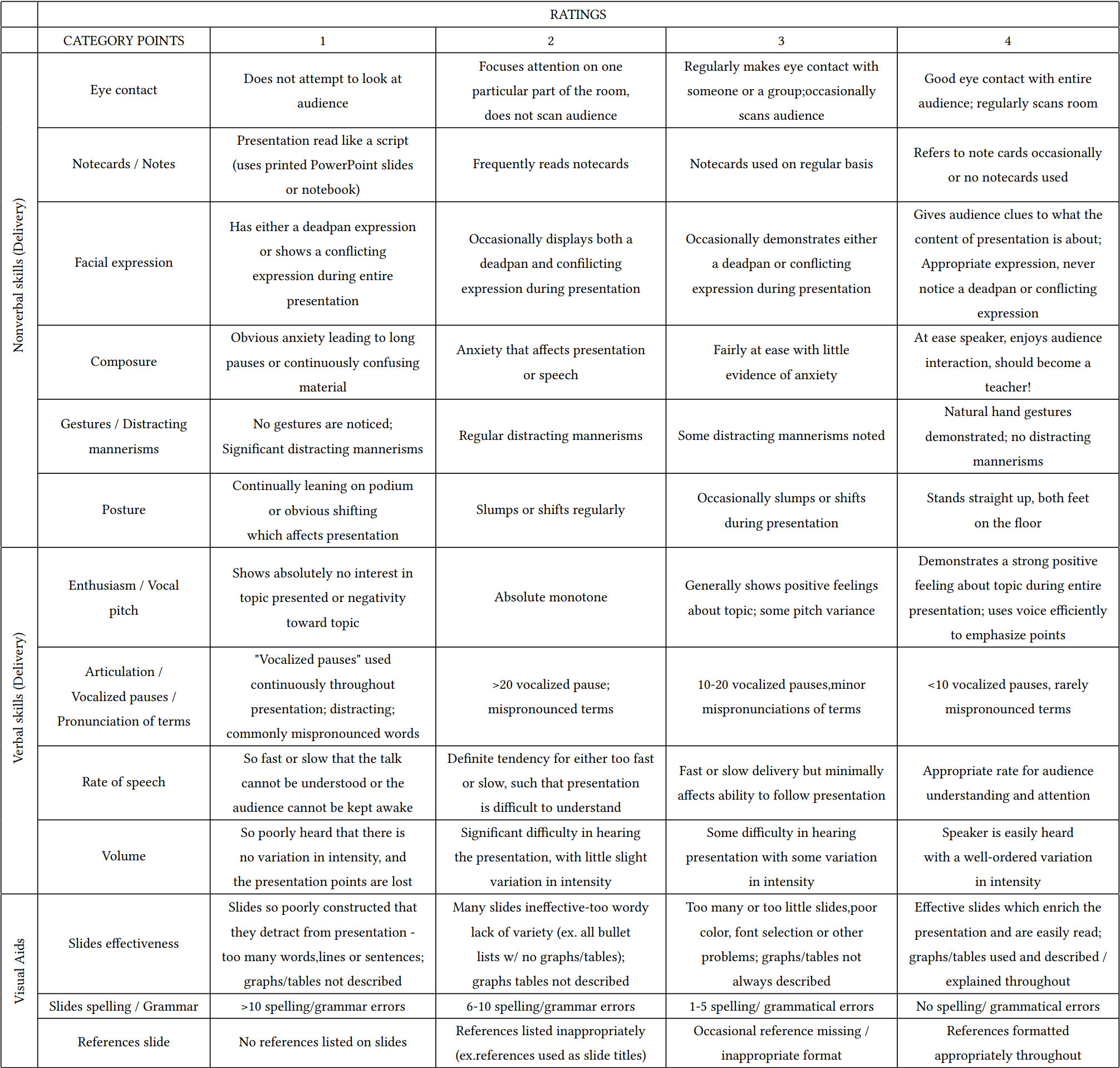}
  \caption{Evaluation form used in the user study.}
  \label{fig:Evaluation_form}
\end{figure*}

\begin{figure*}[h]
  \centering
  \includegraphics[angle=90, height=0.85\textheight]{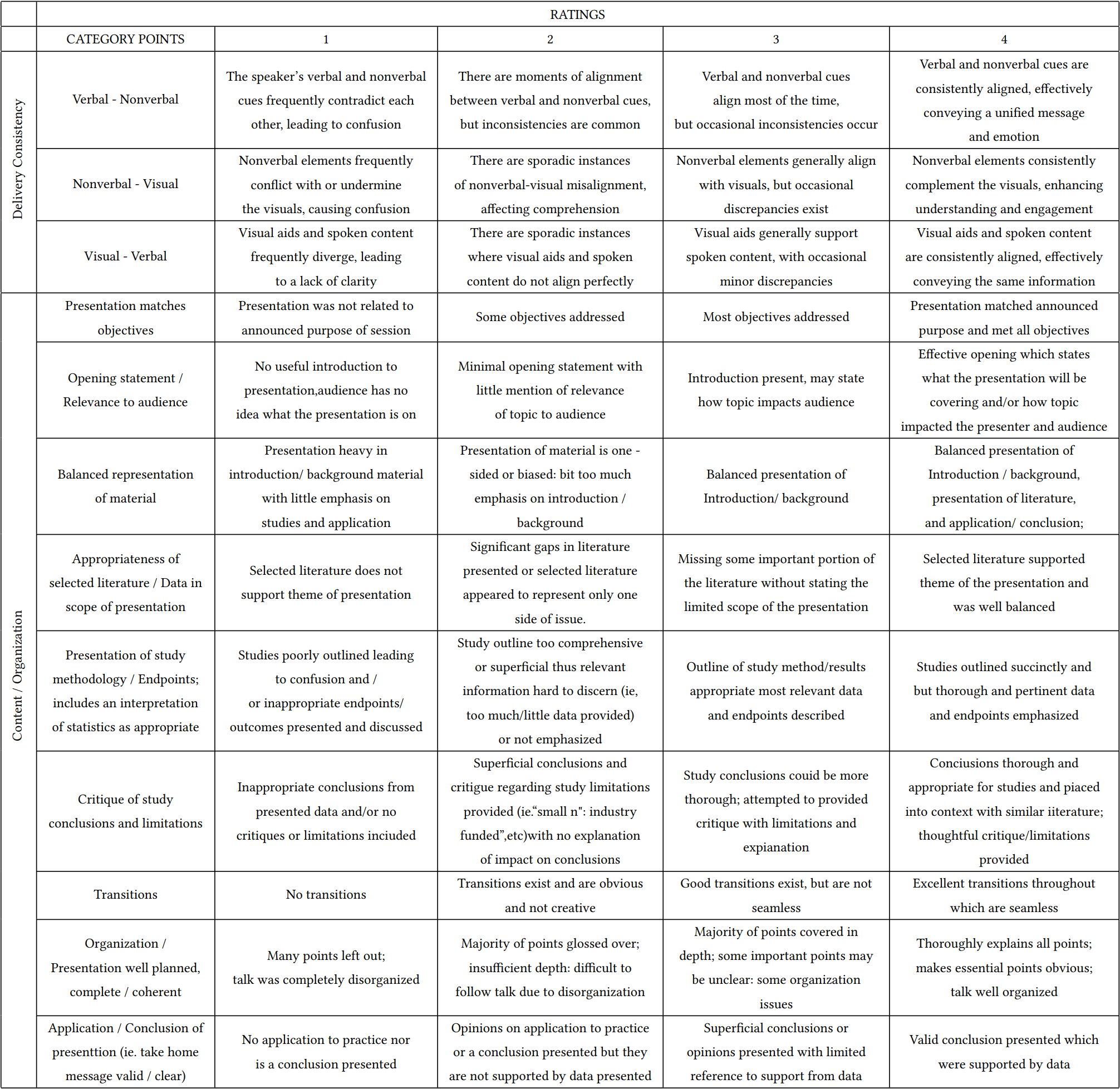}
  \caption{(continued) Evaluation form used in the user study.}
  \label{fig:Evaluation_form_2}
\end{figure*}

\begin{figure*}
  \centering
  \includegraphics[width=\textwidth]{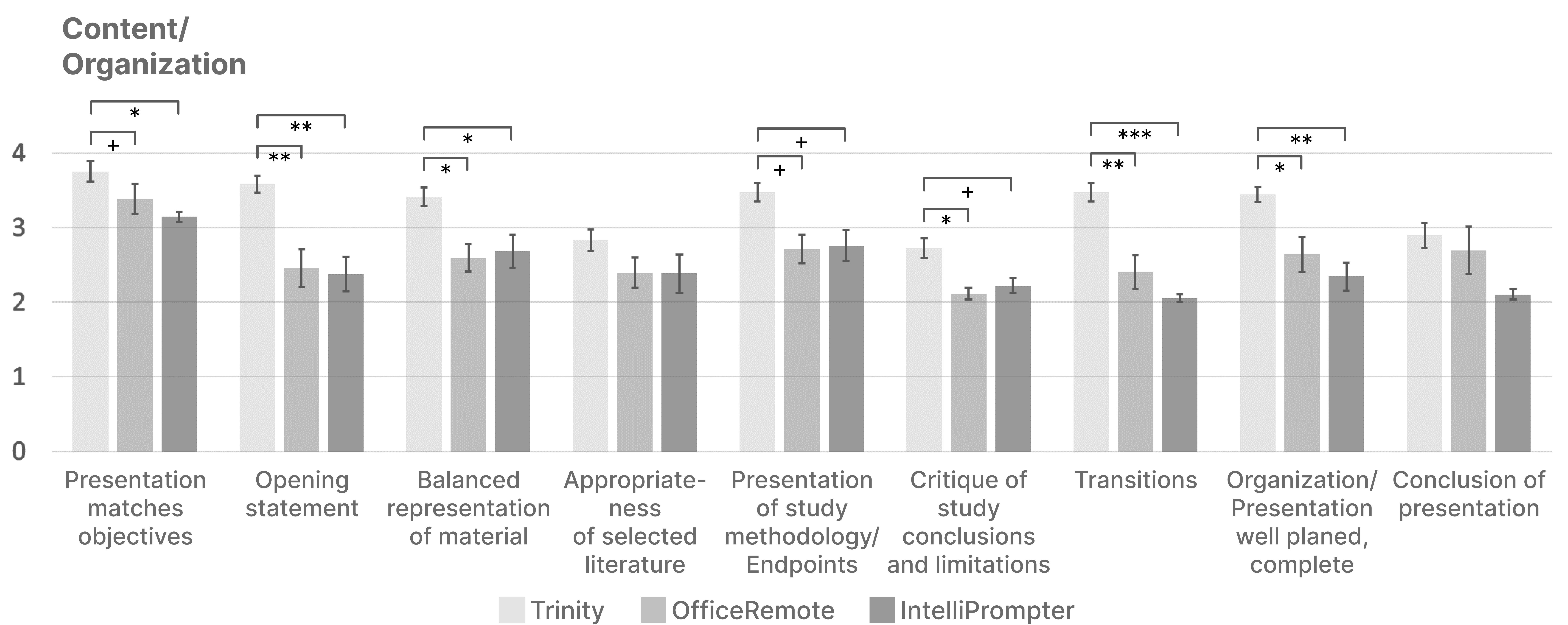}
    \vspace{-6mm}
  \caption{Audience's ratings on content and organization of students' presentations. The error bars indicate standard errors. ($+:p<.1$; $*:p<.05$; $**:p<.01$; $***:p<.001$)}
  \label{fig:Content_Organization}
  \vspace{-3mm}
\end{figure*}

\section{Minor Findings from User Study}
\label{appendix:Minor_findings}

\par \textbf{\prefix{\textsf{Minor F1}}: Presenters verify the polished script against their original manuscript.} \textit{Trinity} generates a polished script from the presenters' manuscript. We noticed that $10$ out of $11$ presenters switched between the manuscript and the polished script for confirmation, with $8$ out of $11$ presenters reviewing the entire polished script:
\begin{center}
\begin{minipage}{0.45\textwidth}
\textsf{\textit{``GPT is indeed powerful, but after all, it can't guarantee to fully align with your intentions, so I definitely need to double-check.'' (S2, male, age: 21)}}
\end{minipage}
\end{center}

\par \noindent \textbf{\prefix{\textsf{Minor F2}}: Presenters frequently check their smartphones during presentations.} Analyzing how presenters use their smartphones during the presentation is crucial. We verified these usage patterns by co-reviewing video replays with student presenters. In general, presenters in both the \textit{Trinity} and \textit{OfficeRemote} conditions frequently glanced at their smartphones, whereas those in the \textit{IntelliPrompter} condition primarily focused on their laptop screens. This suggests that \textit{Trinity} did not substantially alter their presentation approach, which primarily relies on scripts:
\begin{center}
\begin{minipage}{0.45\textwidth}
\textsf{\textit{``I know it's (referring scripts) not the best way, but scripts are essential for my speeches because if I speak in English, I might not be able to talk coherently without relying on them.'' (S3, male, age: 22)}}
\end{minipage}
\end{center}

\par \noindent \textbf{\prefix{\textsf{Minor F3}}: Students generally trust the system but not completely.} All presenters in the \textit{Trinity} condition rated their trust above the neutral point (i.e., $4$) on the 7-point Likert scale. They expressed trust in the system because there were no compelling reasons to doubt it and no obviously questionable or unexpected information, which aligns with findings from psychological literature known as the \textit{Truth-Default Theory}~\cite{levine2014truth}. However, their trust in the system was constrained by their limited knowledge of best practices and professional certification.
\begin{center}
\begin{minipage}{0.45\textwidth}
\textsf{\textit{``The polished script and provided delivery prompts seem to make sense, but I'm not sure if this is the best way to express or to do. So, I would trust the system more if it's endorsed by professionals.'' (S3, male, age: 22)}}
\end{minipage}
\end{center}


\end{document}